\documentclass[aps,prd,twocolumn,preprintnumbers,showpacs,nofootinbib,amssymb]{revtex4}
\usepackage{graphicx}
\usepackage{amsmath}
\usepackage{amssymb}
\usepackage{amsfonts}
\usepackage{bm}
\usepackage{color}

\def\be{\begin{equation}}
\def\ee{\end{equation}}
\def\bea{\begin{eqnarray}}
\def\eea{\end{eqnarray}}

\begin{document}

\title{Is the Universe logotropic?}
\author{Pierre-Henri Chavanis}
\email{chavanis@irsamc.ups-tlse.fr}
\affiliation{Laboratoire de Physique Th\'eorique,
Universit\'e Paul Sabatier, 118 route de Narbonne  31062 Toulouse, France}

\begin{abstract}
We consider the possibility that the  universe is made of a single dark
fluid  described by a logotropic equation  of state
$P=A\ln(\rho/\rho_*)$,
where $\rho$ is the rest-mass density, $\rho_*$ is a reference density, and $A$
is the logotropic temperature. The energy density $\epsilon$ is the sum of two
terms: a rest-mass energy term $\rho
c^2$ that mimics dark matter and an internal energy term
$u(\rho)=-P(\rho)-A$ that
mimics dark energy. This decomposition leads to a natural, and physical,
unification of dark matter and dark energy, and elucidates their mysterious
nature.  In
the early universe, the dark fluid
behaves as pressureless dark
matter ($P\simeq 0$, $\epsilon\propto a^{-3}$) and, in the late universe, it
behaves as dark energy ($P\sim -\epsilon$, $\epsilon\propto \ln a$). The
logotropic model depends on a single parameter
$B=A/\rho_{\Lambda}c^2$ (dimensionless logotropic temperature) where
$\rho_{\Lambda}=6.72\times 10^{-24}\, {\rm g}\, {\rm
m}^{-3}$ is the cosmological density. For $B=0$, we recover the $\Lambda$CDM
model with a different
justification. For $B>0$, we can describe deviations from the $\Lambda$CDM
model. Using cosmological constraints, we find that $0\le B\le 0.09425$.  We
consider the possibility that dark matter halos are described by the same
logotropic equation of state. When $B>0$,
pressure gradients prevent gravitational collapse and provide halo density cores
instead of cuspy density profiles, in agreement with the observations. The
universal rotation curve of logotropic dark matter halos is consistent with the
observational Burkert profile up to the halo
radius.  It decreases as $r^{-1}$ at large
distances,
similarly to the profile of dark matter halos close to the core radius
[Burkert, arXiv:1501.06604].  Interestingly,
if we assume that all the dark matter halos have the same logotropic temperature
$B$, we find that their surface density
$\Sigma=\rho_0 r_h$ is constant. This result is in agreement with the
observations [Donato {\it et al.}, MNRAS {\bf 397}, 1169 (2009)] where it is
found that $\Sigma_0=141\, M_{\odot}/{\rm pc}^2$ for
dark matter halos differing by several orders of magnitude in size. Using this
observational result, we obtain $B=3.53\times 10^{-3}$. Then, we show that the
mass enclosed within a sphere of fixed radius $r_{u}=300\, {\rm pc}$ has the same value
$M_{300}=1.93\times 10^{7}\, M_{\odot}$ for all the dwarf halos, in agreement
with the
observations [Strigari {\it et al.}, Nature {\bf 454}, 1096 (2008)].  Finally,
assuming that $\rho_*=\rho_P$, where $\rho_P=5.16\times 10^{99}\, {\rm g}\, {\rm
m}^{-3}$ is the Planck density, we {\it predict} $B=3.53\times 10^{-3}$,
in perfect agreement with the value obtained from the
observations. We approximately have $B\simeq
1/\ln(\rho_P/\rho_{\Lambda})\simeq 1/[123\ln(10)]$ where $123$ is the famous
number occurring in the ratio $\rho_P/\rho_{\Lambda}\sim 10^{123}$ between the
Planck density  and the
cosmological density. This
value of $B$ is sufficiently
low to satisfy the cosmological bound $0\le B\le 0.09425$ and sufficiently large
to differ from CDM ($B=0$) and avoid density cusps in dark matter halos.
It leads to a Jeans length at the beginning of the matter era of the order of
$\lambda_J=40.4\, {\rm pc}$ which is consistent with the minimum size of dark
matter halos observed in the universe. Therefore, a logotropic
equation of state is a good candidate to account both for galactic and
cosmological observations. This may be a hint that dark matter and dark energy
are the manifestation of a single dark fluid.  If we assume that the dark fluid
is made of a self-interacting scalar field, representing  for example
Bose-Einstein condensates, we find that the logotropic equation of state arises
from the Gross-Pitaevskii equation with an inverted quadratic potential, or from
the Klein-Gordon equation with a logarithmic potential. We
also relate the logotropic equation of state to Tsallis
generalized thermodynamics and to the Cardassian model (motivated by the
existence of extra-dimensions).

\end{abstract}


\maketitle

\section{Introduction}
\label{sec_intro}

The cosmological constant has an interesting history. Its pre-history dates back
to 1896, when von Seeliger \cite{seeliger} and Neumann \cite{neumann} introduced
an attenuation factor in the Newtonian potential, $\Phi=-G m e^{-\lambda r}/r$,
in order to resolve the non-convergence of the gravitational force in Newtonian
cosmology for an infinite homogeneous distribution of matter \cite{norton}. In
1917, Einstein
\cite{einsteincosmo} introduced a cosmological constant $\Lambda$ in the
equations of general relativity in order to obtain an infinite homogeneous static
universe.\footnote{As a preamble of his paper, in order to motivate the cosmological constant, Einstein
\cite{einsteincosmo} considered the Newtonian approximation and replaced the
Poisson equation by an equation of the form $\Delta\Phi-(\Lambda/c^2)\Phi=4\pi
G\rho$, so as to obtain a homogeneous static solution $\Phi=-4\pi G\rho c^2/\Lambda$. In
Einstein's belief, the Newtonian effect of the cosmological constant was to
shield the gravitational interaction on a distance $c/\sqrt{\Lambda}$, similarly to the proposal of Seeliger and Neumann. This is
actually  incorrect. Lema\^itre \cite{lemaitrecosmo} was the first to understand
that the cosmological constant can be interpreted as a force of ``cosmic
repulsion'' (see also \cite{eddington,trautman,spiegel,kjs}). In the Newtonian context, the  modified Poisson equation including
the cosmological constant is $\Delta\Phi=4\pi G\rho-\Lambda$, leading to the homogeneous
static solution $\rho=\Lambda/4\pi G$.} After
the discovery  of the expansion of the universe, predicted theoretically by
Friedmann \cite{friedmann1,friedmann2} and Lema\^itre \cite{lemaitre1,lemaitre2}, and ascertained by
the observations of Hubble \cite{hubble}, Einstein considered the cosmological
constant as ``the biggest blunder of his life'' \cite{blunder}, and banished it.
In 1932, Einstein and de Sitter \cite{eds} published a short paper describing
the expansion of a pressureless universe without cosmological constant. This is
the so-called Einstein-de Sitter (EdS) model. In this model, the universe
undergoes a decelerating expansion. The discovery of the acceleration of the
expansion of the universe  in recent years \cite{novae} revived the interest
in the cosmological constant since a positive cosmological constant can
precisely account for such an acceleration.\footnote{The effect of the
cosmological constant in an empty universe ($\rho=0$) was first considered by de
Sitter \cite{deSitter1,deSitter2} in the context of a {\it static} universe. It was
understood later by Lema\^itre \cite{lemaitre1925} that de Sitter's solution
actually describes a universe expanding exponentially rapidly with time.} At
present, the standard model of cosmology is based on a pressureless dark matter
fluid ($P=0$) with a positive cosmological constant ($\Lambda>0$). This is the
cold dark matter ($\Lambda$CDM) model. From the observations, the
cosmological constant has the value $\Lambda=1.13\times 10^{-35}\, {\rm
s}^{-2}$. The
cosmological constant $\Lambda$
is equivalent to a constant energy density $\epsilon_{\Lambda}=\Lambda c^2/8\pi
G=6.04\times 10^{-7}\, {\rm g}\,
{\rm m}^{-1}{\rm s}^{-2}$, called dark energy, associated with an equation
of state $P=-\epsilon$
involving a negative
pressure.
Some authors \cite{lemaitre1934,sakharov,zeldovichV} have proposed to interpret
the  cosmological constant in terms of the energy of the vacuum.
However, when
one tries to explain the cosmological constant in relation to the vacuum energy,
one is confronted to the
so-called ``cosmological constant problem'' \cite{weinbergcosmo,paddycosmo}
because
quantum field theory predicts that the vacuum energy density should be of
the order of the Planck density $\rho_P=c^5/\hbar G^2=5.16\times 10^{99}\, {\rm
g}\, {\rm
m}^{-3}$ which is $123$
orders of magnitude larger than the value
of the cosmological density $\rho_{\Lambda}=\epsilon_{\Lambda}/c^2=6.72\times
10^{-24}\, {\rm g}\, {\rm m}^{-3}$ deduced from the observations.
To circumvent this problem, some authors have proposed to abandon the
cosmological constant
$\Lambda$ and to explain the acceleration of the universe in terms of a dark
energy with a time-varying density 
associated to a scalar field called
``quintessence''
\cite{quintessence}. This mysterious concept of ``dark energy'' adds to the
other mysterious concept of ``dark matter'' necessary to account
for the missing mass of the galaxies inferred from the virial theorem
\cite{zwicky} and to explain their flat
rotation curves \cite{rubin,persic}. As an alternative to
quintessence, Kamenshchik {\it et al.}
\cite{kamenshchik} proposed a heuristic unification of dark matter and dark energy
in terms of an exotic fluid with an  equation of state $P=-A/\epsilon$, called
the  Chaplygin gas. This equation of state provides a model of universe that
behaves as a pressureless fluid (dark matter) at early times, and as a fluid
with a constant energy density  (dark energy) at late times, yielding an
exponential acceleration similar to the effect of the cosmological constant.
However, in the
intermediate regime of interest, this model does not give a good agreement with
the observations \cite{sandvik,zhu}. Therefore, generalizations of the Chaplygin
gas
have been considered \cite{chaplygingen}. We note
that an equation of state
corresponding to a constant negative pressure $P=-\epsilon_{\Lambda}$ also
provides a heuristic unification of dark matter and dark energy, and is simpler
\cite{cosmopoly1,cosmopoly2,cosmopoly3,aip}.
Furthermore, this equation of state is consistent with the observations since it
gives the same results as the  $\Lambda$CDM model although its physical
justification is different. Therefore, we can realize a simple unification of
dark matter and dark energy by assuming that the universe is filled with a
single dark fluid with a constant pressure. However, this model has two
drawbacks. Since it has no free parameter, it cannot account for (small)
deviations from the $\Lambda$CDM model. On the other hand, a constant equation
of state cannot describe dark matter halos. Indeed, since there is no pressure
gradient ($\nabla P={\bf 0}$), this model behaves similarly to the CDM model
where $P=0$, and generates cuspy density profiles at the center
of the halos  \cite{nfw}.
This is problematic because observations of dark matter halos favor density cores
instead of cusps \cite{burkert,observations}. Cores can be naturally
explained by assuming that
dark matter halos are collisional and that they are described by an equation of
state $P(\rho)$ that strictly
increases with the density so that pressure gradients can counterbalance the
gravitational attraction. Therefore, if we want to realize a successful
unification of dark matter and dark energy, the idea is to consider a dark fluid
with an equation of state $P(\rho)$ increasing slowly with the density.

A natural idea is to consider a polytropic equation of state where the pressure
is a power of the density with a small index $\gamma\simeq 0$. There are two
types of polytropic equations of state in general relativity, as first realized
by Tooper \cite{tooper1,tooper2} in the context of general relativistic
stars.\footnote{The
cosmological implications of these two types
of polytropic models have been recently discussed by the author in
\cite{mlbec,stiff}. Similar arguments have been developed independently in
\cite{spyrou} without reference to the work of Tooper.} Model I corresponds to
an equation of state of the form $P=K\epsilon^{\gamma}$ with $\gamma=1+1/n$,
where the pressure is a power of the energy density. This model has been studied
in detail in \cite{cosmopoly1,cosmopoly2,cosmopoly3,aip} in a cosmological
context. It can be viewed as a generalization of the Chaplygin gas. In this
model, when $K<0$ and $\gamma<1$ (i.e. $n<0$) \cite{cosmopoly2}, the dark fluid
behaves as a pressureless fluid (dark matter) in the early universe and as a
fluid with a constant energy density (dark energy)  in the late
universe.\footnote{When $K<0$ and $\gamma>1$ (i.e. $n>0$) \cite{cosmopoly1},
this polytropic model describes a phase of early inflation followed by a
decelerating expansion (modeling, e.g., radiation). In this model, the
temperature is initially very low, increases exponentially rapidly during the
inflation up to the Planck temperature (GUT scale), then decreases
algebraically. By contrast, in the usual inflationary scenario
\cite{guth1,guth2,guth3,linde}, the temperature decreases during the inflation
and one must advocate a phase of re-heating (not very well understood) to
account for the observations. On the other hand, the polytropic model generates
a value of the entropy as large as $S/k_B=5.04\times 10^{87}$ \cite{cosmopoly1}.
It is also
possible to introduce a generalized polytropic model based on a quadratic
equation of state that describes simultaneously the early inflation, the
intermediate decelerating expansion, and the late acceleration of the universe
\cite{quadratic}.} However, this distinction is only asymptotic and, in the
intermediate regime, it is not possible to separate these two components
unambiguously because they do not appear additively in the energy density
$\epsilon$ unless $\gamma=0$, which is  equivalent to the $\Lambda$CDM model.
Model II
corresponds to an equation of state of the form $P=K\rho^{\gamma}$ with
$\gamma=1+1/n$, where the pressure is a power of the rest-mass density. In this
model, the energy density $\epsilon$ is the sum of two terms: a rest-mass energy
$\rho c^2$ and an internal energy $u(\rho)=P(\rho)/(\gamma-1)$. Since
$\rho\propto a^{-3}$, the
rest mass term mimics dark matter. When $K<0$ and
$\gamma<1$
(i.e. $n<0$), the internal energy term mimics dark energy. This decomposition
leads to a natural, and physical, unification of dark matter and dark energy,
and elucidates their mysterious nature. In order to account for the
observations, these polytropic models must be relatively close to the
$\Lambda$CDM model, equivalent to a constant negative pressure. This  implies
that the polytropic index $\gamma$ must be close to zero. For  $\gamma=0$ (i.e.
$n=-1$), we recover the $\Lambda$CDM model exactly. For $\gamma$ different from
zero, we can study deviations from the $\Lambda$CDM model. The polytropic models
have been confronted to the observations in order to constrain the values of the
admissible index $\gamma$ or $n$. For Model I, it is found that
$n=-1.05^{+0.15}_{-0.16}$ \cite{karami}. For Model II, it is found that
$-0.089<\gamma\le 0$ (corresponding to $-1\le n<-0.92$) \cite{spyrou}. Although
these polytropic models are interesting in themselves, we anticipate a problem
in order to {\it justify} them from a fundamental theory. Actually, it is
possible to justify a polytropic equation of state by assuming that the dark
fluid is made of a self-interacting scalar field representing, for example,
Bose-Einstein condensates (BECs). A polytropic equation of state with
$\gamma\neq 0$
arises from the Gross-Pitaevskii (GP) equation with a power-law potential
$|\psi|^{2(\gamma-1)}$ or from the Klein-Gordon (KG) equation with a power-law
potential $|\phi|^{2\gamma}$. However, it is  hard to imagine how one can
justify from fundamental arguments why the universe is described by a polytropic
equation of state with a ``curious'' index equal, for example, to
$\gamma=-0.089$. The most natural value of the polytropic index is $\gamma=0$
corresponding to a constant pressure but, in that case, the GP and KG equations
degenerate. Furthermore, a polytropic equation of state with $\gamma=0$ has no
pressure gradient, which is problematic to account for the flat core of dark
matter halos as previously mentioned.

In this paper, we develop another idea that was sketched in Appendix B of
\cite{stiff}. We propose to describe the universe by a single dark fluid with a
logotropic equation of state of the form $P=A\ln(\rho/\rho_*)$, where $\rho$ is
the rest-mass density, $\rho_*$ is a reference density, and  $A$ is the
logotropic temperature. This equation of state was introduced phenomenologically in astrophysics 
by McLaughlin and  Pudritz \cite{pud} to describe the internal structure and the
average properties of molecular clouds and clumps. It was also studied by
Chavanis and Sire \cite{logo} in the context of Tsallis generalized
thermodynamics \cite{tsallisbook} where it was shown to correspond to a
polytropic equation of state with $\gamma\rightarrow 0$ and $K\rightarrow \infty$ in such a way
that $A=\gamma K$ is finite.\footnote{In the framework of generalized
thermodynamics, one can really interpret $A$ as a generalized temperature
associated with a generalized entropy called Log-entropy
\cite{logo} (see Appendix \ref{sec_log}).} This
limit is very relevant to the present cosmological context because we precisely
require an equation of state with a polytropic exponent close to $\gamma=0$.
{\it Therefore, the logotropic equation of state is a natural, and serious,
candidate for the unification of dark matter and dark energy.} Since the
pressure must increase with the density, the logotropic temperature must be
strictly positive ($A>0$) which is satisfying from a thermodynamical point of
view. In the early universe, the dark fluid behaves as pressureless dark matter
($P\simeq 0$, $\epsilon\propto a^{-3}$) and, in the late universe, it behaves as 
dark energy ($P\sim -\epsilon$, $\epsilon\propto \ln a$). The logotropic model
depends on a single parameter $B=A/\rho_{\Lambda}c^2$ (dimensionless
logotropic
temperature), where $\rho_{\Lambda}$ is the cosmological density. For $B=0$, we
recover the $\Lambda$CDM model with a different
justification. For $B>0$, we can describe deviations from the $\Lambda$CDM
model. Using cosmological constraints, we find that $0\le B\le 0.09425$. For
$B=0.09425$, the universe is normal for $t<20.7\, {\rm Gyrs}$ (the energy
density decreases as the scale factor increases) and becomes phantom afterwards
(the energy density increases with the scale factor). Actually, the logotropic
model breaks down before entering in the phantom regime because the velocity of
sound exceeds  the velocity of light when $t>17.3\, {\rm Gyrs}$. However, this
moment is quite remote in the future.

Since the logotropic equation of state is expected to provide a unification of
dark energy and dark matter, it should also describe the structure of dark
matter halos. In this respect, the logotropic equation of state has several nice
properties. First of all, since the pressure increases with the density,
pressure gradients prevent gravitational collapse and provide halo density cores
instead of cuspy density profiles, in agreement with the
observations \cite{burkert,observations}. On the other hand, the universal
rotation
curve of logotropic dark matter halos is consistent with the observational
Burkert profile \cite{burkert} up to the halo radius.\footnote{The
logotropic
rotation curve continues to increase at larger distances while the Burkert
rotation curve decreases. This feature comes from the fact that the logotropic
density decreases as $r^{-1}$ while the Burkert density
decreases as $r^{-3}$. Interestingly, we note that the profile of dark matter
halos close to the core radius decreases as $r^{-1}$
\cite{burkertnew}  in agreement with the logotropic profile. On the other hand,
some galaxies such as low surface brightness (LSB) galaxies have rotation curves that strictly  increase with the distance. These galaxies are particularly well
isolated and little affected by tidal effects. In other cases, tidal
effects or complex physical processes (e.g. incomplete relaxation) have to be
taken into account. This can affect the behavior of the density profile and of
the rotation curve at large distances and explain how we pass from a $r^{-1}$
behavior to a $r^{-3}$ behavior.} Finally, and very interestingly, if we assume
that all the dark matter halos have the same logotropic temperature $B$ (a
consequence of our model), we find that their surface density $\Sigma=\rho_0
r_h$ is constant. This result is in agreement with the observations, where it
is found that  $\Sigma_0$ has the same value $\Sigma_0=141\, M_{\odot}/{\rm
pc}^2$ for dark matter halos differing by several orders of magnitude in size
\cite{kormendy,spano,donato}, a result unexplained so far. Using this
observational result, we obtain $B=3.53\times 10^{-3}$.  Then, we show that the
mass enclosed within a sphere of fixed radius $r_{u}=300\, {\rm pc}$ has the same value
$M_{300}=1.93\times 10^{7}\, M_{\odot}$ for all the dwarf  halos, in agreement
with the
observations of Strigari {\it et al.} \cite{strigari}. This so far unexplained
result is a direct consequence of the logotropic equation of state. Finally,
assuming that $\rho_*=\rho_P$, where $\rho_P$ is
the Planck density, we {\it predict} $B=3.53\times 10^{-3}$,
in perfect agreement with the value  obtained from the
observations. We approximately have $B\simeq
1/\ln(\rho_P/\rho_{\Lambda})\simeq 1/[123\ln(10)]$ where $123$ is the famous
number occurring in the ratio $\rho_P/\rho_{\Lambda}\sim 10^{123}$ between the
Planck density $\rho_P=5.16\times 10^{99}\, {\rm g}\, {\rm m}^{-3}$  and the
cosmological density $\rho_{\Lambda}=6.72\times 10^{-24}\, {\rm g}\, {\rm
m}^{-3}$. This value of $B$ is sufficiently low to satisfy the cosmological bound $0\le B\le
0.09425$ and sufficiently large to differ from CDM ($B=0$) and avoid density
cusps in dark matter halos. Therefore, a logotropic equation of state is a good candidate to account
both for galactic and cosmological observations. This may be a hint that dark
matter and dark energy are the manifestation of a single dark fluid.

Finally, it may be less problematic to theoretically justify the logotropic
model than a polytropic model with an index $\gamma=-0.089$ for example
\cite{spyrou}. Indeed, the logotropic equation of state has a universal
functional form (logarithmic) while the polytropic model depends on an unknown
index $\gamma$. As shown in detail below, the logotropic model removes the
degeneracy of the polytropic model when $\gamma=0$. Therefore, there is some
hope to justify the logotropic model from a fundamental field theory. If we
assume that the dark fluid is made of a self-interacting scalar field,
representing for example BECs, we find that the logotropic
equation of state arises from the GP equation with an inverted quadratic
potential $|\psi|^{-2}$ or from the KG equation with a logarithmic potential
$\ln|\phi|$. To our knowledge, these equations have not been considered
previously. They are, of course, interesting in themselves and our approach
strongly suggests that they may have important applications in cosmology.

The paper is organized as follows. In Sec. \ref{sec_cosm}, we review different
cosmological models that have been proposed in the past. We start with
well-known models (Einstein-de Sitter, $\Lambda$CDM, Chaplygin) then discuss
less well-known ones (polytropic models of type I and II, Cardassian model) that
are important in our study. In Sec. \ref{sec_logojust}, we introduce the
logotropic model. In Sec. \ref{sec_lc}, we study the logotropic equation of
state in a cosmological setting, as a unification of dark matter and dark
energy, and obtain a cosmological bound $0\le B\le 0.09425$ for the normalized
logotropic temperature $B$. In Sec. \ref{sec_ldm}, we apply the logotropic
equation of state to the structure of dark matter halos and obtain $B=3.53\times
10^{-3}$ from observations. In Sec. \ref{sec_pred}, we  predict $B=3.53\times
10^{-3}$ from theoretical considerations. In Appendix \ref{sec_gr}, we provide
general results concerning the thermodynamics of a relativistic fluid.
In Appendix \ref{sec_sa}, we provide a simple argument to justify the logotropic
equation of state and mention its relation to a fifth force. In Appendix
\ref{sec_ft}, we develop a field theory for the dark fluid and
determine the form of the GP and KG equations that generate a logotropic
equation of state, as well as the corresponding generalized entropy.

{\it Remark:} Secs.  \ref{sec_cosm} and \ref{sec_logojust} show the close
connections between different cosmological models (many of these connections 
having not been discussed before) and explain the physical reasons why we
consider a logotropic equation of state. These sections are important to
motivate our study. They also introduce our general formalism. However,
the reader just interested by the results may skip these sections and go
directly to Sec. \ref{sec_lc} for the application of the logotropic equation of
state to cosmology and to Sec. \ref{sec_ldm} for the application of the
logotropic equation of state to dark matter halos.

\section{Cosmological models}
\label{sec_cosm}

\subsection{The Friedmann equations}
\label{sec_friedmann}

We assume that the universe is homogeneous and isotropic, and contains a uniform perfect fluid of energy density $\epsilon(t)$ and isotropic pressure $P(t)$. The radius of curvature of the $3$-dimensional space, or scale factor, is noted $a(t)$ and the curvature of space is noted $k$. The universe is closed if $k>0$, flat if $k=0$, and open if $k<0$. We assume that the universe is flat ($k=0$) in agreement with the observations of the cosmic microwave background (CMB) \cite{bt}. In that case, the Einstein equations can be written as \cite{weinbergbook}:
\begin{equation}
\frac{d\epsilon}{dt}+3\frac{\dot a}{a}(\epsilon+P)=0,
\label{f1}
\end{equation}
\begin{equation}
\frac{\ddot a}{a}=-\frac{4\pi G}{3c^2}\left(\epsilon+3P\right )+\frac{\Lambda}{3},
\label{f2}
\end{equation}
\begin{equation}
H^2=\left (\frac{\dot a}{a}\right )^2=\frac{8\pi G}{3c^2}\epsilon+\frac{\Lambda}{3},
\label{f3}
\end{equation}
where we have introduced the Hubble parameter $H=\dot a/a$ and accounted for a possible non-zero cosmological constant $\Lambda$.

Eqs. (\ref{f1})-(\ref{f3}) are the well-known Friedmann equations describing a non-static universe. Among these three equations, only two are independent. The first equation can be viewed as an equation of continuity. For a given barotropic equation of state $P=P(\epsilon)$, it determines the relation between the energy density $\epsilon$ and the scale factor $a$. Then, the evolution of the scale factor $a(t)$ is given by Eq. (\ref{f3}).

Introducing the equation of state parameter $w=P/\epsilon$, and assuming $\Lambda=0$, we see from Eq. (\ref{f2}) that the universe is decelerating if $w>-1/3$ (strong energy condition) and accelerating if $w<-1/3$. On the other hand, according to Eq. (\ref{f1}), the energy density decreases with the scale factor if $w>-1$ (null dominant energy condition) and increases with the scale factor if $w<-1$. The latter case corresponds to a ``phantom'' universe \cite{phantom}.

\subsection{Einstein-de Sitter model}
\label{sec_eds}

We consider a universe made of pressureless matter (dust) with an equation of state
\begin{equation}
P=0,
\label{eds1}
\end{equation}
and we assume that $\Lambda=0$. This is the Einstein-de Sitter model \cite{eds}. The equation of continuity (\ref{f1}) can be integrated into
\begin{equation}
\epsilon=\epsilon_0\left (\frac{a_0}{a}\right )^3,
\label{eds2}
\end{equation}
where the subscript $0$ refers to present-day values. The Friedmann equation (\ref{f3}) writes
\begin{equation}
\left (\frac{\dot a}{a}\right )^2=\frac{8\pi G\epsilon_0}{3c^2} \left (\frac{a_0}{a}\right )^3,
\label{eds3}
\end{equation}
and it can be integrated into
\begin{equation}
\frac{a}{a_0}=\left (\frac{3}{2}H_0 t\right )^{2/3},\qquad \epsilon=\frac{c^2}{6\pi G t^2},
\label{eds4}
\end{equation}
where $H_0=(8\pi G\epsilon_0/3c^2)^{1/2}$ is the present value of the Hubble
constant. From observations $H_0=70.2 \, {\rm km}\,  {\rm s}^{-1}\, {\rm
Mpc}^{-1}=2.275\, 10^{-18} \, {\rm s}^{-1}$ yielding
$\epsilon_0=8.32\times 10^{-7}\, {\rm g}\, {\rm m}^{-1}{\rm
s}^{-2}$. In this model, the universe is
always decelerating and its age is
\begin{equation}
t_0^{\rm EdS}=\frac{2}{3H_0}.
\label{eds5}
\end{equation}
Numerically, $t_0^{\rm EdS}=9.29\, {\rm Gyrs}$.

\subsection{$\Lambda$CDM model}
\label{sec_lcdm}

We still consider a universe made of pressureless matter (dust) with $P=0$ but, in order to account for the acceleration of the expansion of the universe,  we now assume that $\Lambda>0$. This is the $\Lambda$CDM model. The Friedmann equation (\ref{f3}) now writes
\begin{equation}
\left (\frac{\dot a}{a}\right )^2=\frac{8\pi G\epsilon_0}{3c^2} \left (\frac{a_0}{a}\right )^3+\frac{\Lambda}{3}.
\label{lcdm1}
\end{equation}
As discussed in the next section, this equation can be interpreted in terms of dark matter and dark energy, and it can be solved analytically.

\subsection{Dark matter and dark energy}
\label{sec_dmde}

As is clear from Eq. (\ref{f3}), the cosmological constant is equivalent to a
fluid with a constant energy density\footnote{This energy was initially
interpreted as the
vacuum energy \cite{lemaitre1934,sakharov,zeldovichV}. However, from particle
physics and quantum field theory, the vacuum energy is expected to be of the
order of the Planck density $\rho_P=5.16\times 10^{99}\, {\rm g}\, {\rm m}^{-3}$
while cosmological observations give $\rho_{\Lambda}=6.72\times 10^{-24}\, {\rm
g}\, {\rm m}^{-3}$. These densities differ by the huge ratio
$\rho_P/\rho_{\Lambda}\sim 10^{123}$. This is the so-called cosmological
constant problem \cite{weinbergcosmo,paddycosmo}.}
\begin{equation}
\epsilon_{\Lambda}=\rho_{\Lambda}c^2=\frac{\Lambda c^2}{8\pi G}.
\label{dmde1}
\end{equation}
According to Eq. (\ref{f1}), the equation of state leading to a constant energy
density is $P=-\epsilon$. Instead of introducing a cosmological constant
$\Lambda$, we can consider that the universe is made of two fluids: a
pressureless dark matter fluid with an equation of state $P=0$ leading to a dark
matter density $\epsilon_m=\epsilon_{m,0}(a_0/a)^3$, and a dark energy fluid
with an equation of state
\begin{equation}
P=-\epsilon
\label{dmde2}
\end{equation}
leading to a constant density
identified with $\epsilon_{\Lambda}=\rho_{\Lambda}c^2=6.04\times 10^{-7}\, {\rm
g}\, {\rm m}^{-1}{\rm s}^{-2}$. In this context, $\epsilon_{\Lambda}$ is called
the dark energy density, or the cosmological density. The total energy density
of the universe ($\epsilon=\epsilon_m+\epsilon_{\Lambda}$) is
therefore\footnote{In this paper, we consider a sufficiently old universe so
that radiation can be neglected. On the other hand, baryonic matter can be
treated as a pressureless fluid ($P_b=0$) whose contribution adds to the
contribution of dark matter. We can take it into account in the previous
expressions by considering that $\epsilon_m$ refers to dark matter {\it and}
baryonic matter.}
\begin{equation}
\epsilon=\epsilon_{0,m}\left (\frac{a_0}{a}\right )^3+\epsilon_{\Lambda}.
\label{dmde3}
\end{equation}
As a result, the Friedmann equation (\ref{f3}) can be written as
\begin{equation}
\frac{H}{H_0}=\sqrt{\frac{\Omega_{m,0}}{(a/a_0)^3}+\Omega_{\Lambda,0}},
\label{dmde5}
\end{equation}
where we have introduced the present fractions of dark matter and dark energy $\Omega_{m,0}=\epsilon_{m,0}/\epsilon_0$ and $\Omega_{\Lambda,0}=\epsilon_{\Lambda}/\epsilon_0$. By construction, $\Omega_{m,0}+\Omega_{\Lambda,0}=1$. From observations $\Omega_{m,0}=0.274$ and
$\Omega_{\Lambda,0}=0.726$. Eq. (\ref{dmde5}) has a well-known analytical solution
\begin{equation}
\frac{a}{a_0}=\left (\frac{\Omega_{m,0}}{\Omega_{\Lambda,0}}\right )^{1/3}\sinh^{2/3}\left (\frac{3}{2}\sqrt{\Omega_{\Lambda,0}}H_0 t\right ),
\label{dmde6}
\end{equation}
\begin{equation}
\frac{\epsilon}{\epsilon_0}=\frac{\Omega_{\Lambda,0}}{\tanh^2 \left (\frac{3}{2}\sqrt{\Omega_{\Lambda,0}}H_0 t\right )}.
\label{dmde7}
\end{equation}
For early times, we recover the EdS solution (with modified coefficients):
\begin{equation}
\frac{a}{a_0}\sim \left (\frac{3}{2}\sqrt{\Omega_{m,0}} H_0 t\right )^{2/3},\qquad \frac{\epsilon}{\epsilon_0}\sim \frac{4}{9H_0^2 t^2},
\label{dmde8}
\end{equation}
and for late times, we recover the de Sitter solution
\begin{equation}
\frac{a}{a_0}\sim \left (\frac{\Omega_{m,0}}{4\Omega_{\Lambda,0}}\right )^{1/3} e^{\sqrt{\Omega_{\Lambda,0}}H_0 t},\qquad \epsilon\simeq  \epsilon_{\Lambda}.
\label{dmde9}
\end{equation}
In the  $\Lambda$CDM model, the age of the universe is
\begin{equation}
t_0=\frac{2}{3H_0\sqrt{\Omega_{\Lambda,0}}}\sinh^{-1}\left\lbrack \left  (\frac{\Omega_{\Lambda,0}}{\Omega_{m,0}}\right )^{1/2}\right\rbrack.
\label{dmde10}
\end{equation}
Numerically, $t_0^{\rm \Lambda CDM}=13.8\, {\rm Gyrs}$.

In this model, the universe is decelerating at early times and accelerating at late times. The moment at which the universe starts accelerating  ($\ddot a\ge 0$) takes place when $\epsilon_c=3\epsilon_{\Lambda}$, i.e. $a_c/a_0=(\Omega_{m,0}/2\Omega_{\Lambda,0})^{1/3}=0.574$.\footnote{This is most easily seen by considering Eq. (\ref{f2}) with $P=0$ and $\Lambda>0$ leading to $\epsilon_m=2\epsilon_{\Lambda}$, hence $\epsilon=3\epsilon_{\Lambda}$.} This corresponds to a time
\begin{equation}
t_c=\frac{2}{3H_0\sqrt{\Omega_{\Lambda,0}}}\sinh^{-1}\left ( \frac{1}{\sqrt{2}}\right ).
\label{dmde11}
\end{equation}
Numerically, $t_c^{\rm \Lambda CDM}=7.18\, {\rm Gyrs}$. The transition between the dark matter era and the dark energy era takes place when $\epsilon_m=\epsilon_{\Lambda}$, hence $\epsilon_2=2\epsilon_{\Lambda}$, i.e. $a_2/a_0=(\Omega_{m,0}/\Omega_{\Lambda,0})^{1/3}=0.723$. This corresponds to a time
\begin{equation}
t_2=\frac{2}{3H_0\sqrt{\Omega_{\Lambda,0}}}\sinh^{-1}(1).
\label{dmde11b}
\end{equation}
Numerically, $t_2^{\rm \Lambda CDM}=9.61\, {\rm Gyrs}$.

\subsection{The Chaplygin gas}
\label{sec_chap}

Kamenshchik {\it et al.} \cite{kamenshchik} have proposed to unify dark matter and dark energy by a single dark fluid described by the equation of state
\begin{equation}
P=-\frac{A}{\epsilon}.
\label{chap1}
\end{equation}
This is called the Chaplygin gas because this equation of state was introduced by Chaplygin \cite{chaphist} as a convenient soluble model to study the lifting force on a plane wing in aerodynamics. This equation of state has also a connection with string theory and admits a supersymmetric generalization (see \cite{kamenshchik} for details). For the equation of state (\ref{chap1}), the equation of continuity (\ref{f1}) can be integrated into
\begin{equation}
\epsilon=\sqrt{A}\left\lbrack \left (\frac{a_*}{a}\right )^{6}+1\right\rbrack^{1/2},
\label{chap2}
\end{equation}
where $a_*$ is a constant of integration. Eq. (\ref{chap2}) can be viewed as a
combination of dark matter and dark energy. For $a\rightarrow 0$, we have
$\epsilon\propto a^{-3}$ which behaves as pressureless dark matter. For
$a\rightarrow +\infty$, we have $\epsilon\rightarrow \sqrt{A}$ which behaves as
dark energy with a constant energy density $\epsilon_{\Lambda}=\sqrt{A}$,
equivalent to a cosmological constant (see Secs. \ref{sec_lcdm}
and \ref{sec_dmde}). Although this model has an important historical
interest (and can be motivated by string theory), it does not give a good
agreement with the observations \cite{sandvik,zhu}. Therefore, generalizations
of
the Chaplygin gas model have been considered.

\subsection{Constant pressure}
\label{sec_cp}

As a first step, we consider a single dark fluid  described by an equation of
state corresponding to a  constant negative pressure
\begin{equation}
P=-\epsilon_{\Lambda}.
\label{cp1}
\end{equation}
The equation of continuity (\ref{f1}) can be integrated into
\begin{equation}
\epsilon=\epsilon_{\Lambda}\left\lbrack \left (\frac{a_*}{a}\right )^{3}+1\right\rbrack,
\label{cp2}
\end{equation}
where $a_*$ is a constant of integration. This model also provides a unification of dark matter and dark energy, and is simpler than the Chaplygin gas \cite{cosmopoly1,cosmopoly2,cosmopoly3,aip}. Actually, this model gives the same result as the  $\Lambda$CDM model of  Secs. \ref{sec_lcdm} and \ref{sec_dmde} although its physical justification is different. As a result, this model agrees with the cosmological observations, unlike the Chaplygin gas. In order to describe (small) deviations to the $\Lambda$CDM model, and in order to describe simultaneously the large scale structure of the universe and  dark matter halos (as discussed in the Introduction), we should consider equations of state that are close to, but different from, a  constant pressure. This is the aim of the polytropic and logotropic equations of state considered in the following sections.

\subsection{Polytropic model of type I}
\label{sec_one}

We assume that the universe is made of a single dark fluid. We can specify its equation of state in the form $P=P(\epsilon)$, where $\epsilon$ is the energy density. We consider a polytropic equation of state
\begin{equation}
P=K \left (\frac{\epsilon}{c^2}\right )^{\gamma},\qquad \gamma=1+\frac{1}{n},
\label{one1}
\end{equation}
in which the pressure is a power of the energy density. This corresponds to the
polytropic equation of state considered in the first paper of Tooper
\cite{tooper1} in the context of general relativistic stars. Therefore, we shall
call it polytropic model of type I or, simply,  Model I. This polytropic model
has been studied in a cosmological context in
\cite{cosmopoly1,cosmopoly2,cosmopoly3,aip}. It can be viewed as a
generalization of the Chaplygin gas model.

For the equation of state (\ref{one1}), the equation of continuity (\ref{f1}) can be integrated analytically. Assuming $w\equiv P/\epsilon\ge -1$ (non-phantom), $K<0$, and $n<0$ (i.e. $\gamma<1$), the relation between the energy density and the scale factor can be written as \cite{cosmopoly2}:
\begin{equation}
\epsilon=\epsilon_{*}\left\lbrack \left (\frac{a_*}{a}\right )^{3/|n|}+1\right\rbrack^{|n|},
\label{one2}
\end{equation}
where $\epsilon_{*}=(|K|/c^2)^{|n|}c^2$ and $a_*$ is a constant of integration. For $n=-1/2$, $\gamma=-1$, and $K=-A/c^2$,  we recover the Chaplygin gas model (\ref{chap1}) and (\ref{chap2}). For $n=-1$, $\gamma=0$, and $K=-\epsilon_{\Lambda}$, we recover the constant pressure model (\ref{cp1}) and (\ref{cp2}), equivalent to the $\Lambda$CDM model. More generally, Eq. (\ref{one2}) can be viewed as a combination of dark matter and dark energy. Indeed, for $a\rightarrow 0$, we have $\epsilon\sim\epsilon_* (a_*/a)^3$ which behaves as pressureless dark matter. For $a\rightarrow +\infty$, we have $\epsilon\rightarrow \epsilon_{*}$ which behaves as dark energy. This allows us to make the identification  $\epsilon_*=\epsilon_{\Lambda}$. In this analogy, using Eq. (\ref{dmde1}) and the relation following Eq. (\ref{one2}), we find that the polytropic constant $K$ is equivalent to a cosmological constant
 \begin{equation}
\Lambda=8\pi G \left (\frac{|K|}{c^2}\right )^{|n|}.
\label{one3}
\end{equation}

Introducing present-day variables, we can rewrite Eq. (\ref{one2}) as
\begin{equation}
\epsilon=\epsilon_{\Lambda}\left\lbrace \left\lbrack \left (\frac{\epsilon_0}{\epsilon_{\Lambda}}\right )^{1/|n|}-1\right\rbrack\left (\frac{a_0}{a}\right )^{3/|n|}+1\right\rbrace^{|n|}.
\label{one4}
\end{equation}
For $a\rightarrow 0$,
\begin{equation}
\epsilon\simeq \epsilon_{\Lambda} \left\lbrack \left (\frac{\epsilon_0}{\epsilon_{\Lambda}}\right )^{1/|n|}-1\right\rbrack^{|n|} \left (\frac{a_0}{a}\right )^{3}.
\label{one5}
\end{equation}
For $a\rightarrow +\infty$,
\begin{equation}
\epsilon\rightarrow \epsilon_{\Lambda}.
\label{one6}
\end{equation}
This model behaves as pressureless dark matter at early times and as dark energy
at late times. However, this decomposition is only asymptotic and, in the
intermediate regime, it is not possible to separate these two components
unambiguously because they do not appear additively in the energy density
unless $n=-1$ (i.e. $\gamma=0$), corresponding to the $\Lambda$CDM model (see
Sec. \ref{sec_cp}). We
also note that the contribution of baryonic matter, considered as a pressureless
fluid ($P_b=0$), must be added in the total energy density as an independent
species.

The polytropic model (\ref{one1}) with $K<0$ and  $n<0$ can be used to discuss
deviations from the $\Lambda$CDM model ($n=-1$). It is found from observations
that $n=-1.05^{+0.15}_{-0.16}$ \cite{karami}. If we impose that the velocity of
sound $c_s$ is real, the condition
$c_s^2=P'(\epsilon)c^2=(K\gamma/c^2)(\epsilon/c^2)^{\gamma-1}\ge 0$ constrains
the index $n$ in the range $-1\le n\le 0$ (i.e. $\gamma\le 0$). Combining this
condition with the observations, we get $-1\le n\le -0.9$ (i.e. $-0.11\le
\gamma\le 0$).

In conclusion, the polytropic model of type I offers a unified picture of dark
matter and dark energy \cite{cosmopoly2,karami}.  In this model, the universe
starts from the matter dominated epoch and approaches a de Sitter phase at late
times. Contrary to the $\Lambda$CDM model, the cosmic coincidence problem 
(namely why the ratio of dark energy  and dark matter densities is of order
unity today)
\cite{ccp} is solved naturally in the polytropic gas scenario
\cite{cosmopoly2,karami}. We finally note that a theoretical justification of a
polytropic equation of state of the form of Eq. (\ref{one1}) for a perfect
relativistic fluid in cosmology has been given
recently by Kazinski  \cite{kazinski} in relation to the quantum gravitational
anomaly. He predicts a polytropic equation of state of the form
$P=K(\epsilon/c^2)^{1+1/n}$ with $n\in \mathbb{N}$.

\subsection{Polytropic model of type II}
\label{sec_two}

As in the previous section, we assume that the universe is made of a single dark
fluid  but, here, we specify its equation of state in the form $P=P(\rho)$,
where $\rho$ is the rest-mass density. In order to solve the Friedmann equation
(\ref{f3}),  we need to determine the energy density $\epsilon$. To that
purpose, we assume that the dark fluid is at  $T=0$, or that the
evolution is adiabatic (which is the case for a perfect fluid
\cite{weinbergbook}). In that case, 
the first law of thermodynamics reduces to (see Appendix \ref{sec_gr}):
\begin{equation}
d\epsilon=\frac{P+\epsilon}{\rho}d\rho.
\label{two1}
\end{equation}
For a given $P(\rho)$ this equation can be solved to give the relation between the energy density and the rest-mass density. This relation can be written as (see Appendix \ref{sec_gr}):
\begin{equation}
\epsilon=\rho c^2+\rho\int^{\rho}\frac{P(\rho')}{{\rho'}^2}\, d\rho'=\rho c^2+u(\rho),
\label{two2}
\end{equation}
where $\rho c^2$ is the rest mass energy of the dark fluid and  $u(\rho)$ is its internal energy. On the other hand, combining the first law of thermodynamics (\ref{two1})  with the continuity equation (\ref{f1}), we get
\begin{equation}
\frac{d\rho}{dt}+3\frac{\dot a}{a}\rho=0.
\label{two3}
\end{equation}
We note that this equation is exact for a fluid at $T=0$, or for a perfect
fluid, and that it does not depend on the explicit form of the equation of state
$P(\rho)$. It expresses the conservation of the rest-mass density $\rho$
(or number density $n=\rho/m$).
It can be integrated into
\begin{equation}
\rho=\rho_0 \left (\frac{a_0}{a}\right )^3,
\label{two4}
\end{equation}
where $\rho_0$ is the present value of the rest-mass density and $a_0$ is the
present
value of the scale factor.
Substituting Eq. (\ref{two4}) in Eq. (\ref{two2}), we obtain
\begin{equation}
\epsilon=\rho_0 c^2\left (\frac{a_0}{a}\right )^3+u\left\lbrack \rho_0 \left (\frac{a_0}{a}\right )^3\right\rbrack.
\label{two5}
\end{equation}

For any barotropic equation of state $P(\rho)$, our treatment shows that the
energy density in Eqs. (\ref{two2}) and (\ref{two5}) is the sum of two terms. The first term corresponds to the rest-mass energy of the fluid and the second term corresponds to its internal energy. In the present approach, there is just {\it one} (relativistic) dark fluid. If we want to relate our approach to the traditional viewpoint where the universe is made of dark matter and dark energy, we note that the first term in Eqs. (\ref{two2}) and (\ref{two5}) is equivalent to the term of Eq. (\ref{eds2}) that describes a pressureless fluid (dark matter). As a result, we can interpret the second term in Eqs. (\ref{two2}) and (\ref{two5}) as the term corresponding to the dark energy. The equation
\begin{equation}
u(\rho)=\rho\int^{\rho}\frac{P(\rho')}{{\rho'}^2}\,
d\rho'
\label{two6}
\end{equation}
clearly shows that the dark energy term arises from pressure effects (i.e.
collisions between particles). If $P=0$,
the internal energy vanishes and we recover the standard EdS model describing
only pressureless dark matter (see Sec. \ref{sec_eds}). Alternatively, if we
assume that the
pressure is constant,
$P=-\epsilon_{\Lambda}$, we find that the internal energy is constant,
$u=\epsilon_{\Lambda}$, and we recover the model of Sec. \ref{sec_cp} that is
equivalent to the $\Lambda$CDM model. Therefore, the rest-mass energy $\rho c^2$
of the dark fluid mimics ``dark matter'' and its internal energy $u(\rho)$
mimics ``dark energy'' (see Appendix B of \cite{stiff}). This decomposition
leads to a natural, and
physical, unification of dark matter and dark energy, and elucidates their
mysterious nature.  We note that, in general, the dark energy term $u(\rho)$ in
Eq. (\ref{two5}) is not constant (it depends on the rest-mass density $\rho$,
and it does not even necessarily tend to a constant for $\rho\rightarrow 0$),
contrary to the standard dark energy term $\epsilon_{\Lambda}$ in Eq.
(\ref{cp2}) corresponding to a constant negative pressure.  Therefore, our
approach can  be used to obtain generalized models of dark energy by considering
different expressions of the equation of state $P(\rho)$. If $P(\rho)$ is
sufficiently close to a constant, the resulting models will be relatively close
to
the $\Lambda$CDM model. Therefore, we can use this approach to study small
deviations from the $\Lambda$CDM model and constrain the possible equations of
state $P(\rho)$ of dark energy from observations.

To be specific, we consider a polytropic equation of state
\begin{equation}
P=K\rho^{\gamma},\qquad \gamma=1+\frac{1}{n}
\label{two7}
\end{equation}
in which the pressure is a power of the rest-mass density. This corresponds to
the polytropic equation of state considered in the second paper of Tooper
\cite{tooper2} in the context of general relativistic stars. Therefore, we shall
call it polytropic model of type II, or simply, Model II. This polytropic model
has been studied in a cosmological context in \cite{mlbec,stiff},
and independently in \cite{spyrou}. 

For the equation of state (\ref{two7}), using Eq. (\ref{two2}), we find that the energy density is related to the rest-mass
density by \cite{mlbec,tooper2}:
\begin{equation}
\epsilon=\rho c^2+K\rho\ln\left (\frac{\rho}{\rho_*}\right ), \qquad (\gamma=1),
\label{two8}
\end{equation}
\begin{equation}
\epsilon=\rho c^2+\frac{K}{\gamma-1}\rho^{\gamma}=\rho c^2+nP(\rho),\qquad
(\gamma\neq 1).
\label{two9}
\end{equation}
Combining Eq. (\ref{two4}) with Eqs. (\ref{two8}) and (\ref{two9}), we obtain for
$\gamma=1$:
\begin{equation}
\epsilon=\rho_0 c^2\left (\frac{a_0}{a}\right
)^3+K\rho_0\left (\frac{a_0}{a}\right
)^3\ln\left\lbrack \frac{\rho_0}{\rho_*} \left
(\frac{a_0}{a}\right )^{3}\right\rbrack
\label{two10}
\end{equation}
and for $\gamma\neq 1$:
\begin{equation}
\epsilon=\rho_0 c^2\left (\frac{a_0}{a}\right
)^3+\frac{K}{\gamma-1}\rho_0^{\gamma}\left
(\frac{a_0}{a}\right )^{3\gamma}.
\label{two11}
\end{equation}
Introducing relevant (see below) notations 
$\Omega_{m,0}=\rho_0 c^2/\epsilon_0$, $\Omega'_{m,0}=K\rho_0/\epsilon_0$ and
$\Omega_{\gamma,0}=K\rho_0^{\gamma}/[(\gamma-1)\epsilon_0]$, the Friedmann
equation (\ref{f3}) with $\Lambda=0$ can be
written as
\begin{equation}
\frac{H}{H_0}=\sqrt{\frac{\Omega_{m,0}}{(a/a_0)^3}+\frac{\Omega'_{m,0}\ln\lbrack (\rho_0/\rho_*)
(a_0/a)^3\rbrack}{(a/a_0)^{3}}}
\label{two12}
\end{equation}
for $\gamma=1$, and as
\begin{equation}
\frac{H}{H_0}=\sqrt{\frac{\Omega_{m,0}}{(a/a_0)^3}+\frac{\Omega_{\gamma,0}}{(a/a_0)^{3\gamma}}
}
\label{two13}
\end{equation}
for $\gamma\neq 1$. By construction,
$\Omega_{m,0}+\Omega'_{m,0}\ln(\rho_0/\rho_*)=1$ and
$\Omega_{m,0}+\Omega_{\gamma,0}=1$. As discussed previously, the rest-mass
energy of the dark fluid mimics dark matter so that $\Omega_{m,0}$ represents
the present fraction of dark matter. The contribution of baryonic matter,
considered as a pressureless fluid ($P_b=0$), must be added in the total energy
density, as an independent species. Since it simply adds to the rest-mass energy
of the dark fluid, we can take it into account by considering that
$\Omega_{m,0}$ actually represents the present {\it total} fraction of matter
(baryonic and dark).

For the polytropic equation of state (\ref{two7}), the energy density in Eqs.
(\ref{two8})-(\ref{two11}) is the sum of two terms. An ordinary term
$\epsilon_m=\rho c^2 \propto a^{-3}$ (rest-mass energy) equivalent to
pressureless  dark matter and a new term
$\epsilon_{\gamma}=u(\rho)=P(\rho)/(\gamma-1)\propto a^{-3\gamma}$ (internal
energy) depending on the polytropic index $\gamma$. The
equation of state of the new term is (see also Appendix
\ref{sec_sa}) $P=(\gamma-1)u=u/n$. When $\gamma>1$ (i.e.
$n>0$), we find that $P\sim \epsilon/n$ and $\epsilon\sim nK\rho^{\gamma}\propto
a^{-3\gamma}$ in the ``early'' universe (small $a$, large $\rho$) and that
$P\sim K(\epsilon/c^2)^{\gamma}$ and $\epsilon\sim \rho c^2\propto a^{-3}$ in
the ``late'' universe (large $a$, small $\rho$). When $\gamma<1$ (i.e. $n<0$),
we find that $P\sim K(\epsilon/c^2)^{\gamma}$ and $\epsilon\sim \rho c^2\propto
a^{-3}$ in the ``early'' universe (small $a$, large $\rho$)
and that $P\sim \epsilon/n$ and $\epsilon\sim
nK\rho^{\gamma}\propto a^{-3\gamma}$ in the ``late'' universe (large $a$, small
$\rho$).\footnote{Since $P\sim (\gamma-1)\epsilon$ in the late universe, we find
that $w\equiv P/\epsilon\rightarrow \gamma-1$ when $a\rightarrow +\infty$.
Therefore, the universe is normal when $\gamma>0$ (the energy density decreases
as the scale factor increases) and phantom when $\gamma<0$ (the energy density
increases as the scale factor increases).} Therefore, the new
term (i.e. the internal
energy of the dark fluid)  dominates in the late universe when $\gamma<1$ (i.e.
$n<0$), while it
dominates in the early universe when $\gamma>1$ (i.e. $n>0$). This remark allows
us to refine the discussion given previously. The new term  can mimic dark
energy
only for $\gamma<1$ (i.e. $n<0$).  For $\gamma=0$  (i.e. $n=-1$), the new term
corresponds to a constant dark energy density ($P=-u$,
$u=\epsilon_{\Lambda}$), equivalent to the $\Lambda$CDM model. By
contrast, when $\gamma>1$ (i.e. $n>0$), the new term can mimic the effect of an
exotic constituent appearing in the early universe. For  $\gamma=2$ (i.e.
$n=1$), the new term is equivalent to stiff matter ($P=u$,
$u=\epsilon_{s}\propto a^{-6}$) as discussed in \cite{stiff} in relation to
self-interacting BECs at $T=0$ and in relation to the cosmological model of
Zel'dovich \cite{zeldocosmo,zeldovich}.  For $\gamma=4/3$ (i.e. $n=3$), the new
term is equivalent to the radiation of an
ultra-relativistic gas ($P=u/3$, $u=\epsilon_{rad}\propto a^{-4}$).

We now restrict ourselves to the case $\gamma<1$ in order to describe dark
energy. If we assume that the energy density of the new term (internal energy)
is positive in order to account for the observations (some energy is {\it
missing} with respect of the EdS model where there is only dark matter), we
conclude from the relation $\epsilon_{\gamma}=P/(\gamma-1)$  that the pressure
must be negative ($K<0$, $P<0$). This is the case, in particular, for the
constant equation of state $P=-\epsilon_{\Lambda}<0$, corresponding to
$\gamma=0$, equivalent to the usual $\Lambda$CDM model. If we assume that the
dark fluid is made of a self-interacting scalar field (e.g. a BEC), it
is shown in Appendix \ref{sec_ft} that a negative pressure can arise from an
attractive potential of interaction. Therefore, an equation of state with a
negative pressure (such as the equation of state of dark energy) can be
justified from a field theory.

The polytropic model (\ref{two11}) with an index $\gamma<1$ can be used to
discuss deviations from the $\Lambda$CDM model ($\gamma=0$). A detailed study
has been done recently in \cite{spyrou}. It is found from cosmological
constraints that $-0.089<\gamma\le 0$ (corresponding to $-1\le
n<-0.92$).\footnote{For $\gamma<0$, the universe becomes phantom at late times
(see footnote 10).}  We note that this range is very close to the one obtained
from Model I (see Sec. \ref{sec_one}) but this may be a coincidence.

Before closing this discussion, we briefly recall the analogies and the
differences between the two polytropic models: (i) Models I and II describe dark
energy provided that $\gamma<1$ (i.e. $n<0$) and $K<0$ (negative pressure). They
are equivalent to the $\Lambda$CDM model when $\gamma=0$ (i.e. $n=-1$). (ii) In
Model II, the energy density (\ref{two11})  is the sum of two terms (rest-mass
energy and internal energy) that can be associated with what has been called
dark matter and dark energy, respectively. In Model I, this distinction is less
clear because the expression (\ref{one4}) of the energy density is non-additive,
so
the distinction between dark matter and dark energy is only asymptotic. (iii) In
model I, the energy density tends to a constant $\epsilon_{\Lambda}$ at late
times (leading to a de Sitter era), which is equivalent to the cosmological
constant or to the ordinary form of dark energy. In model II, the dark energy
term is not constant (i.e. it depends on time), even asymptotically.

\subsection{Cardassian model}
\label{sec_card}

There is an interesting consequence of the ideas developed
in the previous section. Using Eq. (\ref{two2}), we can write the Friedmann
equation
(\ref{f3}) with $\Lambda=0$ in terms of the rest-mass density $\rho$ as
\begin{equation}
H^2=\frac{8\pi
G}{3}\rho+\nu(\rho),
\label{card}
\end{equation}
where $\rho=\rho_0(a_0/a)^3$ and $\nu(\rho)=(8\pi G/3c^2)u(\rho)$.
This equation is similar to the modified Friedmann equation introduced by Freese
and Lewis \cite{freese} in the so-called Cardassian model.\footnote{The name
Cardassian refers to a humanoid race in Star Trek whose goal is to take over
the universe \cite{freese}.} For the polytropic
equation of state (\ref{two7}), it exactly concides with the power law
Cardassian model. Eq. (\ref{card}) was originally \cite{freese} justified by the
fact that the Friedmann equation is modified as a consequence of embedding our
universe as a three-dimensional surface ($3$-brane) in higher dimensions.
Hence, the modified Friedmann equation (\ref{card}) may result from the
existence of extra-dimensions. Our approach provides another, simpler,
justification of this equation from the ordinary four dimensional Einstein
equations. In particular, starting from the usual Friedmann equation
(\ref{f3}) and considering a dark fluid at $T=0$, or an adiabatic fluid, the
``new'' term 
$\nu(\rho)$ in the ``modified'' Friedmann equation (\ref{card}) can be
interpreted as the internal
energy (\ref{two6}) of the dark fluid while the ``ordinary'' term $(8\pi
G/3)\rho$ corresponds to its rest-mass density.

\section{Logotropic equation of state}
\label{sec_logojust}

In Sec.
\ref{sec_two}, we have argued that dark matter and dark energy may be the
manifestation of a unique dark fluid characterized by an equation of state
$P(\rho)$. If this claim is correct, the equation of state $P(\rho)$ should
describe both the cosmological evolution of the universe and the structure of
the dark matter halos. Let us therefore consider the possibility that dark
matter halos are described by an equation of state $P(\rho)$, the same as the
one governing the cosmological evolution of the universe.  In that case, they
satisfy the classical condition of hydrostatic equilibrium between pressure and
gravity\footnote{For dark matter halos, we can use Newtonian gravity.}
\begin{equation}
\nabla P+\rho\nabla\Phi={\bf 0}.
\label{logojust1}
\end{equation}
We have seen that a dark fluid with a constant pressure
($P=-\epsilon_{\Lambda}$) generates a cosmological model equivalent to the
$\Lambda$CDM model. Therefore, this equation of state seems to be suitable at
the cosmological scale. However, when this equation of state is applied to dark
matter halos, we find that $\nabla P={\bf 0}$, so there is no pressure gradient
to balance the gravitational force. Therefore, as in the classical CDM model
where $P=0$, the dark matter halos described by a constant equation of state
display cuspy density profiles while
observations tend to favor a constant density core (cusp/core
problem). This observational result is a strong argument against a constant
equation of state, thus against the $\Lambda$CDM model, because it is not
suitable at the scale of dark matter halos.  Therefore, we turn next to a
polytropic equation of state of the form of Eq. (\ref{two7}). The  condition of
hydrostatic equilibrium (\ref{logojust1}) becomes 
\begin{equation}
K\gamma \rho^{\gamma-1}\nabla \rho+\rho\nabla\Phi={\bf 0}.
\label{logojust2}
\end{equation}
Because of pressure gradients when $K\neq 0$ and $\gamma\neq 0$, the dark matter
halos described by a polytropic equation of state display a central core instead
of a cusp.\footnote{For the pressure gradient to counterbalance the
gravitational attraction, we need $K\gamma>0$. Since we have shown from
cosmological considerations in Sec. \ref{sec_two} that $K<0$, we must have
$\gamma<0$. This is consistent with the cosmological bound $\gamma\le 0$
obtained in \cite{spyrou}. } If we want this equation of state to describe both
dark matter halos and the cosmological evolution of the universe,  $\gamma$ must
be close to zero (i.e. the equation of state must be close to a constant
pressure, equivalent to the $\Lambda$CDM model). But, in that case, we encounter
the problem of the justification. Indeed, it seems difficult to theoretically
justify why $\gamma$ should take a value like $-0.089$ for example
\cite{spyrou}. Therefore,
we consider another possibility. We assume that $\gamma\rightarrow 0$ and
$K\rightarrow \infty$ in such a way that $A=K\gamma$ is finite \cite{logo}. In
that limit, Eq. (\ref{logojust2}) becomes
\begin{equation}
A \frac{\nabla \rho}{\rho}+\rho\nabla\Phi={\bf 0}.
\label{logojust3}
\end{equation}
Comparing Eq. (\ref{logojust3}) with Eq. (\ref{logojust1}), we see that the pressure term corresponds to the logotropic equation of state \cite{pud,logo}:
\begin{equation}
P=A\ln\rho+C,
\label{logojust4}
\end{equation}
where $A$ and $C$ are constants. As shown in Appendix \ref{sec_log}, the
parameter $A$ can be interpreted as a logotropic temperature. We note that $A$
must be positive if we want that the pressure forces balance the gravitational
attraction. This is satisfying from a thermodynamical point of view. In the
following sections, we successively apply the logotropic equation of state to
the evolution of the universe and to the structure of dark matter halos.

\section{Logotropic cosmology}
\label{sec_lc}

\subsection{The equation of state}
\label{sec_est}

We assume that the universe is made of a single dark fluid described by a
logotropic equation of state
\begin{equation}
P=A\ln\left (\frac{\rho}{\rho_*}\right ),
\label{ed1}
\end{equation}
where $\rho$ is the rest-mass density, $A$ is the logotropic temperature  (we
assume $A\ge 0$, see Sec. \ref{sec_logojust}), and $\rho_*$ is a reference
density.\footnote{The meaning of this density will be unraveled in Sec.
\ref{sec_pred} but we leave it unspecified for the moment.} 
It will be called the logotropic dark fluid
(LDF). According to Eq.
(\ref{two2}), the relation between the energy density and the rest-mass density
is
\begin{equation}
\epsilon=\rho c^2-A\ln \left (\frac{\rho}{\rho_*}\right )-A=\rho c^2+u(\rho).
\label{ed2}
\end{equation}
The pressure is related to the internal energy by $P=-u-A$. Using Eq.
(\ref{two4}), we obtain
\begin{equation}
\epsilon=\rho_0 c^2\left (\frac{a_0}{a}\right
)^3-A\ln\left\lbrack \frac{\rho_0}{\rho_*}\left (\frac{a_0}{a}\right
)^3\right\rbrack -A.
\label{ed4}
\end{equation}
As explained in Sec. \ref{sec_two}, the first term (rest-mass energy) in Eqs.
(\ref{ed2}) and (\ref{ed4}) mimics  dark matter and the second term (internal
energy) mimics dark energy. The evolution of the energy density with the
scale factor is plotted in Fig. \ref{Reps} and the relation between the energy
density and the rest-mass density is plotted in Fig. \ref{rhoeps}. The universe
starts at $a=0$ with an
infinite rest-mass density ($\rho\rightarrow +\infty$) and an infinite energy density
($\epsilon\rightarrow +\infty$). The rest-mass density decreases as $a$ increases, see Eq.
(\ref{two4}). The energy density decreases as $a$ increases (i.e. $\rho$ decreases), reaches a
minimum $\epsilon_M=-A\ln ({A}/{\rho_* c^2})$ at $a_M=a_0(\rho_0c^2/A)^{1/3}$
(i.e. $\rho_M={A}/{c^2}$),
increases as $a$ increases (i.e. $\rho$ decreases) further, and tends to
$\epsilon\rightarrow +\infty$ as $a\rightarrow +\infty$ (i.e. $\rho\rightarrow
0$). The branch $a\le a_M$ (i.e. $\rho\ge \rho_M$)  corresponds to a normal
behavior in which the energy density decreases as the scale factor increases.
The branch $a\ge a_M$ (i.e. $\rho\le \rho_M$)  corresponds to a phantom
behavior in which the energy density increases as the scale factor increases. We
note that $A$ is equal to the rest-mass energy at the point where the universe
becomes phantom.

Combining Eqs.
(\ref{ed1}) and (\ref{ed2}), we
obtain the following relation between the pressure and the energy density
\begin{equation}
\epsilon=\rho_* c^2 e^{P/A}-P-A.
\label{ed4c}
\end{equation}
This determines the equation of state $P=P(\epsilon)$ under the form
$\epsilon=\epsilon(P)$. The pressure decreases as $a$ increases (i.e. $\rho$
decreases).
It starts from $P\rightarrow +\infty$ at $a=0$ (i.e. $\rho\rightarrow +\infty$, $\epsilon\rightarrow +\infty$), achieves
the value
$P_M=-\epsilon_M$ at $a_M$ (i.e. $\rho_M$, $\epsilon_M$) and tends to $P\rightarrow -\infty$ when
$a\rightarrow +\infty$ (i.e. $\rho\rightarrow 0$, $\epsilon\rightarrow
+\infty$), see Fig. \ref{RP}. The equation of state $P(\epsilon)$ is defined
for $\epsilon\ge \epsilon_M$ and has two branches corresponding to a normal
universe
($P\ge P_M$) and a phantom universe ($P\le P_M$), as shown in Fig. \ref{epsp}.
Therefore,
the equation of state $P(\epsilon)$  is multi-valued.  

In the early universe ($a\rightarrow 0$, $\rho\rightarrow
+\infty$),  the
rest-mass energy (dark matter) dominates, and  we have
\begin{equation}
\epsilon\sim \rho c^2\sim \rho_0 c^2 \left (\frac{a_0}{a}\right
)^{3},\qquad P\sim A\ln \left (\frac{\epsilon}{\rho_* c^2}\right ).
\end{equation}
For very small values of the scale factor, we recover the results of the CDM
model ($P=0$) since $\epsilon\propto a^{-3}$. In
the late universe ($a\rightarrow +\infty$, $\rho\rightarrow 0$), the internal
energy (dark energy) dominates, and we have
\begin{equation}
\epsilon\sim
-A\ln \left (\frac{\rho}{\rho_*}\right )\sim 3A \ln a,\qquad P\sim -\epsilon.
\label{ed4b}
\end{equation}
We note that the equation of state $P(\epsilon)$ reduces to $P\sim
-\epsilon$ for $\epsilon\rightarrow +\infty$, which is similar to the usual
equation of state (\ref{dmde2}) of the dark energy.\footnote{It is
interesting to note that the equation of state
$P=-\epsilon$ can be obtained as an asymptotic limit of the logotropic equation
of state (\ref{ed1}). This was not apparent at first sights.}  However, this is only an asymptotic result, different
from the usual equation of state of the dark energy where $P=-\epsilon$ exactly,
implying a constant energy density $\epsilon=\epsilon_{\Lambda}$ (see Sec.
\ref{sec_dmde}). As a result, in our model, the energy density does not tend to
a constant $\epsilon_{\Lambda}$ for $a\rightarrow +\infty$, but slowly increases
as $\ln
a$. This corresponds to a  phantom regime where  $P/\epsilon<-1$ at high energy
densities (we have $P+\epsilon\rightarrow -A<0$ when $\epsilon\rightarrow
+\infty$ on the phantom branch). As we shall see, this leads to a super de
Sitter behavior. 
We also emphasize that a polytropic
equation of state  with an index  $\gamma=0$ (or $n=-1$), corresponding  to a
constant pressure $P=-\epsilon_{\Lambda}$, leads to different results since the
energy density tends to a constant value $\epsilon_{\Lambda}$ for $a\rightarrow
+\infty$ (see Sec. \ref{sec_cp}), which is different from Eq. (\ref{ed4b}).

\subsection{The energy density}
\label{sec_ed}

Since, in our model, the rest-mass energy of the dark fluid in Eq. (\ref{ed4})
mimics dark
matter, we
identify $\rho_0 c^2$ with the present energy density of dark matter. We set
\begin{equation}
\rho_0 c^2=\Omega_{m,0}\epsilon_0,
\label{ed5}
\end{equation}
where $\epsilon_0$ is the present energy density of the universe and
$\Omega_{m,0}$ is the present fraction of dark matter.\footnote{As
explained in Sec. \ref{sec_two}, we also include baryonic matter in $\rho_0$ so
that $\Omega_{m,0}$ actually represents the present total fraction of mass
(baryonic and dark).} As a result, the present internal energy of the dark
fluid $u_0=\epsilon_0-\rho_0c^2$ is identified  with the present dark energy
density $\epsilon_{\Lambda}=
(1-\Omega_{m,0})\epsilon_0$  where
$\Omega_{\Lambda,0}=1-\Omega_{m,0}$ is the present
fraction of dark energy. From the observations, one
has $H_0=70.2 \, {\rm km}\,  {\rm s}^{-1}\, {\rm Mpc}^{-1}=2.275\, 10^{-18} \,
{\rm s}^{-1}$, $\Omega_{m,0}=0.274$ and $\Omega_{\Lambda,0}=0.726$, yielding
$\epsilon_0={3H_0^2c^2}/{8\pi G}=8.32\times 10^{-7}\, {\rm g}\, {\rm m}^{-1}{\rm
s}^{-2}$ and $\epsilon_0/c^2=9.26\times 10^{-24}\, {\rm g}\, {\rm m}^{-3}$.
Therefore, $\rho_0=2.54\times 10^{-24}\, {\rm g}\, {\rm m}^{-3}$ and
$\rho_{\Lambda}=\epsilon_{\Lambda}/c^2=6.72\times 10^{-24}\, {\rm g}\, {\rm
m}^{-3}$.

Applying Eq. (\ref{ed4}) with Eq. (\ref{ed5}) at $a=a_0$, we get
\begin{equation}
\epsilon_0=\Omega_{m,0}\epsilon_0-A\ln\left (
\frac{\Omega_{m,0}\epsilon_0}{\rho_* c^2}\right ) -A,
\label{ed6}
\end{equation}
which determines $\rho_*$ as a function of $A$, for known (measured) values of $\epsilon_0$ and $\Omega_{m,0}$.  Using Eqs. (\ref{ed5}) and (\ref{ed6}), the relation (\ref{ed4}) between the energy density and the scale factor can be rewritten as
\begin{equation}
\frac{\epsilon}{\epsilon_0}=\frac{\Omega_{m,0}}{(a/a_0)^3}+(1-\Omega_{m,0}
)\left\lbrack 1+\frac{3A}{\epsilon_0(1-\Omega_{m,0})}\ln\left (\frac{a}{a_0}\right )\right \rbrack.
\label{ed7}
\end{equation}
It is convenient to introduce the dimensionless logotropic temperature
\begin{equation}
B=\frac{A}{\epsilon_{\Lambda}}=\frac{A}{\epsilon_0(1-\Omega_{m,0})}
\label{ed8}
\end{equation}
and the normalized scale factor
\begin{equation}
R=\frac{a}{a_0}.
\label{ed10}
\end{equation}
In terms of these variables, the characteristic density $\rho_*$ is given from
Eq. (\ref{ed6}) by 
\begin{equation}
\frac{\rho_* c^2}{\epsilon_0 \Omega_{m,0}}=e^{1+1/B},
\label{ed11}
\end{equation}
and Eq. (\ref{ed7}) becomes
\begin{equation}
\frac{\epsilon}{\epsilon_0}=\frac{\Omega_{m,0}}{R^3}+(1-\Omega_{m,0})(1+3B\ln
R).
\label{ed9}
\end{equation}
The dimensionless logotropic temperature $B$ is the only free parameter of the model.

\begin{figure}[!ht]
\includegraphics[width=0.98\linewidth]{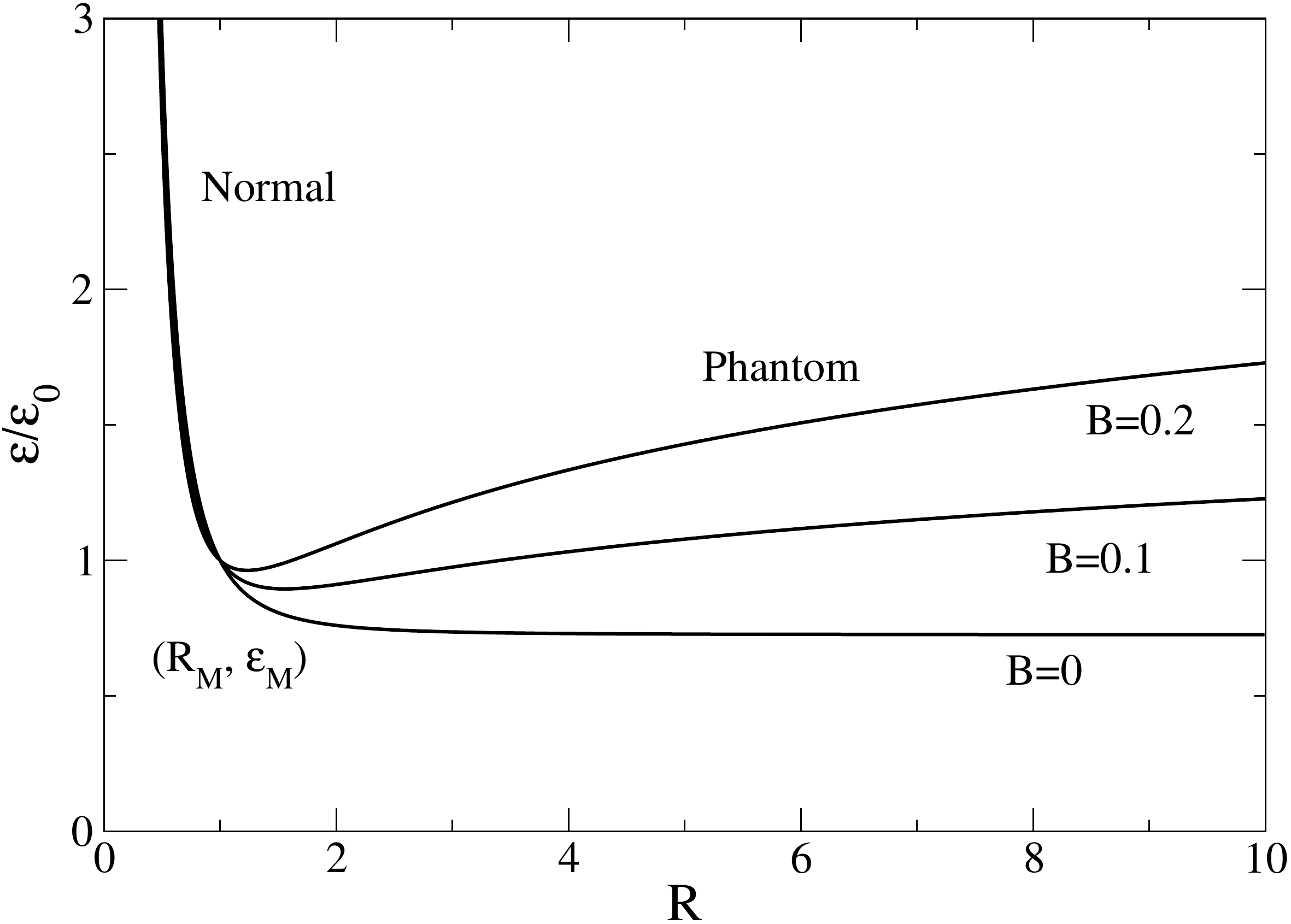}
\caption{Evolution of the energy density as a function of the scale factor.
\label{Reps}}
\end{figure}

For $B=0$, Eq. (\ref{ed9}) reduces to
\begin{equation}
\frac{\epsilon}{\epsilon_0}=\frac{\Omega_{m,0}}{R^3}+1-\Omega_{m,0}.
\label{ed9b}
\end{equation}
This returns the $\Lambda$CDM model with a different justification.
This is because when $B\rightarrow 0$, using Eqs. (\ref{ed8}) and (\ref{ed11}),
the equation of state (\ref{ed1}) reduces to a constant negative pressure
$P=-A/B=-\epsilon_0(1-\Omega_{m,0})=-\epsilon_{\Lambda}$ equivalent to Eq.
(\ref{cp1}).\footnote{More precisely, using Eqs. (\ref{ed8}) and (\ref{ed11}),
we can write the logotropic equation of state (\ref{ed1}) under the form
\begin{equation}
P=B\epsilon_{\Lambda}\ln\rho-\epsilon_{\Lambda}\left\lbrack 1+B+B\ln\left (\frac{\epsilon_0\Omega_{m,0}}{c^2}\right )\right\rbrack.
\label{ed14}
\end{equation}
For $B=0$, we get $P=-\epsilon_{\Lambda}$ which is equivalent to the
$\Lambda$CDM model (see Sec. \ref{sec_cp}).}

For $B\neq 0$, our model differs from the $\Lambda$CDM model. The relation between the energy density and the scale factor  is represented in Fig. \ref{Reps} for different values of $B$. For $R\rightarrow 0$,
\begin{equation}
\frac{\epsilon}{\epsilon_0}\sim \frac{\Omega_{m,0}}{R^3}
\label{ed12}
\end{equation}
and, for $R\rightarrow +\infty$,
\begin{equation}
\frac{\epsilon}{\epsilon_0}\sim 3B(1-\Omega_{m,0})\ln
R,\qquad (B\neq 0).
\label{ed13}
\end{equation}
When $B>0$, the curve $\epsilon(R)$ presents a minimum at
($R_M,\epsilon_M$), as detailed in the next section.  
When $R<R_M$, the energy density
decreases with the scale factor, which corresponds to a ``normal'' universe.
When $R>R_M$, the energy density increases with the scale factor, which
corresponds to a ``phantom'' universe.

\begin{figure}[!ht]
\includegraphics[width=0.98\linewidth]{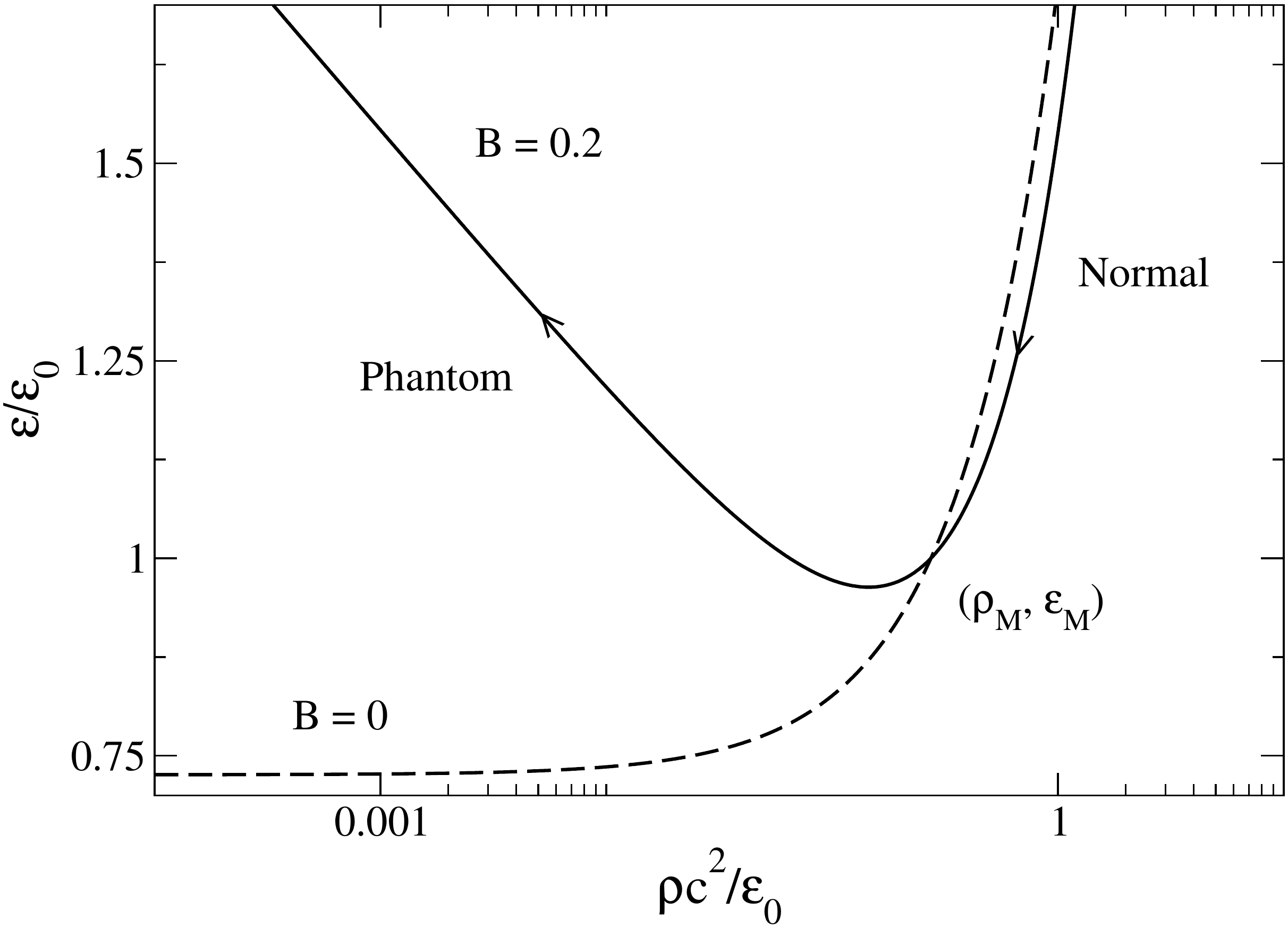}
\caption{Relation between the energy density and the rest-mass density (for
illustration we have taken $B=0.2$). \label{rhoeps}}
\end{figure}

It is useful to write the relation between the
energy density and the rest-mass density using dimensionless variables.
According to Eqs. (\ref{two4}), (\ref{ed5}) and (\ref{ed10}), we have
\begin{equation}
\label{edadd1}
\rho=\frac{\Omega_{m,0}\epsilon_0}{c^2}\frac{1}{R^3}.
\end{equation}
Eqs. (\ref{ed9}) and (\ref{edadd1}) determine the relation
$\epsilon(\rho)$ in parametric
form. Eliminating $R$ between these two relations, we explicitly obtain
\begin{equation}
\frac{\epsilon}{\epsilon_0}=\frac{\rho
c^2}{\epsilon_0}+(1-\Omega_{m,0})\left\lbrack 1+B\ln\left
(\frac{\Omega_{m,0}\epsilon_0}{\rho c^2}\right )\right\rbrack,
\end{equation}
as represented in Fig. \ref{rhoeps}. For $B=0$, this equation reduces to
$\epsilon=\rho c^2+\epsilon_{\Lambda}$ equivalent to
the $\Lambda$CDM model.

\subsection{The point of minimum energy density}
\label{sec_me}

\begin{figure}[!ht]
\includegraphics[width=0.98\linewidth]{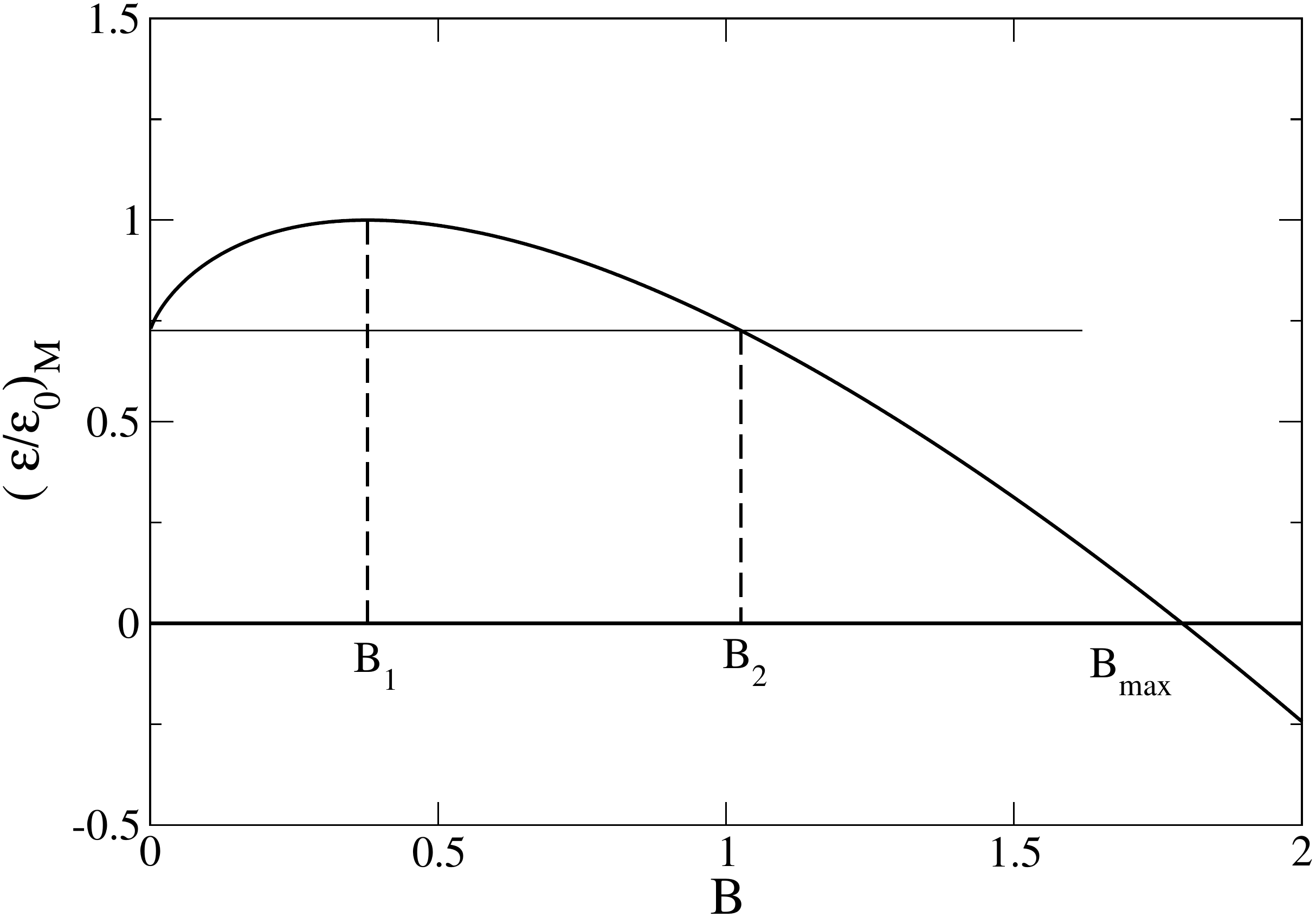}
\caption{Minimum energy density $\epsilon_M(B)$ as a function of the logotropic
temperature $B$. \label{Bepsetoile}}
\end{figure}

The minimum of the curve $\epsilon(R)$ defined by Eq. (\ref{ed9}) is given by
\begin{equation}
R_M=\left \lbrack \frac{\Omega_{m,0}}{B(1-\Omega_{m,0})}\right\rbrack^{1/3},
\label{me1}
\end{equation}
\begin{equation}
\left (\frac{\epsilon}{\epsilon_0}\right )_M=(1-\Omega_{m,0})\left\lbrack B+1+B
\ln \left(\frac{\Omega_{m,0}}{1-\Omega_{m,0}}\right )-B\ln B\right\rbrack.
\label{me2}
\end{equation}
The function $\epsilon_M(B)$ is represented in Fig. \ref{Bepsetoile}. For the $\Lambda$CDM model, corresponding to $B=0$, we have
\begin{equation}
\left (\frac{\epsilon}{\epsilon_0}\right )_M(0)=1-\Omega_{m,0}=0.726.
\label{me3}
\end{equation}
In that case, $R_M\rightarrow +\infty$, so the minimum
$\epsilon_M=\epsilon_{\Lambda}$ is rejected at infinity. For $B\rightarrow 0$,
\begin{equation}
\left (\frac{\epsilon}{\epsilon_0}\right )_M=(1-\Omega_{m,0})(1-B\ln
B),
\label{me3b}
\end{equation}
and for $B\rightarrow +\infty$,
\begin{equation}
\left (\frac{\epsilon}{\epsilon_0}\right )_M=-(1-\Omega_{m,0})B\ln B\rightarrow
-\infty.
\label{me4}
\end{equation}
We note that the function $\epsilon_M(B)$ first increases with $B$ before decreasing. Its maximum is located at
\begin{equation}
B_1=\frac{\Omega_{m,0}}{1-\Omega_{m,0}}=0.377, \qquad \left
(\frac{\epsilon}{\epsilon_0}\right )_M(B_1)=1.
\label{me5}
\end{equation}
The function $\epsilon_M(B)$ recovers its ``initial'' value $\epsilon_M(0)$ at the point
\begin{equation}
B_2=\frac{\Omega_{m,0}}{1-\Omega_{m,0}}e=1.03,
\label{me7}
\end{equation}
\begin{equation}
\left
(\frac{\epsilon}{\epsilon_0}\right )_M(B_2)=1-\Omega_{m,0}=0.726.
\label{me8}
\end{equation}
Finally, the minimum energy density $\epsilon_M(B)$ vanishes at $B_{\rm
max}$ given by the implicit equation
\begin{equation}
\ln(B_{\rm max})-\frac{1}{B_{\rm max}}=1+\ln\left
(\frac{\Omega_{m,0}}{1-\Omega_{m,0}}\right ).
\label{me9}
\end{equation}
Solving this equation numerically, we find $B_{\rm max}=1.79$. In the following,
we assume $0\le B< B_{\rm max}=1.79$ so the energy density $\epsilon(R)$ remains
always positive (which is of course a necessary condition). This puts a first
constraint on the allowable values of $B$. If
we demand that the present-day universe is not
phantom, corresponding to the condition $R_M>1$, using Eqs. (\ref{me1}) and
(\ref{me5}), we obtain the more stringent constraint $0\le B<B_1=0.377$. This
gives a physical interpretation to $B_1$.

\subsection{The pressure}
\label{sec_pr}

\begin{figure}[!ht]
\includegraphics[width=0.98\linewidth]{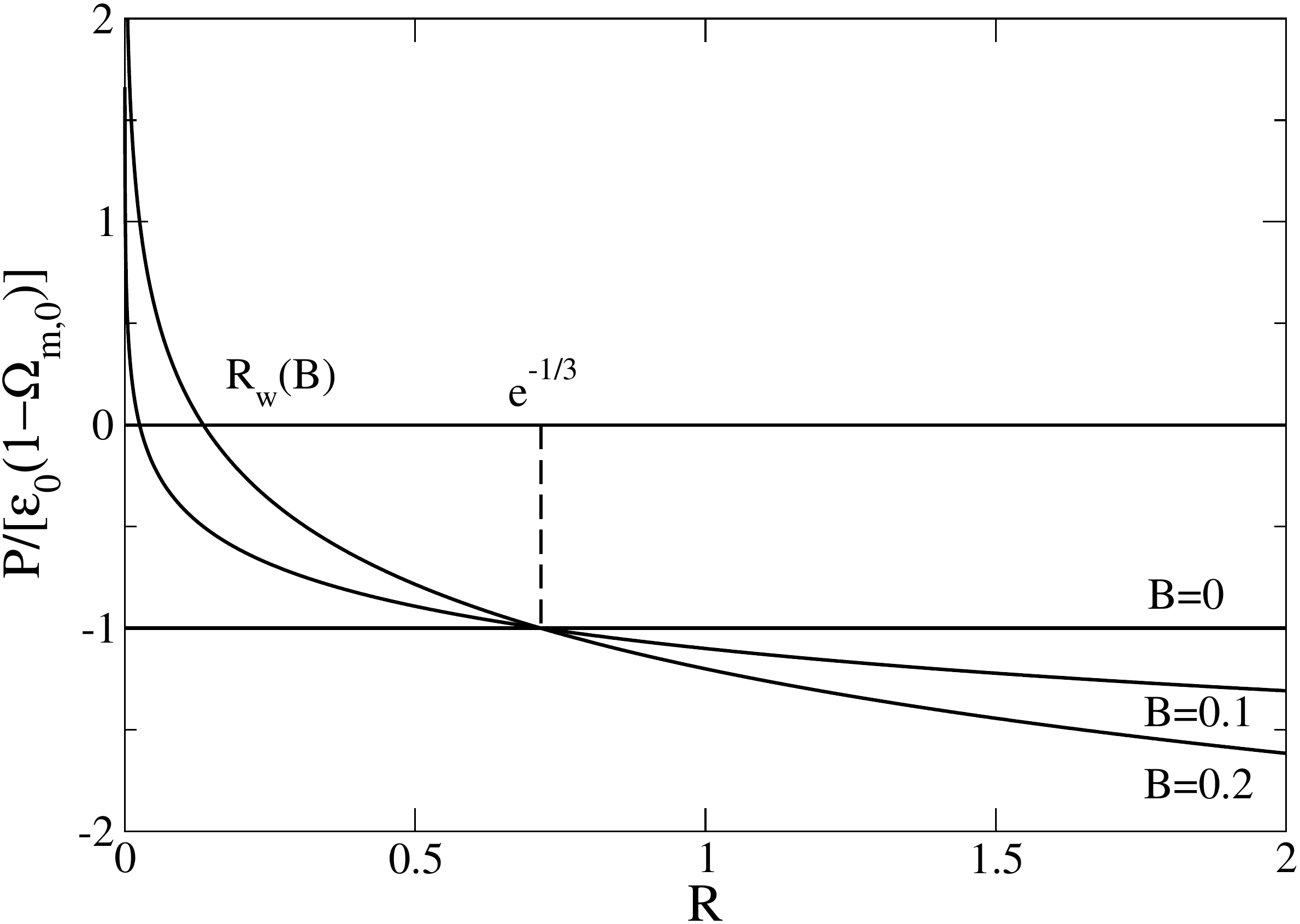}
\caption{Evolution of the pressure as a function of the scale factor. 
\label{RP}}
\end{figure}

Using Eqs. (\ref{two4}), (\ref{ed5}), (\ref{ed8}), (\ref{ed10}) and
(\ref{ed11}), the pressure (\ref{ed1}) can be expressed in terms of the scale
factor as
\begin{equation}
P=-\epsilon_0(1-\Omega_{m,0})(B+1+3B\ln
R).
\label{pr1}
\end{equation}
For $B=0$, we obtain a constant pressure $P=-\epsilon_0(1-\Omega_{m,0})=-\epsilon_{\Lambda}$, equivalent to the $\Lambda$CDM model. For $B\neq 0$, our model differs from the $\Lambda$CDM model. For $R\rightarrow 0$ and for $R\rightarrow +\infty$,
\begin{equation}
P\sim -3B\epsilon_0(1-\Omega_{m,0})\ln R.
\label{pr1b}
\end{equation}
The pressure vanishes at
\begin{equation}
R_w=e^{-(B+1)/3B}.
\label{pr2}
\end{equation}
At that point, according to Eq. (\ref{ed9}), the energy density is
\begin{equation}
\left (\frac{\epsilon}{\epsilon_0}\right )_w=\Omega_{m,0}
e^{(B+1)/B}-(1-\Omega_{m,0})B.
\label{pr3}
\end{equation}
The pressure is positive for $R<R_w$ and negative for $R>R_w$. Since $R_w<1$,
the pressure is always negative in  the present-day universe, for any
$B\ge 0$. Its value  is $P_0=-(B+1)\epsilon_{\Lambda}$. The relation between the
pressure and the scale factor is plotted in Fig. \ref{RP} for different values
of $B$. We note the ``exceptional'' point $R_e=e^{-1/3}=0.7165$ at which the
pressure takes the same value $P_e=-\epsilon_{\Lambda}=-6.04\times 10^{-7}\,
{\rm g}\, {\rm m}^{-1}\, {\rm s}^{-2}$, independently of $B$.

\begin{figure}[!ht]
\includegraphics[width=0.98\linewidth]{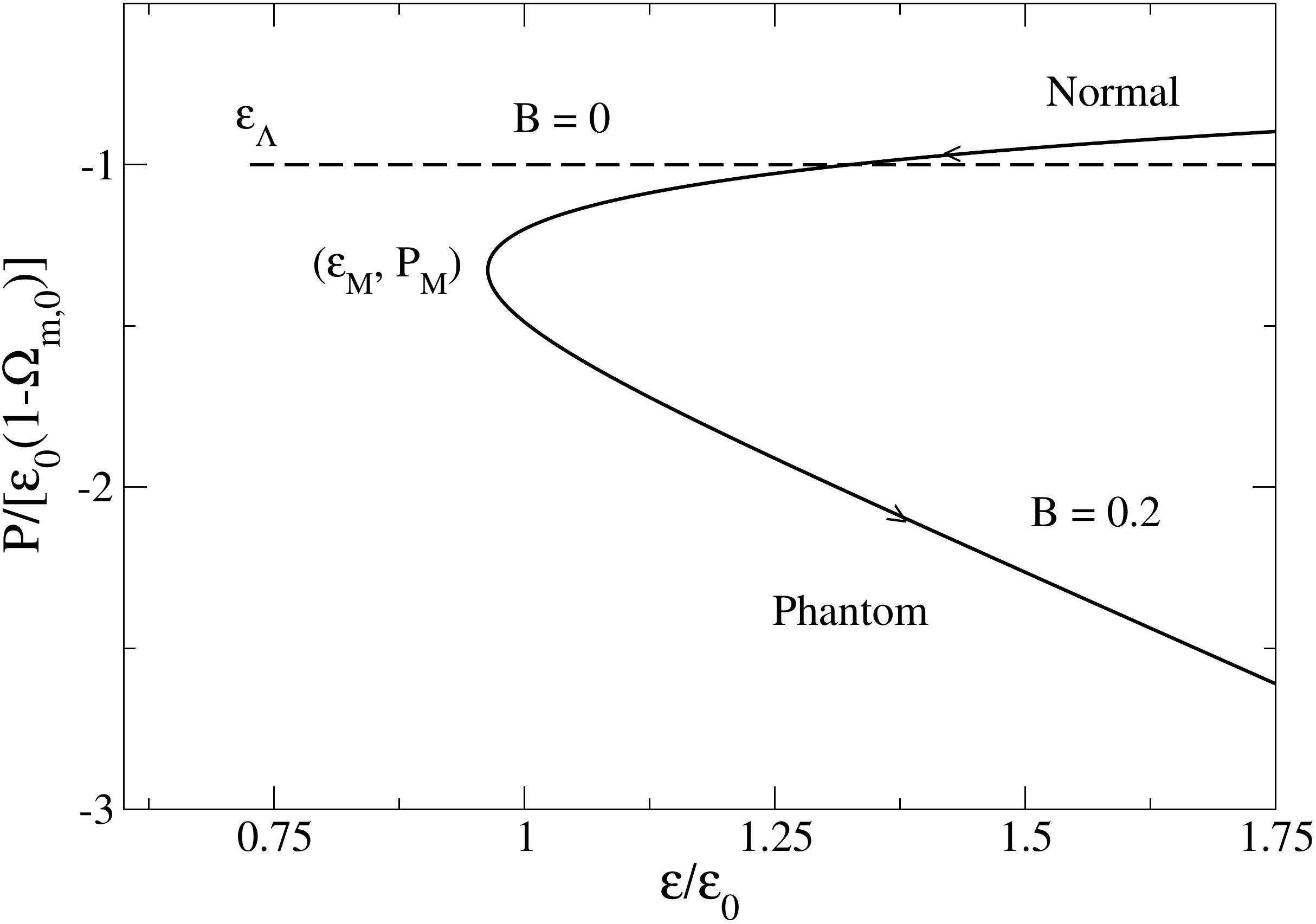}
\caption{Equation of state $P(\epsilon)$ giving the pressure as a function of
the energy density (for illustration we have taken $B=0.2$).\label{epsp}}
\end{figure}

The equation of state $P(\epsilon)$ is given in parametric form by Eqs.
(\ref{ed9}) and (\ref{pr1}). Using Eqs. (\ref{ed4c}), (\ref{ed8}) and
(\ref{ed11}), we also have
\begin{eqnarray}
\frac{\epsilon}{\epsilon_0}=\Omega_{m,0}e^{(B+1)/B}e^{P/[\epsilon_0(1-\Omega_{m,0})B]}\nonumber\\
-(1-\Omega_{m,0})\left \lbrack \frac{P}{\epsilon_0 (1-\Omega_{m,0})}+B\right \rbrack
\label{pr4}
\end{eqnarray}
which determines $\epsilon(P)$ in dimensionless form. The equation of state $P(\epsilon)$ presents two branches corresponding to a normal universe and a phantom universe (see Fig. \ref{epsp}).
For $\epsilon\rightarrow +\infty$, we get
\begin{eqnarray}
P\sim B\epsilon_0(1-\Omega_{m,0})\ln\epsilon
\label{pr4b}
\end{eqnarray}
on the normal branch ($R\rightarrow 0$) and
\begin{eqnarray}
P\sim -\epsilon
\label{pr4bb}
\end{eqnarray}
on the phantom branch ($R\rightarrow +\infty$). The pressure at the point of
minimum energy $\epsilon_M$ is
\begin{equation}
P_M=-\epsilon_0(1-\Omega_{m,0})\left\lbrack B+1+B
\ln \left(\frac{\Omega_{m,0}}{1-\Omega_{m,0}}\right )-B\ln B\right\rbrack.
\label{pr5}
\end{equation}
Comparing Eqs. (\ref{me2}) and (\ref{pr5}), we find that $P_M=-\epsilon_M$ (see also Sec. \ref{sec_est}).

\subsection{The equation of state parameter $w$}
\label{sec_w}

The equation of state parameter $w$ is defined by
\begin{equation}
P=w\epsilon.
\label{w1}
\end{equation}
According to Eqs. (\ref{ed9}) and (\ref{pr1}), it can be expressed in terms of the scale factor as
\begin{equation}
w=\frac{-(1-\Omega_{m,0})(B+1+3B\ln
R)}{\frac{\Omega_{m,0}}{R^3}+(1-\Omega_{m,0})(1+3B\ln
R)}.
\label{w2}
\end{equation}
For $R\rightarrow 0$,
\begin{equation}
w\sim
-3B\frac{1-\Omega_{m,0}}{\Omega_{m,0}}R^3\ln
R.
\label{w3}
\end{equation}
For $R\rightarrow +\infty$,
\begin{equation}
w\rightarrow -1.
\label{w4}
\end{equation}
More precisely, $w+1\sim -1/(3\ln R)$ for $R\rightarrow +\infty$. We note that $w>0$ for $R<R_w$ and $w<0$ for $R>R_w$. Therefore, the pressure is positive for $R<R_w$ and negative for $R>R_w$, in agreement with the results of Sec. \ref{sec_pr}. We also note that $w>-1$ for $R<R_M$ and $w<-1$ for $R>R_M$. Therefore, the universe is normal for $R<R_M$ (the energy density decreases as the scale factor increases) and phantom for $R>R_M$ (the energy increases as the scale factor increases), in agreement with the results of Sec. \ref{sec_ed}.

\begin{figure}[!ht]
\includegraphics[width=0.98\linewidth]{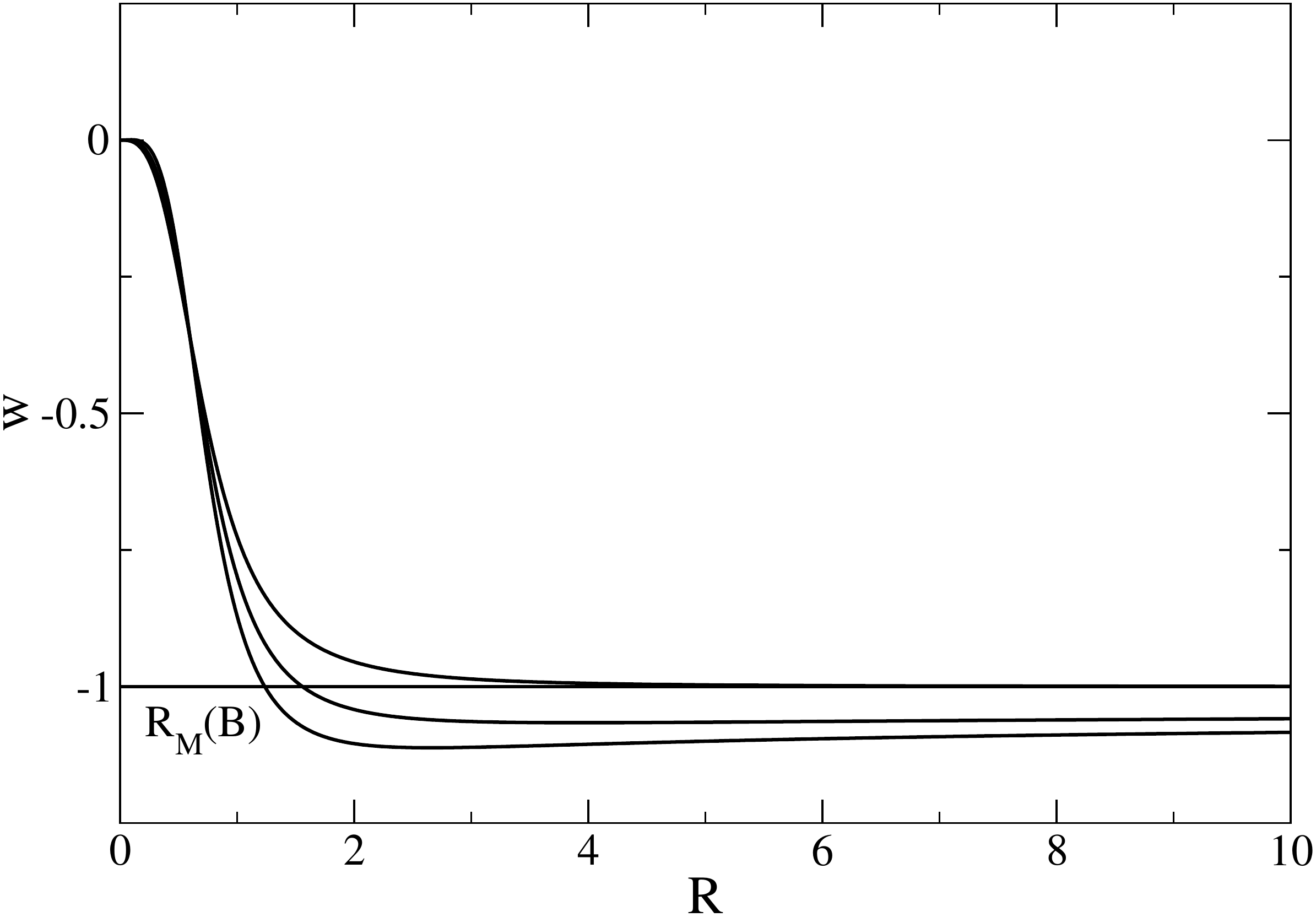}
\caption{The curve $w(R)$ for $B=0$, $B=0.1$, and $B=0.2$
(top to bottom).  \label{Rw}}
\end{figure}

\begin{figure}[!ht]
\includegraphics[width=0.98\linewidth]{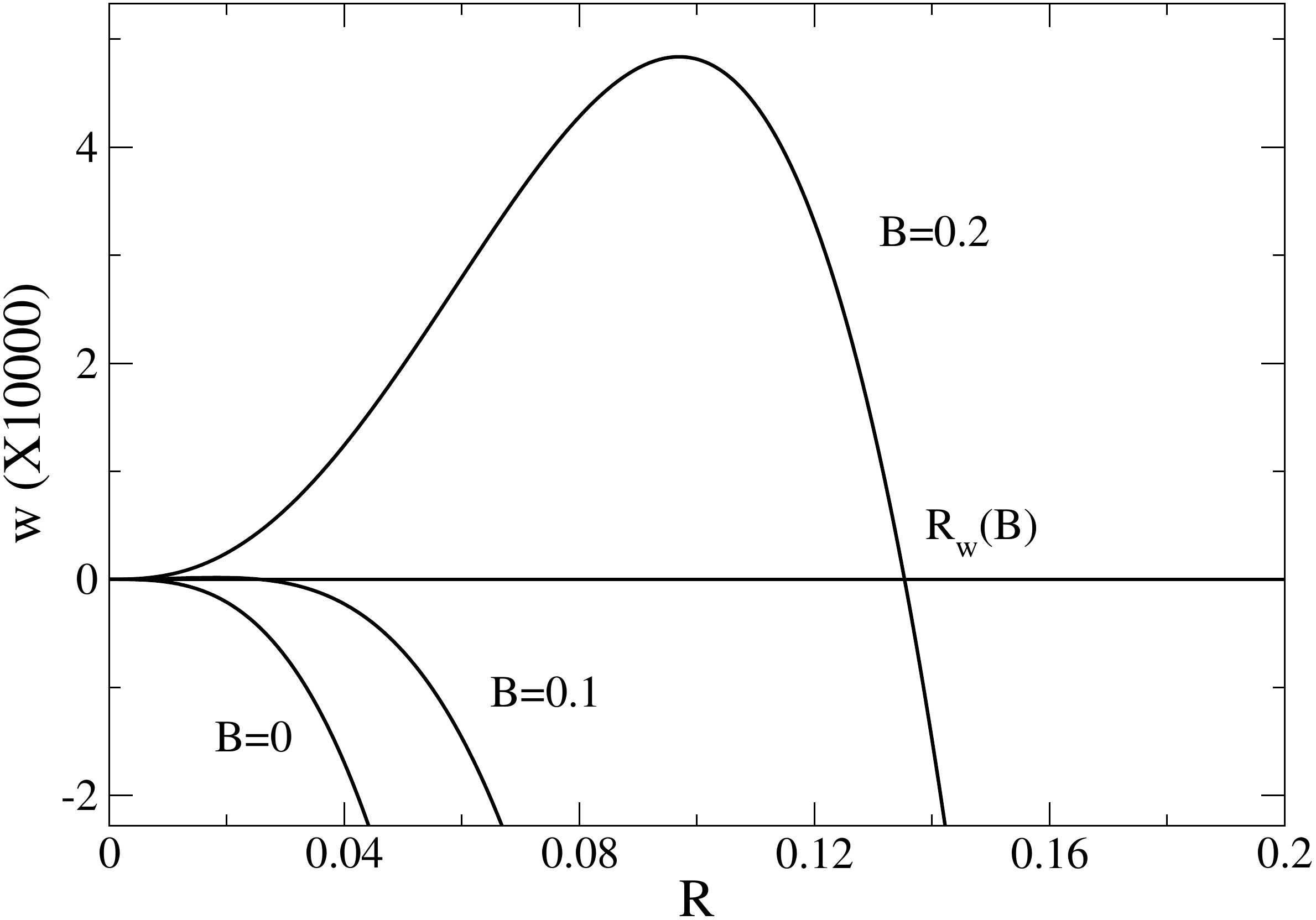}
\caption{Zoom of Fig. \ref{Rw} close to $R=0$ showing the  maximum
of the function $w(R)$.\label{RwZOOM}}
\end{figure}

The curve $w(R)$  is plotted in Fig. \ref{Rw}. It starts from $w=0$ at $R=0$,
increases, reaches a maximum (see Fig. \ref{RwZOOM}), decreases, becomes
negative after $R=R_w$, passes below $-1$ after $R=R_M$, reaches a minimum,
increases, and tends to $-1$ as $R\rightarrow +\infty$.

For $B=0$, corresponding to the $\Lambda$CDM model, the function $w(R)$ is given by
\begin{equation}
w=\frac{-(1-\Omega_{m,0})}{\frac{\Omega_{m,0}}{R^3}+1-\Omega_{m,0}}.
\label{w2b}
\end{equation}
It starts from $w=0$ at $R=0$, decreases monotonically, and tends to $-1$ as $R\rightarrow +\infty$.

The present value of the equation of state parameter is
\begin{equation}
w_0=-(1-\Omega_{m,0})(B+1).
\label{w5}
\end{equation}
For $B=0$, corresponding to the $\Lambda$CDM model, $w_0=-(1-\Omega_{m,0})=-0.726$.

\subsection{The velocity of sound}
\label{sec_vs}

\begin{figure}[!ht]
\includegraphics[width=0.98\linewidth]{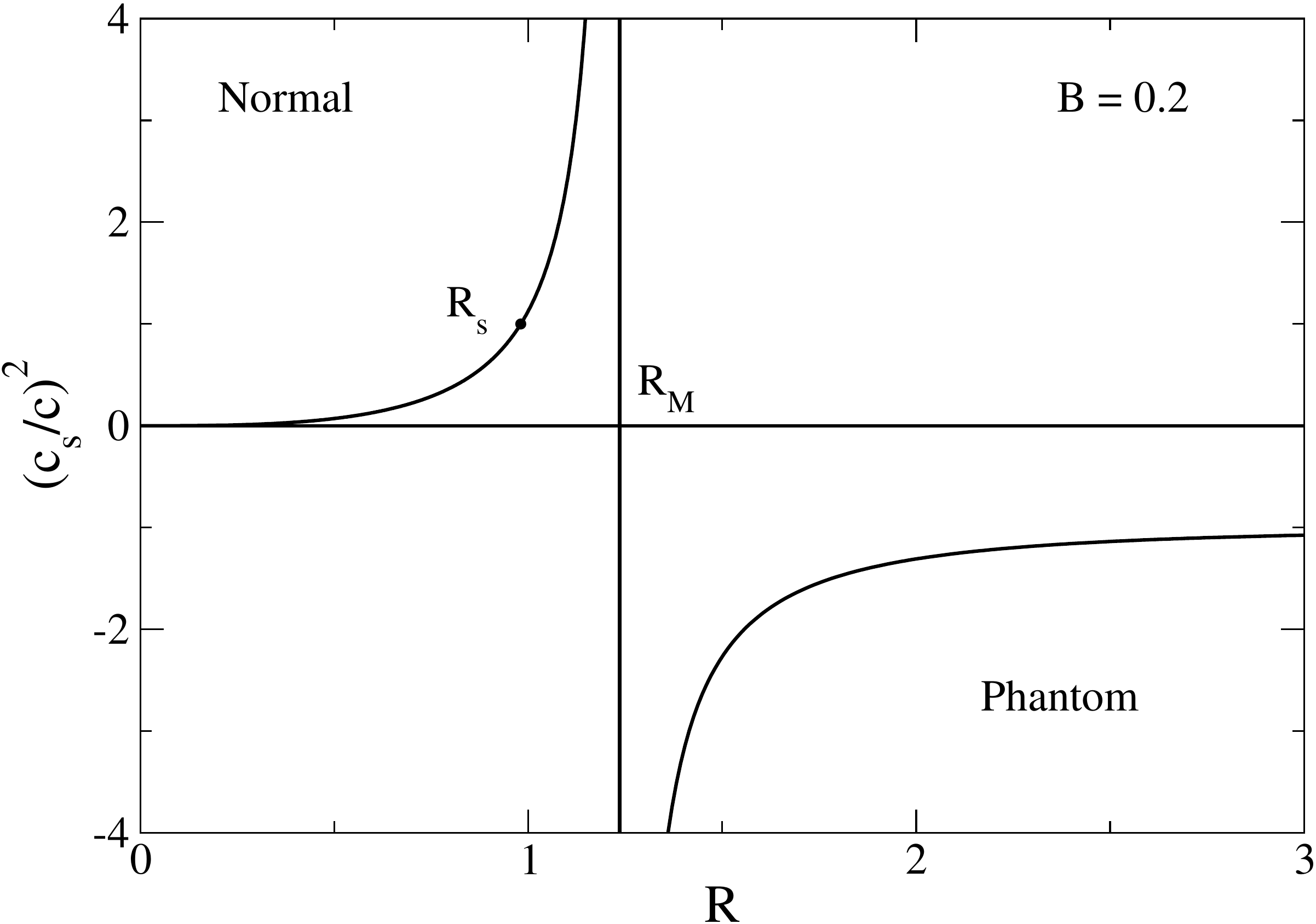}
\caption{Evolution of the square of the velocity of sound as a function of
the scale factor (for illustration we have taken $B=0.2$).
\label{Rc}}
\end{figure}

The velocity of sound $c_s$ is defined by $c_s^2=P'(\epsilon)c^2$.  Taking the derivative of Eq. (\ref{ed4c}) with respect to $\epsilon$, we obtain
\begin{equation}
\frac{c_s^2}{c^2}=\frac{1}{\frac{\rho c^2}{A}-1}.
\label{vs2}
\end{equation}
This relation requires that $A\ge 0$, hence $B\ge 0$, otherwise the velocity of
sound would always be imaginary ($c_s^2\le 0$). The constraint $A\ge 0$ is
consistent with the results of Sec. \ref{sec_logojust}. It also reinforces the
interpretation of $A$ as a (logotropic) temperature.\footnote{For a classical 
isothermal gas with $P=\rho k_B T/m$ and $c_s^2=P'(\rho)=k_B T/m$, the velocity
of sound is real at positive temperatures and imaginary at negative
temperatures, implying that negative temperatures are (usually) forbidden.} 
Substituting Eqs. (\ref{two4}), (\ref{ed8}) and (\ref{me1}) in Eq. (\ref{vs2}),
we get
\begin{equation}
\frac{c_s^2}{c^2}=\frac{1}{\left (\frac{R_M}{R}\right )^3-1}.
\label{vs3}
\end{equation}
For $B=0$, corresponding to the $\Lambda$CDM model, $c_s^2=0$ since the
pressure is constant ($P=-\epsilon_{\Lambda}$). For
$B>0$, the velocity of sound is real for $R<R_M$ (i.e. when the universe is
normal) and imaginary for $R>R_M$ (i.e. when the universe is phantom). The
relation between the velocity of sound  and the scale factor is plotted in Fig.
\ref{Rc}.

The velocity of sound in the present-day universe is
\begin{equation}
\left (\frac{c_s^2}{c^2}\right )_0=\frac{1}{R_M^3-1}.
\label{vs4}
\end{equation}
If we require that $(c_s^2)_0\ge 0$, we must have $R_M>1$, hence $B<B_1=0.377$. The
condition that the present-day velocity of sound is real coincides with the condition that
the present-day universe is non-phantom.

More stringent constraints on $B$ can be obtained from the following arguments. In order to have $(c_s^2/c^2)_0\le 1$, the dimensionless logotropic temperature must satisfy
\begin{equation}
B\le \frac{\Omega_{m,0}}{2(1-\Omega_{m,0})}=\frac{B_1}{2}=0.1885.
\label{vs5}
\end{equation}
This ensures that the velocity of sound in the present universe is less than the speed of light. In order to have $(c_s^2/c^2)_0\le 1/3$, the dimensionless logotropic temperature must satisfy
\begin{equation}
B\le \frac{\Omega_{m,0}}{4(1-\Omega_{m,0})}=\frac{B_1}{4}=0.09425.
\label{vs6}
\end{equation}
This ensures that the present universe is non-relativistic (i.e. the velocity of sound in the present universe is less than the velocity of sound in a radiation-dominated universe for which $P=\epsilon/3$ and $c_s^2/c^2=1/3$). We shall take the value $B_1/4=0.09425$ as an upper bound for the allowable values of $B$.

For a given $B$, the velocity of sound becomes larger than the speed of light ($c_s>c$) when  $R>R_s$ with
\begin{equation}
R_s=\left \lbrack
\frac{\Omega_{m,0}}{2B(1-\Omega_{m,0})}\right\rbrack^{1/3}=\frac{R_M}{2^{1/3}}.
\label{vs7}
\end{equation}
Since $R_s<R_M$, the logotropic model may break down before the universe reaches
the phantom regime at $R=R_M$.

{\it Remark:} If we ompute the velocity of sound from the relation
${c}_s^2=P'(\rho)$, we obtain 
\begin{equation}
c_s^2=\frac{A}{\rho}=\left (\frac{R}{R_M}\right )^3c^2.
\label{vs8}
\end{equation}
This approximation is valid when $\rho\gg A$ or $R\ll R_M$.

\subsection{The deceleration parameter}
\label{sec_dp}

In a flat universe without cosmological constant ($k=\Lambda=0$), the
deceleration parameter $q=-\ddot a a/\dot a^2$  is given by (see Eqs.
(\ref{f2}), (\ref{f3}) and (\ref{w1})):
\begin{equation}
q=\frac{1+3w}{2}.
\label{dp1}
\end{equation}
Using the expression of $w$ given by Eq. (\ref{w2}), we obtain
\begin{equation}
q=\frac{\frac{\Omega_{m,0}}{R^3}-(1-\Omega_{m,0})(3B+2+6B\ln
R)}{2\left \lbrack \frac{\Omega_{m,0}}{R^3}+(1-\Omega_{m,0})(1+3B\ln
R)\right\rbrack}.
\label{dp2}
\end{equation}
For $R\rightarrow 0$,
\begin{equation}
q\sim \frac{1}{2}\left \lbrack
1-9B\frac{1-\Omega_{m,0}}{\Omega_{m,0}}R^3\ln R\right\rbrack.
\label{dp3}
\end{equation}
For $R\rightarrow +\infty$,
\begin{equation}
q\rightarrow -1.
\label{dp4}
\end{equation}
The deceleration parameter $q(R)$ vanishes at  $R=R_c$ determined in the next section. When $R<R_c$, the universe is decelerating ($q>0$) and when $R>R_c$, the universe is accelerating ($q<0$).

\begin{figure}[!ht]
\includegraphics[width=0.98\linewidth]{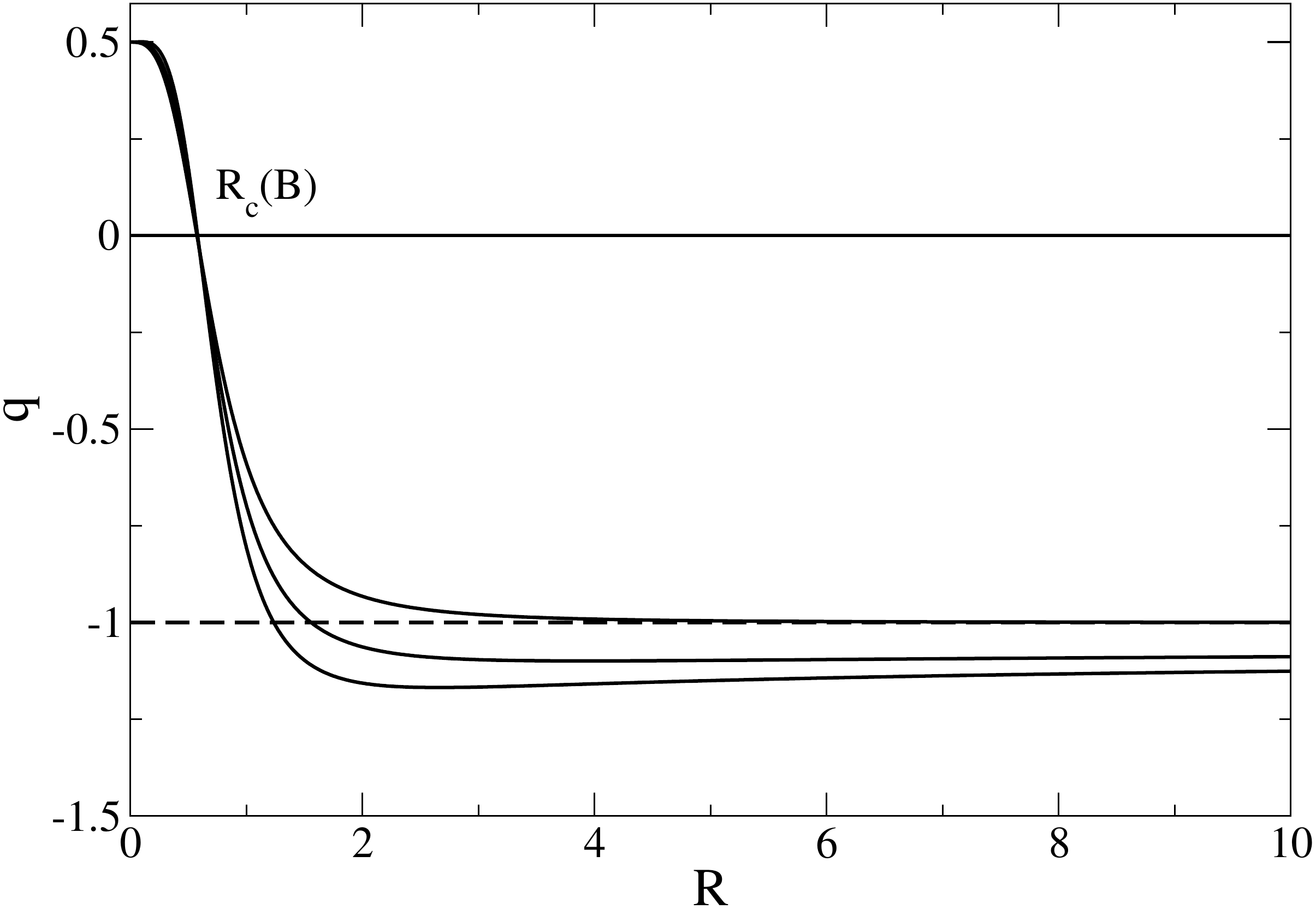}
\caption{Evolution of the deceleration parameter as a function of the scale
factor for $B=0$, $B=0.1$, and $B=0.2$ (top to bottom).
\label{Rq}}
\end{figure}

The curve $q(R)$ is plotted in Fig. \ref{Rq}. Its behavior follows the one of $w(R)$ described in Sec. \ref{sec_w}.

For $B=0$, corresponding to the $\Lambda$CDM model, the function $q(R)$ is given by
\begin{equation}
q=\frac{\frac{\Omega_{m,0}}{R^3}-2(1-\Omega_{m,0})}{2\left \lbrack
\frac{\Omega_{m,0}}{R^3}+1-\Omega_{m,0}\right\rbrack}.
\label{dp5}
\end{equation}

\subsection{The point at which the universe accelerates}
\label{sec_a}

\begin{figure}[!ht]
\includegraphics[width=0.98\linewidth]{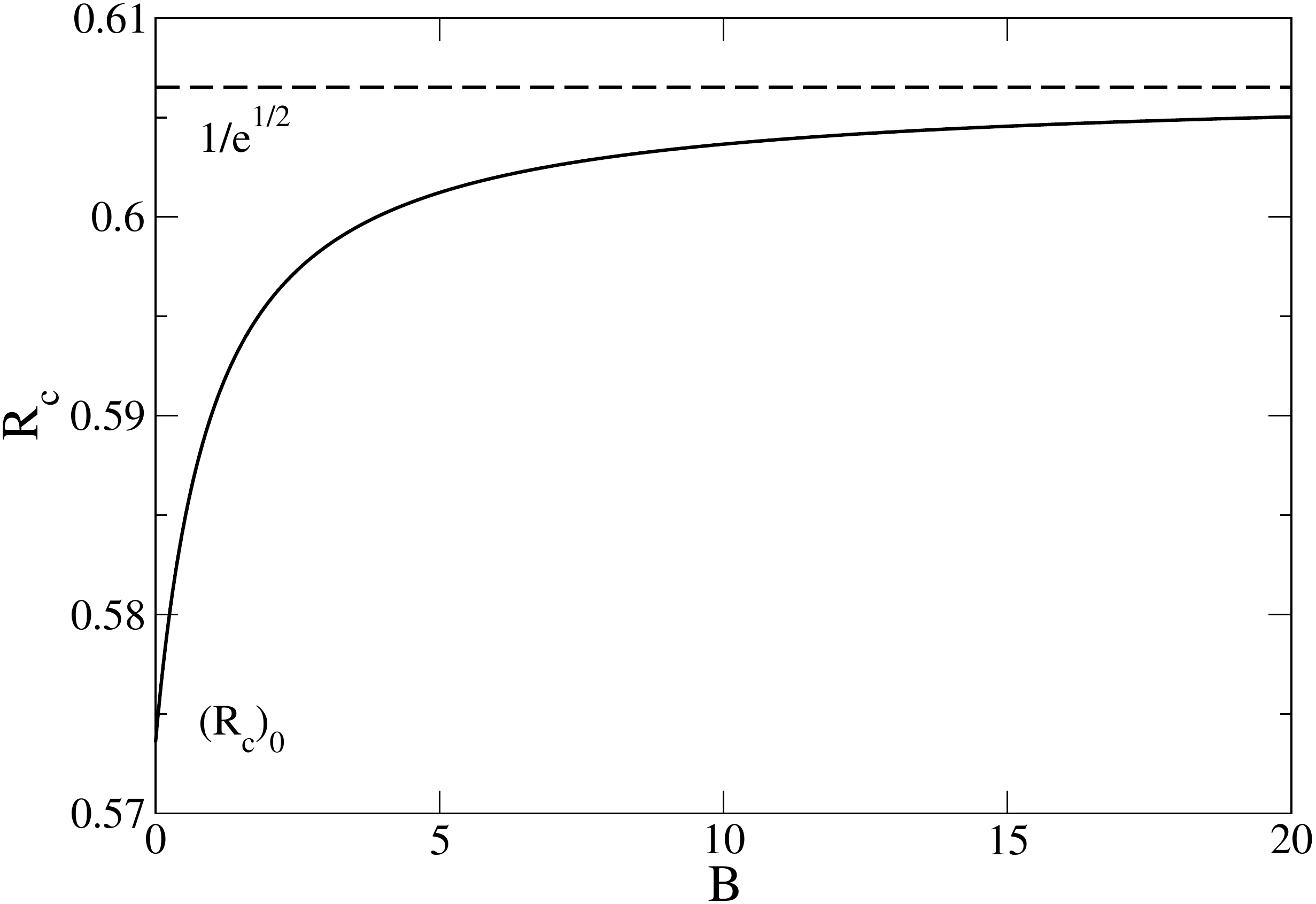}
\caption{The scale factor $R_c$ at which the universe accelerates as a function of $B$. \label{BRc}}
\end{figure}

According to Eq. (\ref{dp2}), the point $R_c(B)$ at which the universe starts accelerating ($q=0$) is determined by the equation
\begin{equation}
B=\frac{\frac{\Omega_{m,0}}{1-\Omega_{m,0}}\frac{1}{R_c^3}-2}{3(1+2\ln R_c)}.
\label{a1}
\end{equation}
For $B=0$, corresponding to the $\Lambda$CDM model,
\begin{equation}
R_c=\left \lbrack
\frac{\Omega_{m,0}}{2(1-\Omega_{m,0})}\right\rbrack^{1/3}=0.574.
\label{a2}
\end{equation}
For $B\rightarrow +\infty$,
\begin{equation}
R_c\rightarrow e^{-1/2}=0.6065.
\label{a3}
\end{equation}
The curve $R_c(B)$ is plotted in Fig. \ref{BRc}. Since $R_c<1$, the present-day
universe is always accelerating, whatever the value of $B$. We also note that,
in the framework of the logotropic model,  $R_c$ lies in the small interval
$[0.574,0.6065]$. Since
$R_c(B)$ increases, we note that when $B>0$ the acceleration of the universe
occurs later than in the $\Lambda$CDM model.

The present value of the deceleration parameter is
\begin{equation}
q_0=\frac{\Omega_{m,0}-(1-\Omega_{m,0})(3B+2)}{2}.
\label{a4}
\end{equation}
For $B=0$, corresponding to the $\Lambda$CDM model,
\begin{equation}
q_0=\frac{3\Omega_{m,0}-2}{2}=-0.589.
\label{a5}
\end{equation}
The condition for the present-day universe to be accelerating
is that $\Omega_{m,0}<2/3$. This inequality is indeed realized by the
observational value $\Omega_{m,0}=0.274$.

\subsection{The transition between matter and dark energy}
\label{sec_trans}

\begin{figure}[!ht]
\includegraphics[width=0.98\linewidth]{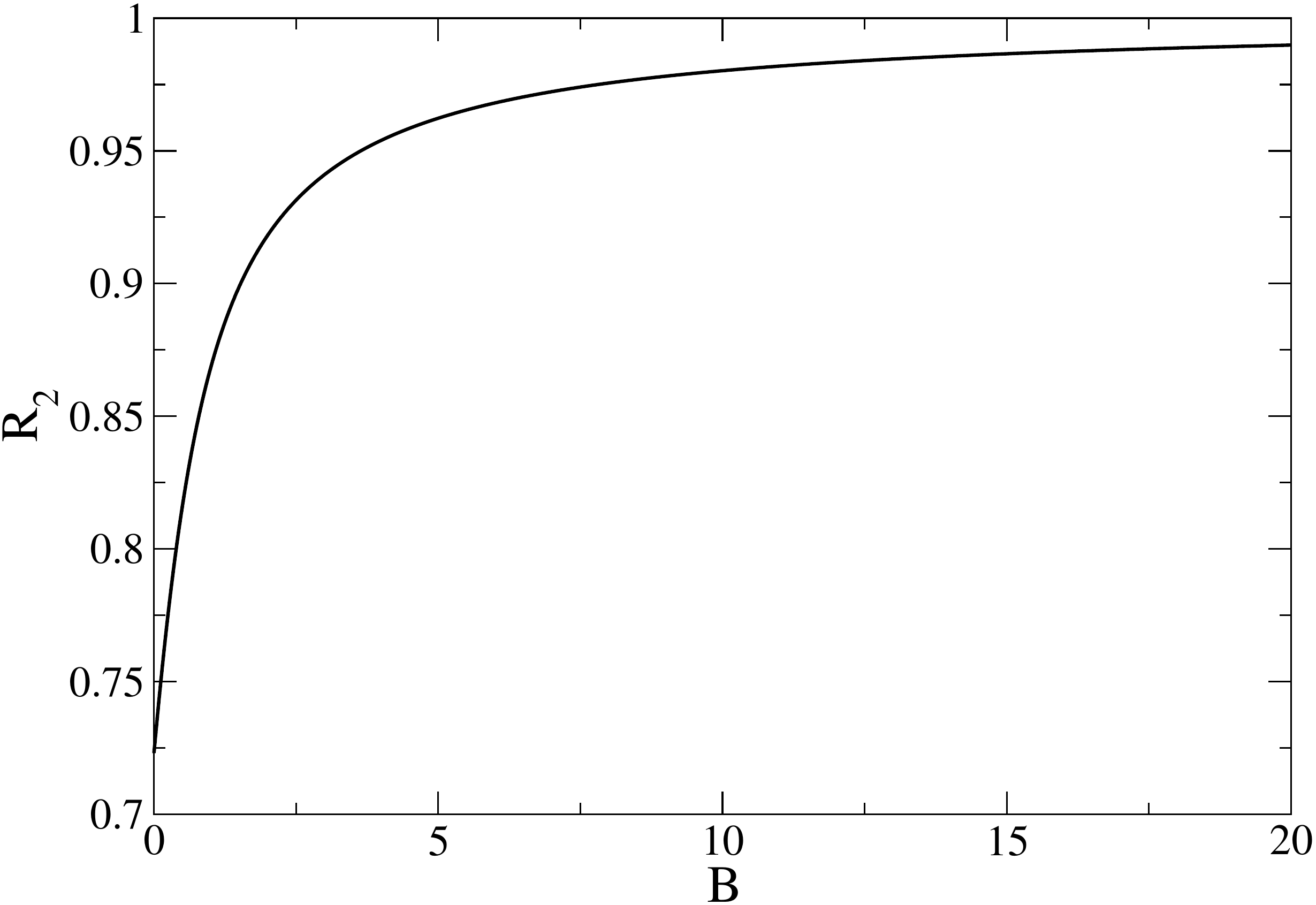}
\caption{The scale factor $R_2$ corresponding to the transition between dark matter and dark
energy as a function of $B$.
\label{BR2}}
\end{figure}

If we interpret in Eq. (\ref{ed9}) the rest-mass density as ``dark matter'' and
the internal energy as ``dark energy'', their ratio evolves with the scale
factor as
\begin{equation}
\frac{\rm DE}{\rm DM}=\frac{u(\rho)}{\rho
c^2}=\frac{1-\Omega_{m,0}}{\Omega_{m,0}} R^3(1+3B\ln R).
\label{trans0}
\end{equation}
The transition between dark matter and
dark energy takes place at a scale factor $R_2$ determined by
 \begin{equation}
\frac{\Omega_{m,0}}{R_2^3}=(1-\Omega_{m,0})(1+3B\ln R_2).
\label{trans1}
\end{equation}
For $B=0$, corresponding to the $\Lambda$CDM model, we get $R_2(0)=0.723$. For
$B\rightarrow +\infty$, we get $R_2\sim {\rm exp}\left\lbrace
(2\Omega_{m,0}-1)/3(1-\Omega_{m,0})B\right\rbrace\rightarrow 1$. More generally,
the function $R_2(B)$ is plotted in Fig. \ref{BR2}.

\subsection{The temporal evolution of the scale factor}
\label{sec_temp}

Using Eq. (\ref{ed9}), the Friedmann equation (\ref{f3}) with $\Lambda=0$ takes
the form
\begin{equation}
H=\frac{\dot R}{R}=H_0 \sqrt{\frac{\Omega_{m,0}}{R^3}+(1-\Omega_{m,0})(1+3B\ln
R)}.
\label{es1}
\end{equation}
The temporal evolution of the scale factor $R(t)$ is given by
\begin{equation}
\int_0^R
\frac{dx}{x\sqrt{\frac{\Omega_{m,0}}{x^3}+(1-\Omega_{m,0})(1+3B\ln
x)}}=H_0 t.
\label{es2}
\end{equation}
For $t\rightarrow 0$,
\begin{equation}
R\sim \left (\frac{3}{2}\sqrt{\Omega_{m,0}}H_0 t\right )^{2/3},\qquad \frac{\epsilon}{\epsilon_0}\sim \frac{4}{9H_0^2 t^2},
\label{es3}
\end{equation}
like in the EdS universe (see Sec. \ref{sec_eds}). For $t\rightarrow +\infty$,
\begin{equation}
R\propto e^{\frac{3B}{4}(1-\Omega_{m,0})H_0^2t^2},\quad \epsilon\sim \left\lbrack \frac{3B}{2}(1-\Omega_{m,0})H_0 t\right\rbrack^2.
\label{es3b}
\end{equation}
This solution, which is valid in the regime where 
the universe is phantom, has a super-de Sitter behavior. However, as indicated
in Sec. \ref{sec_vs}, the logotropic model may break down before the universe
enters in this regime because the velocity of sound exceeds the speed of light
when $R>R_s$ with $R_s<R_M$.

\begin{figure}[!ht]
\includegraphics[width=0.98\linewidth]{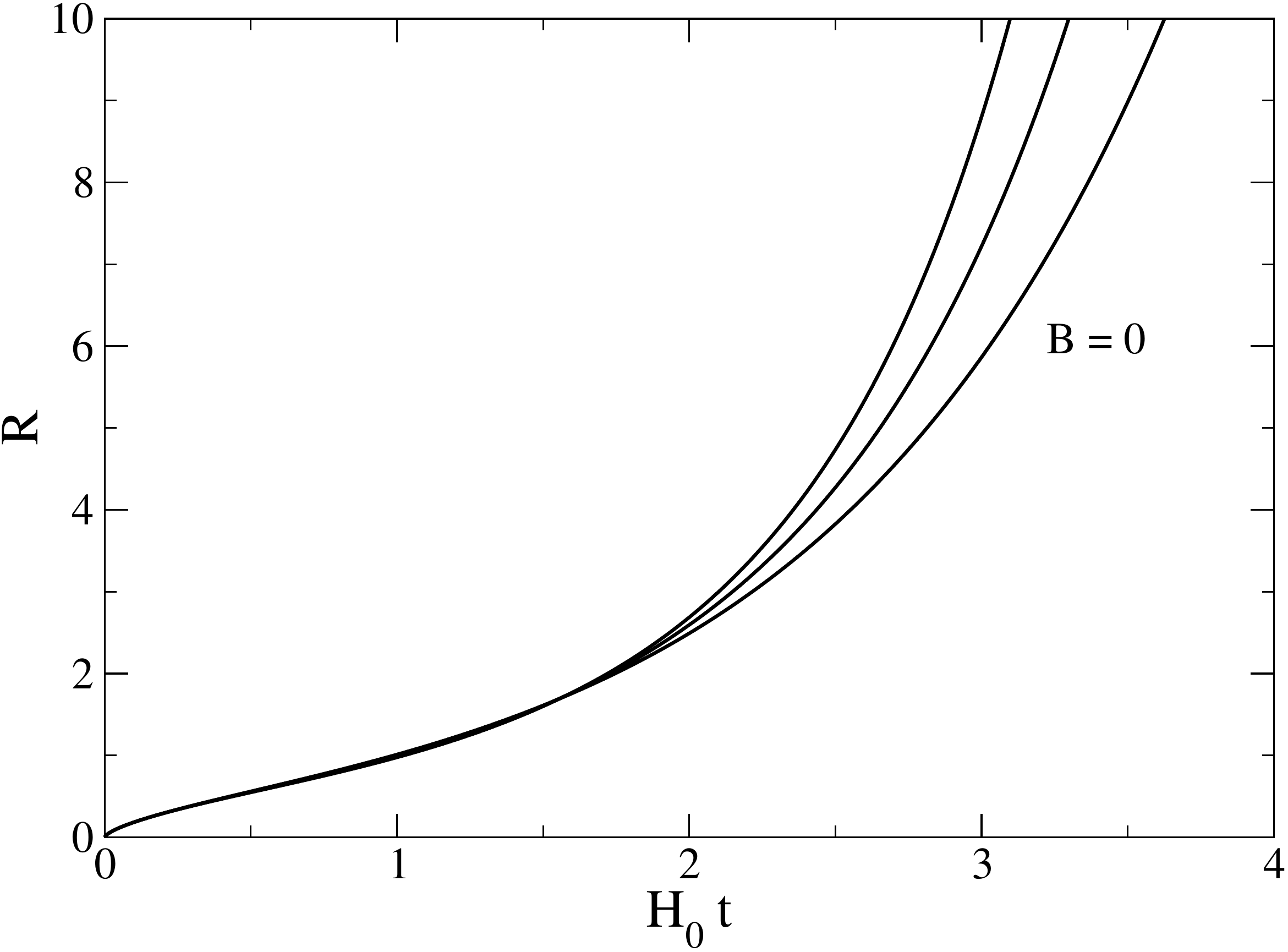}
\caption{Temporal evolution of the scale factor for $B=0$, $B=0.1$, and $B=0.2$ (bottom to top).
\label{tR}}
\end{figure}

The curve $R(t)$ is plotted in Fig. \ref{tR}. For
$B=0$, corresponding to
the $\Lambda$CDM model, we recover the analytical solution of Eq. (\ref{dmde6}).

\subsection{The age of the universe}
\label{sec_es}

Taking $R=1$ in Eq. (\ref{es2}), the age of the universe  is given as a function
of the logotropic temperature by
\begin{equation}
t_0(B)=\frac{1}{H_0}\int_0^{R_0=1}
\frac{dx}{x\sqrt{\frac{\Omega_{m,0}}{x^3}+(1-\Omega_{m,0})(1+3B\ln
x)}}.
\label{agew1}
\end{equation}

\begin{figure}[!ht]
\includegraphics[width=0.98\linewidth]{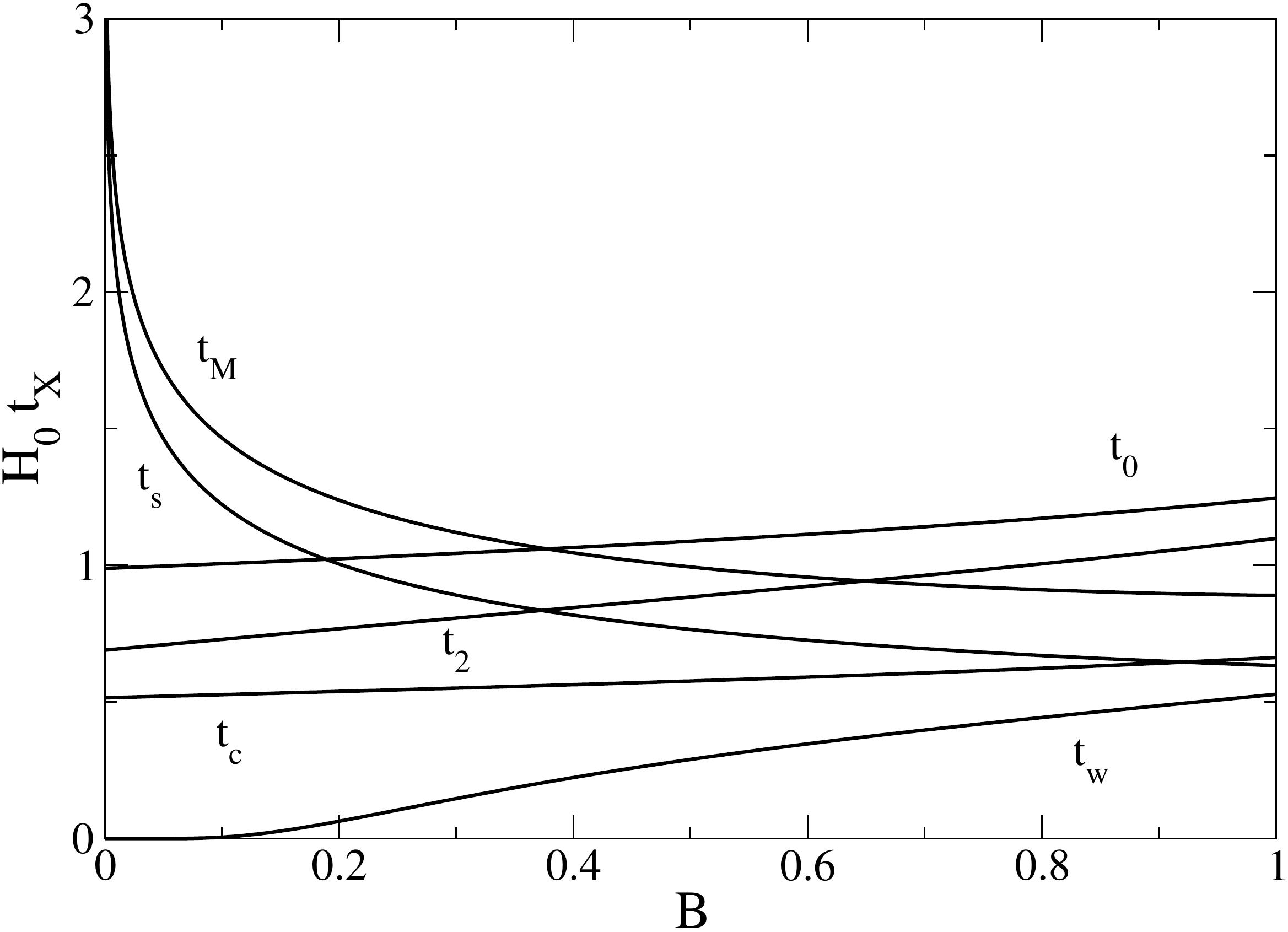}
\caption{The characteristic times of the logotropic model as a function of $B$
(see the text for details). 
\label{h0t}}
\end{figure}

For $B=0$, corresponding to the $\Lambda$CDM model, we recover Eq.
(\ref{dmde10}) leading to the standard value $t_0(0)=13.8\, {\rm Gyrs}$. More
generally, the function $t_0(B)$ is plotted in Fig. \ref{h0t}. We
also plot the functions $t_w(B)$, $t_c(B)$, $t_2(B)$, $t_s(B)$ and $t_M(B)$ 
corresponding to the time at which (i) the pressure becomes negative, (ii) the
universe accelerates, (iii) the transition from dark matter to dark energy occurs, 
(iv) the velocity of sound becomes higher than the speed
of light, (v) the universe enters in the phantom regime. They are obtained by replacing $R_0=1$ in
Eq. (\ref{es2}) by $R_w$, $R_c$, $R_2$, $R_s$ and $R_M$ given by Eqs.
(\ref{pr2}), (\ref{a1}), (\ref{trans1}), (\ref{vs7}), and (\ref{me1}). For
$B=0$, we recover the  standard values $t_c(0)=7.18\, {\rm Gyrs}$ and
$t_2(0)=9.61\, {\rm Gyrs}$ of the $\Lambda$CDM model. For $B\rightarrow 0$, we
have the 
asymptotic behavior
\begin{equation}
t_w(B)\sim \frac{2}{3H_0\sqrt{\Omega_{m,0}}}e^{-\frac{1}{2B}}\rightarrow 0.
\label{agew2}
\end{equation}

\subsection{The Hubble diagram}
\label{sec_dl}

\begin{figure}[!ht]
\includegraphics[width=0.98\linewidth]{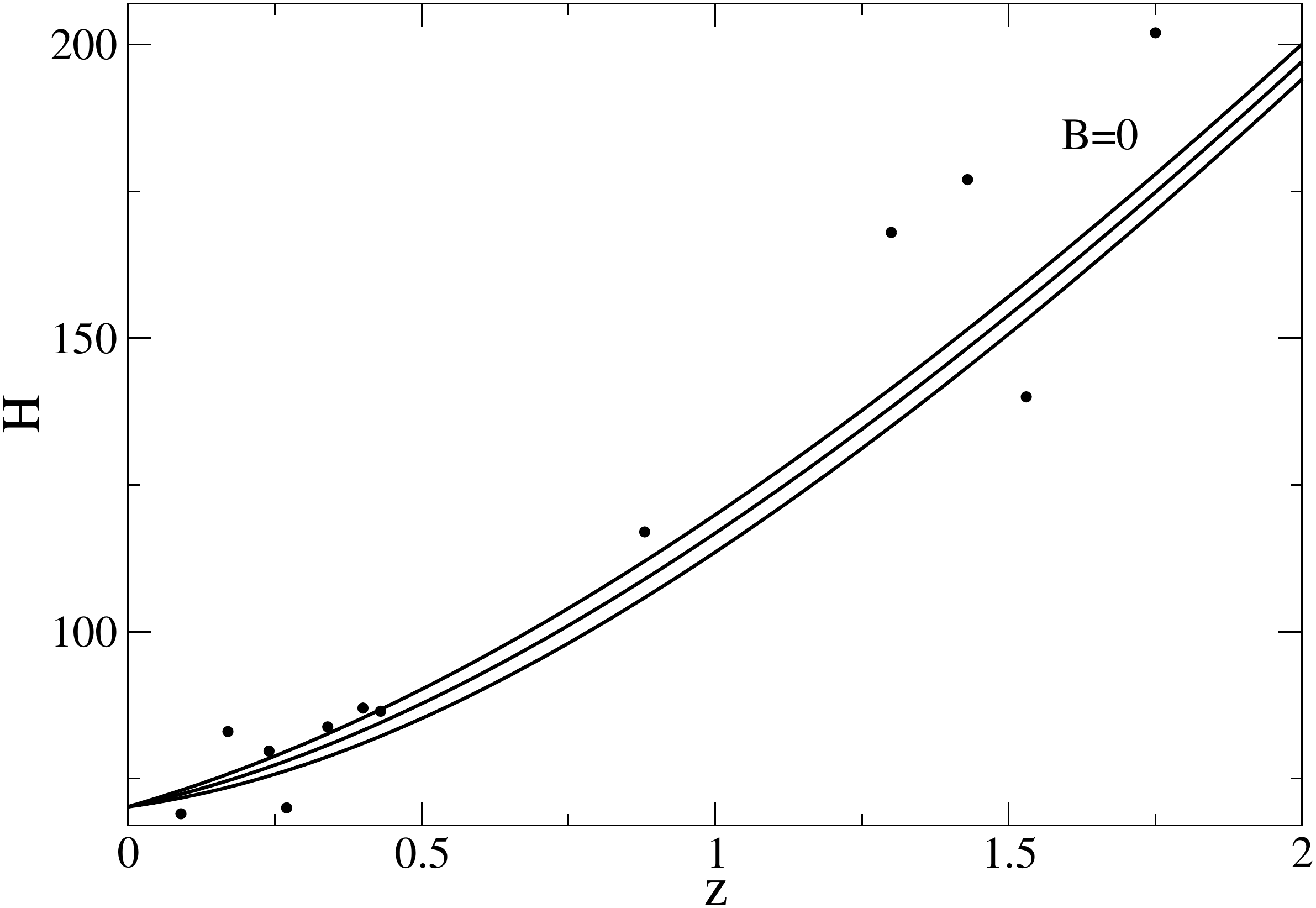}
\caption{The Hubble function $H(z)$ corresponding to the logotropic model with
$B=0$, $B=0.1$ and $B=0.2$ (top to bottom) compared with observational data.
\label{zH}}
\end{figure}

In terms of the redshift parameter
\begin{equation}
z+1=\frac{a_0}{a},
\label{dl1}
\end{equation}
the Hubble function (\ref{es1}) can be written as
\begin{equation}
H(z)=H_0
\sqrt{\Omega_{m,0}(z+1)^3+(1-\Omega_{m,0})[1-3B\ln(z+1)]}.
\label{dl2}
\end{equation}
The Hubble function $H(z)$ is plotted in Fig. \ref{zH} for different values of $B$ and compared with the observational data of Refs. \cite{Hdata1,Hdata2}.

\begin{figure}[!ht]
\includegraphics[width=0.98\linewidth]{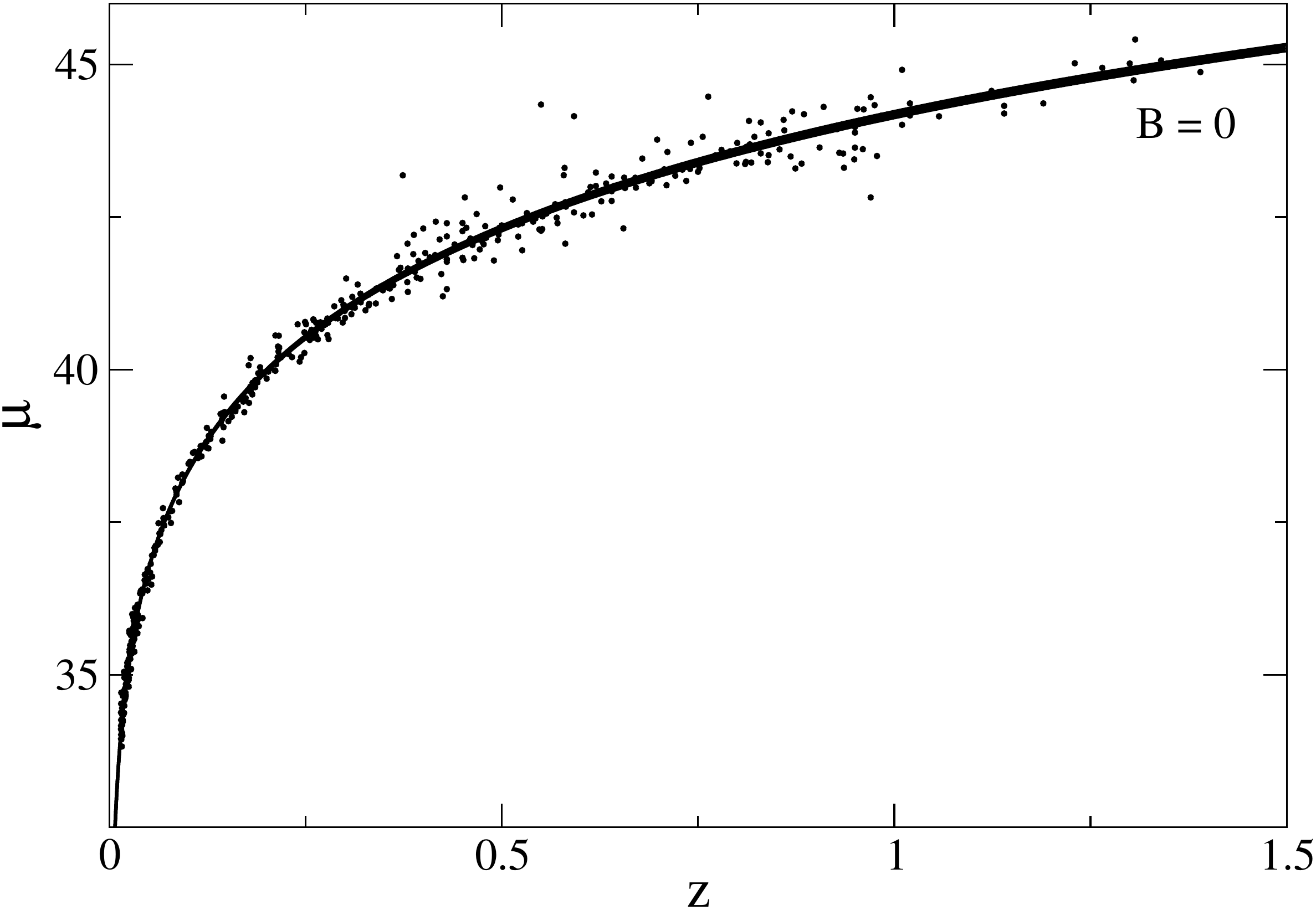}
\caption{Hubble ($\mu$ versus $z$) diagram corresponding to the logotropic model
with $B=0$, $B=0.1$ and $B=0.2$ (bottom to top) compared with observational
data. \label{zmu}}
\end{figure}

The history of the universe expansion is revealed by the Hubble diagram giving
the distance modulus $\mu(z)=m-M$ as a function of the redshift $z$, where $m$
is the apparent luminosity and $M$ the absolute luminosity of a light-emitting
source  (standard candle) such as supernovae of Type Ia. The procedure to obtain
this curve is explained in Ref. \cite{spyrou}. In a spatially flat universe, the
luminosity distance of a source with redshift $z$ is \cite{peacock}:
\begin{equation}
\frac{d_L(z)}{\rm Mpc}=c (1+z)\int_0^z \frac{dz'}{H(z')}.
\label{dl3}
\end{equation}
The $K$-corrected distance modulus of a light-emitting source is then given by
\cite{narlikar}:
\begin{eqnarray}
\mu(z)=5 \log \left \lbrack \frac{d_L(z)}{\rm Mpc}\right \rbrack+25.
\label{dl5}
\end{eqnarray}
The function $\mu(z)$ corresponding to the logotropic model 
is plotted in Fig. \ref{zmu} for
different values of $B$, and  compared with observational data of the Union 2.1
Compilation \cite{suzuki}. Even for values of $B$ as large as $0.2$, the curves
are very close
to the $\Lambda$CDM model and can hardly be distinguished. They are all consistent with the observational results.

\subsection{The shift parameter from CMB}
\label{sec_shift}

\begin{figure}[!ht]
\includegraphics[width=0.98\linewidth]{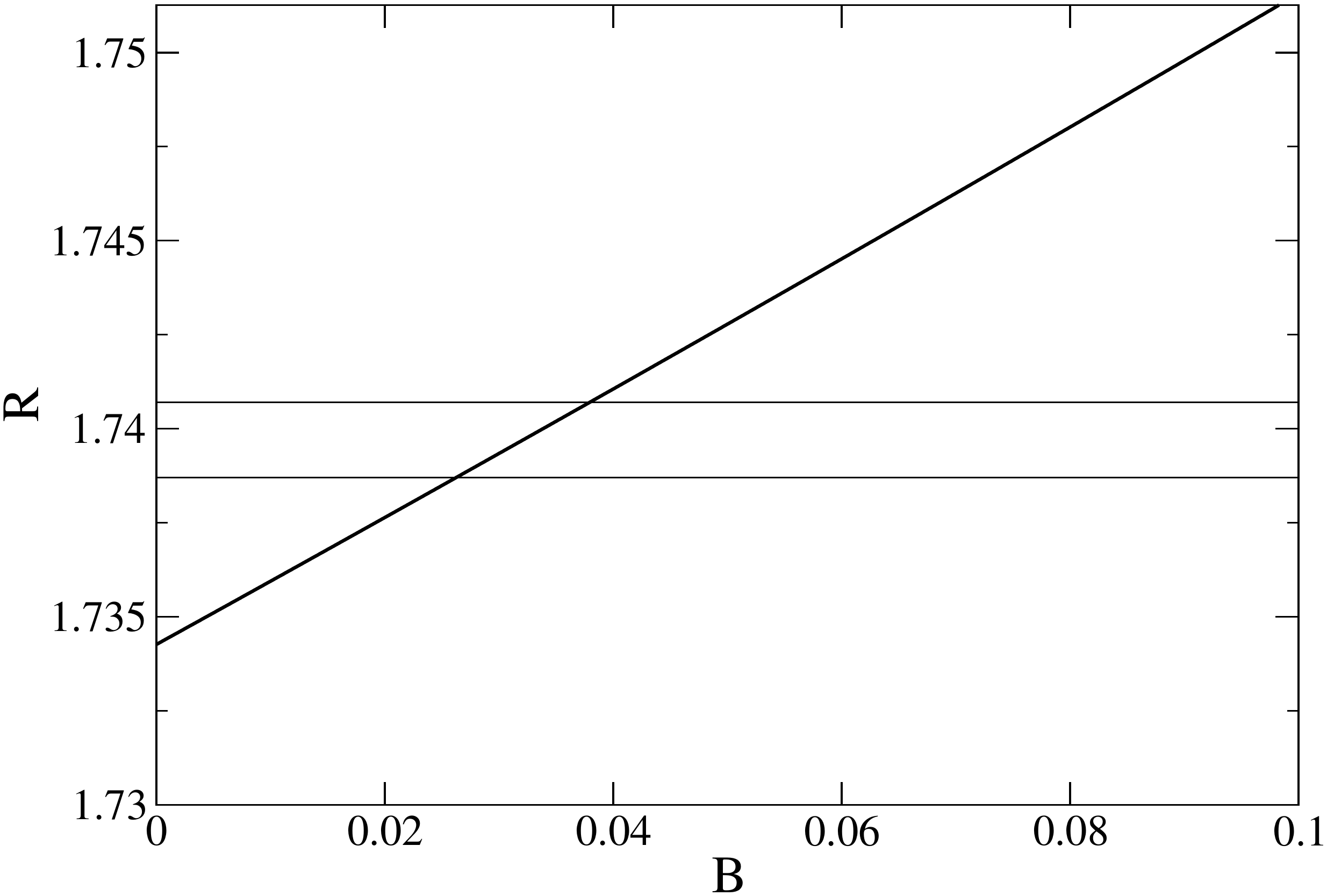}
\caption{The CMB shift parameter $R$ as a function of the logotropic temperature
$B$. The horizontal lines correspond to the upper bound $R_{\rm obs}=1.7387$ of
the $R$-value  obtained from the nine-year WMAP survey final results, and to the
value $R'_{\rm obs}=1.7407$ arising from the combination of the Planck
first-data release with those of the WMAP survey.
\label{BRR}}
\end{figure}

The shift parameter $R$, which is related to the position of the
first acoustic peak in the power spectrum of the temperature anisotropies
\cite{bond}, provides an information from the CMB \cite{spyrou}. It is defined
by
\begin{equation}
R=\sqrt{\Omega_{m,0}}\int_0^{z_*} \frac{dz}{H(z)/H_0},
\label{sp1}
\end{equation}
where $z_*$ is the value of the cosmological redshift at photon decoupling.
We adopt the nine-year WMAP survey final result
$z_*=1091.64\pm 0.47$ \cite{bennett}. The value of the shift parameter obtained
by these authors is $R_{\rm obs}=1.7329\pm 0.0058$. The $\Lambda$CDM model gives
   $R_{\Lambda {\rm CDM}}=1.7342$.  On the other hand,
using a combination of the Planck first-data release with those of the WMAP
survey \cite{ww}, the value $R'_{\rm obs}=1.7407\pm 0.0091$ is obtained.
The curve $R(B)$ obtained from the logotropic model is plotted in Fig.
\ref{BRR}. From the observational value $R_{\rm obs}$ we find the constraint
$B\le 0.0262$, and from the observational value $R'_{\rm obs}$ we find the
constraint $B\le 0.0379$. These values are about $3$ times smaller than the
bound $B_1/4=0.09425$ obtained in Sec. \ref{sec_vs}.

\section{Logotropic dark matter halos}
\label{sec_ldm}

In the previous section, we have described the 
evolution of the universe as a whole, assuming that it is made of a single dark
fluid with a logotropic equation of state. We have argued that this
equation of state describes simultaneously dark matter and dark energy. In this
section, we consider dark matter halos which form during the nonlinear regime
following the gravitational instability (Jeans instability) of the uniform
cosmological background. We consider the possibility that dark matter halos are
also described by the logotropic equation of state (\ref{ed1}), with the same
value of the logotropic temperature $A$, and we investigate the consequences of
this assumption.

For dark matter halos, we can use Newtonian gravity.
Combining the  condition of hydrostatic equilibrium
\begin{equation}
\nabla P+\rho\nabla\Phi={\bf 0}
\label{ldm1}
\end{equation}
with the Poisson equation
\begin{equation}
\Delta\Phi=4\pi G\rho,
\label{ldm2}
\end{equation}
we obtain the differential equation
\begin{equation}
\nabla\cdot \left (\frac{1}{\rho}\nabla P\right )=-4\pi G\rho.
\label{ldm3}
\end{equation}
For the logotropic equation of state (\ref{ed1}), using Eq. (\ref{ldm1}), we
find that the density is related to the gravitational potential by the
Lorentzian-type distribution
\begin{equation}
\rho({\bf r})=\frac{1}{\alpha+\frac{1}{A}\Phi({\bf r})},
\label{ldm3b}
\end{equation}
where $\alpha$ is a constant.\footnote{A self-gravitating system described by a
distribution function of the
form $f({\bf r},{\bf v})=f(\epsilon)$, where $\epsilon=v^2/2+\Phi$ is the
individual energy of the particles, has a barotropic  equation of state
$P(\rho)$ obtained by eliminating $\Phi({\bf r})$ from the relations $\rho=\int
f\, d{\bf v}=\rho(\Phi)$ and  $P=(1/3)\int f v^2\, d{\bf v}=P(\Phi)$ \cite{bt}.
For example, a system described by a polytropic distribution function has a
polytropic equation of state \cite{bt}. In a sense, the logotropic equation of
state (\ref{ed1}) is associated with a Lorentzian distribution function
\cite{logo}. However, the connection is not direct because the Lorentzian is not
normalizable in $d=3$, except if we introduce a maximum bound on the velocity.
Therefore,
rigorously speaking, the logotropic equation of state cannot be derived from a
distribution function and, consequently, $P/\rho$ does not represent a velocity
dispersion $\sigma^2$. It is more likely that the logotropic equation of state
comes from a field theory at $T=0$ as discussed in Appendix \ref{sec_ft}.} On
the other hand, Eq. (\ref{ldm3}) takes the form
\begin{equation}
\Delta \left (\frac{1}{\rho}\right )=\frac{4\pi G}{A}\rho.
\label{ldm4}
\end{equation}
Assuming spherical symmetry, and introducing the notations
\begin{equation}
\theta=\frac{\rho_0}{\rho},\qquad \xi=\frac{r}{r_0},
\label{ldm5}
\end{equation}
where $\rho_0$ is the central density and 
\begin{equation}
r_0=\left (\frac{A}{4\pi G\rho_0^2}\right )^{1/2},
\label{ldm6}
\end{equation}
is a typical core radius, we obtain
\begin{equation}
\frac{1}{\xi^2}\frac{d}{d\xi}\left (\xi^2\frac{d\theta}{d\xi}\right )=\frac{1}{\theta},
\label{ldm7}
\end{equation}
with
\begin{equation}
\theta(0)=1,\qquad \theta'(0)=0.
\label{ldm8}
\end{equation}
This equation coincides with the Lane-Emden equation of index $n=-1$ \cite{chandra}. We note that the Lane-Emden equation of index $n=-1$ cannot be obtained from a polytropic equation of state of index $n=-1$ (which corresponds to a constant pressure $P=K$) because, when $P$ is a constant, there is no pressure gradient in Eq. (\ref{ldm1}) to counterbalance gravity, and the system collapses. This is the
reason for the occurrence of density cusps in the CDM  model for which $P=0$. By
contrast, a logotropic equation of state can sustain gravity and develops flat
density cores instead of cusps. The fact that the logotropic equation of state
(\ref{ed1}) yields a Lane-Emden equation  of index $n=-1$ clearly demonstrates
that it can be interpreted as the limit of a polytropic equation of state with
$\gamma\rightarrow 0$ (i.e. $n\rightarrow -1$) and $K\rightarrow \infty$ such
that $K\gamma=A$ is finite \cite{logo}. In a sense, the logotropic equation of
state is a
``regularization'' of the polytropic equation of state with index $n=-1$.

\begin{figure}[!ht]
\includegraphics[width=0.98\linewidth]{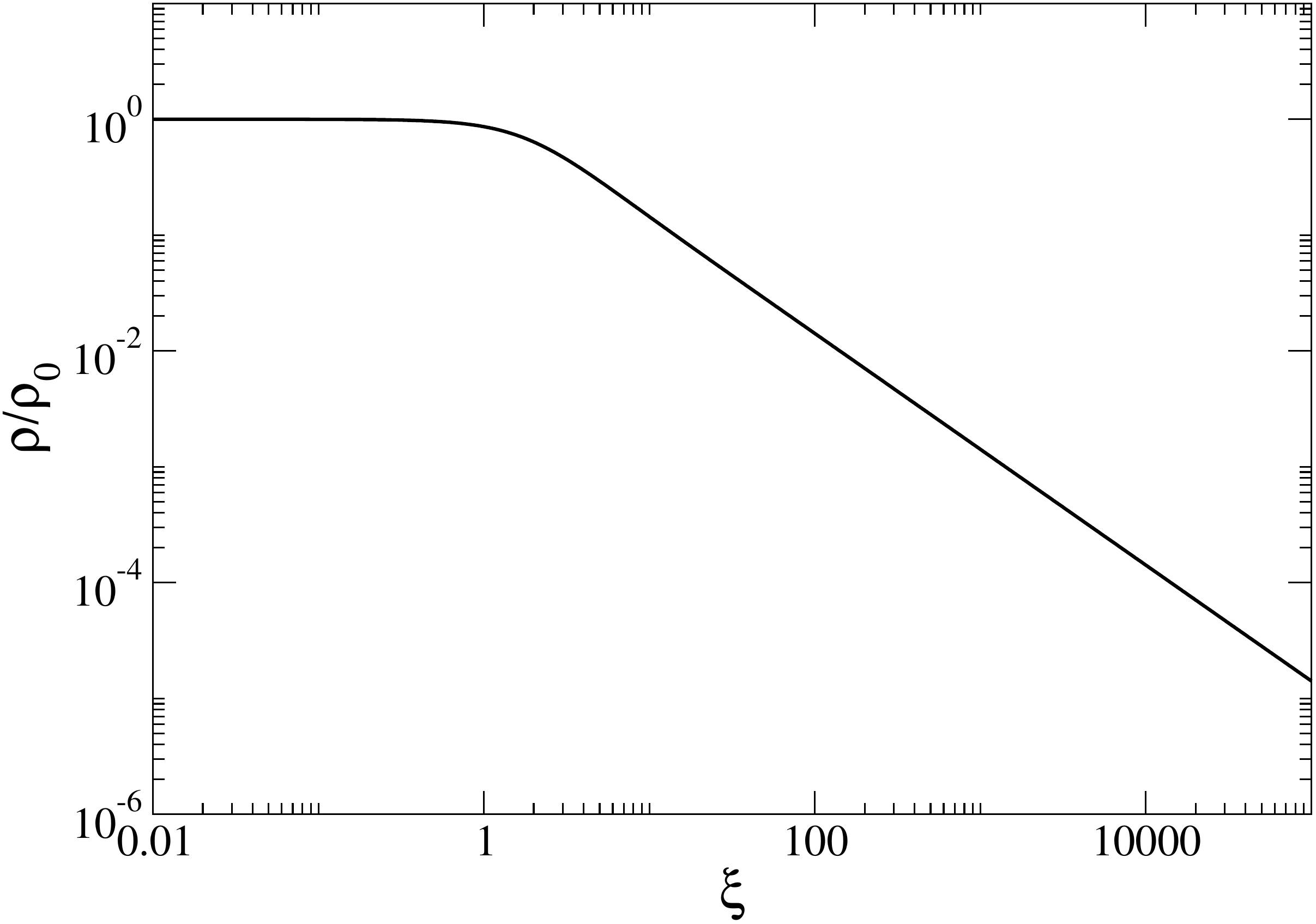}
\caption{Normalized density profile of a logotropic sphere. It decays as
$\xi^{-1}$ when $\xi\rightarrow +\infty$.
\label{xitheta}}
\end{figure}

Since the logotropic spheres are homologous \cite{logo}, they generate a {\it
univeral}
density profile. Indeed, if we rescale the density by the central density
$\rho_0$ and the radial distance by the core radius $r_0$, we get an invariant
profile $1/\theta(\xi)$. The normalized density profile of a logotropic sphere
is represented in Fig. \ref{xitheta}. The density of a logotropic sphere
decreases as $r^{-1}$ at large distances \cite{logo}. This can be compared to
the $r^{-2}$ behavior of the isothermal sphere \cite{chandra}.\footnote{The
singular logotropic sphere is
$\rho_s=(A/8\pi G)^{1/2} r^{-1}$ \cite{logo} and the singular isothermal sphere
is
$\rho_s=(1/2\pi G\beta m) r^{-2}$ \cite{chandra}.} Because of this slow decay,
the total mass
of a logotropic sphere is infinite. This implies that the logotropic equation of
state cannot hold at large distances. We note that the empirical Burkert density
profile, that provides a good fit of many dark matter halos, decreases at large
distances as $r^{-3}$ \cite{burkert}. This is substantially steeper than
the
$r^{-1}$ behavior of a logotropic halo. In practice, the finite mass and finite
radius of dark matter halos result from tidal effects or from incomplete violent
relaxation. Tidal effects and incomplete violent relaxation alter the density
profile at large distances and steepen it. A famous example of distribution
functions taking tidal effects into account is the King model \cite{king}. The
King model at the limit of microcanonical stability can be approximated by the
modified Hubble profile that has a core and that decreases at large distances as
$r^{-3}$, similarly to the Burkert profile \cite{clm}. Since tidal effects, or
other complicated effects such as incomplete relaxation, are not taken into
account in the logotropic model, we should not give too much credit on the
asymptotic behavior of its density profile at very large distances.\footnote{In
the context of
hierarchical clustering (small gravitationally bound clumps of dark matter
halos merge to form progressively larger objects), we suggest that the
core of the objects after merging remains logotropic while their halo
develops a density profile decreasing as $r^{-3}$ due to tidal effects and
incomplete relaxation.} However, it is relevant
to mention that, in a recent paper, Burkert \cite{burkertnew} observes that the
slope of the density profile of dark matter halos close to the core radius is
approximately equal to $-1$ (see the upper right panel of his Fig. 1) which is
precisely the density exponent of a logotrope. This is a first hint that a
logotropic equation of state may be relevant in the case of dark matter halos.

Using the Lane-Emden equation (\ref{ldm7}), the mass profile $M(r)=\int_0^r \rho(r')\, 4\pi {r'}^2\, dr'$ is given by
\begin{equation}
M(r)=4\pi \rho_0 r_0^3 \xi^2 \theta'(\xi).
\label{ldm9}
\end{equation}
The circular velocity defined by $v_c^2(r)={GM(r)}/{r}$ can be expressed as
\begin{equation}
v_c^2(r)=4\pi G \rho_0 r_0^2 \xi \theta'(\xi).
\label{ldm10}
\end{equation}
We define the halo radius $r_h$ as the radius at which
$\rho/\rho_0=1/4$. The dimensionless halo radius is the solution of the equation
$\theta(\xi_h)=4$. We numerically find $\xi_h=5.8458$ and
$\theta'(\xi_h)=0.69343$.
Then, $r_h=\xi_h r_0$. The normalized halo mass at the halo radius is given by
\begin{equation}
\frac{M_h}{\rho_0 r_h^3}=\frac{v_c^2(r_h)}{G \rho_0 r_h^2}=4\pi \frac{\theta'(\xi_h)}{\xi_h}=1.49.
\label{ldm11}
\end{equation}
This value is  relatively close to the value ${M_h}/{\rho_0
r_h^3}=1.60$ \cite{vss,clm} obtained with the Burkert profile defined by 
\begin{equation}
\frac{\rho(r)}{\rho_0}=\frac{1}{(1+x)(1+x^2)},\qquad x=\frac{r}{r_h},
\label{dm1}
\end{equation}
\begin{equation}
\frac{v_c(r)}{v_c(r_h)}=\frac{1.98}{\sqrt{x}}\left\lbrack \ln(1+x)-\arctan x+\frac{1}{2}\ln(1+x^2)\right \rbrack^{1/2}.
\label{dm3}
\end{equation}

\begin{figure}[!ht]
\includegraphics[width=0.98\linewidth]{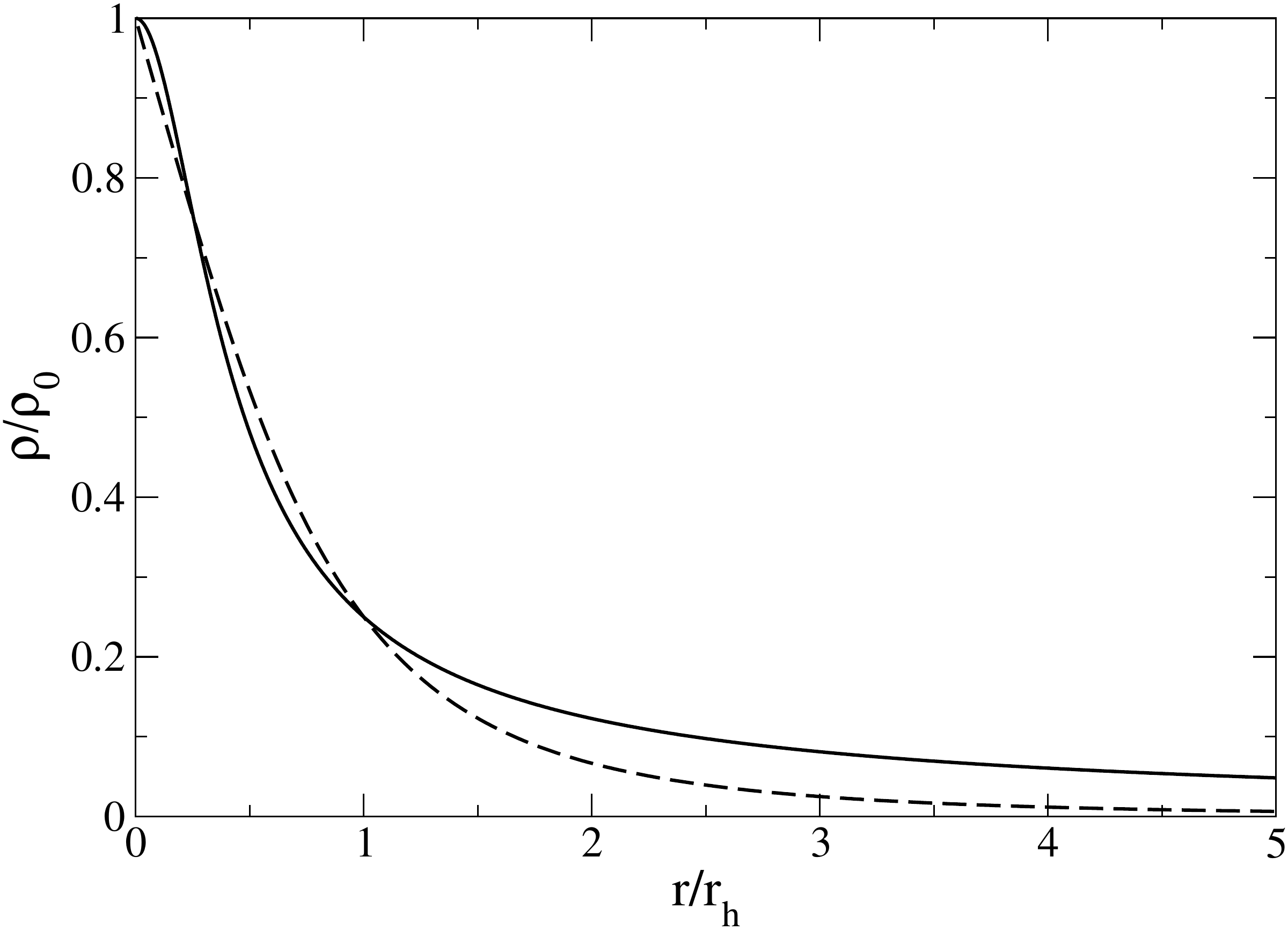}
\caption{Normalized density profile of a logotropic sphere (solid line) compared with the Burkert profile (dashed line).
\label{densLOGO}}
\end{figure}

\begin{figure}[!ht]
\includegraphics[width=0.98\linewidth]{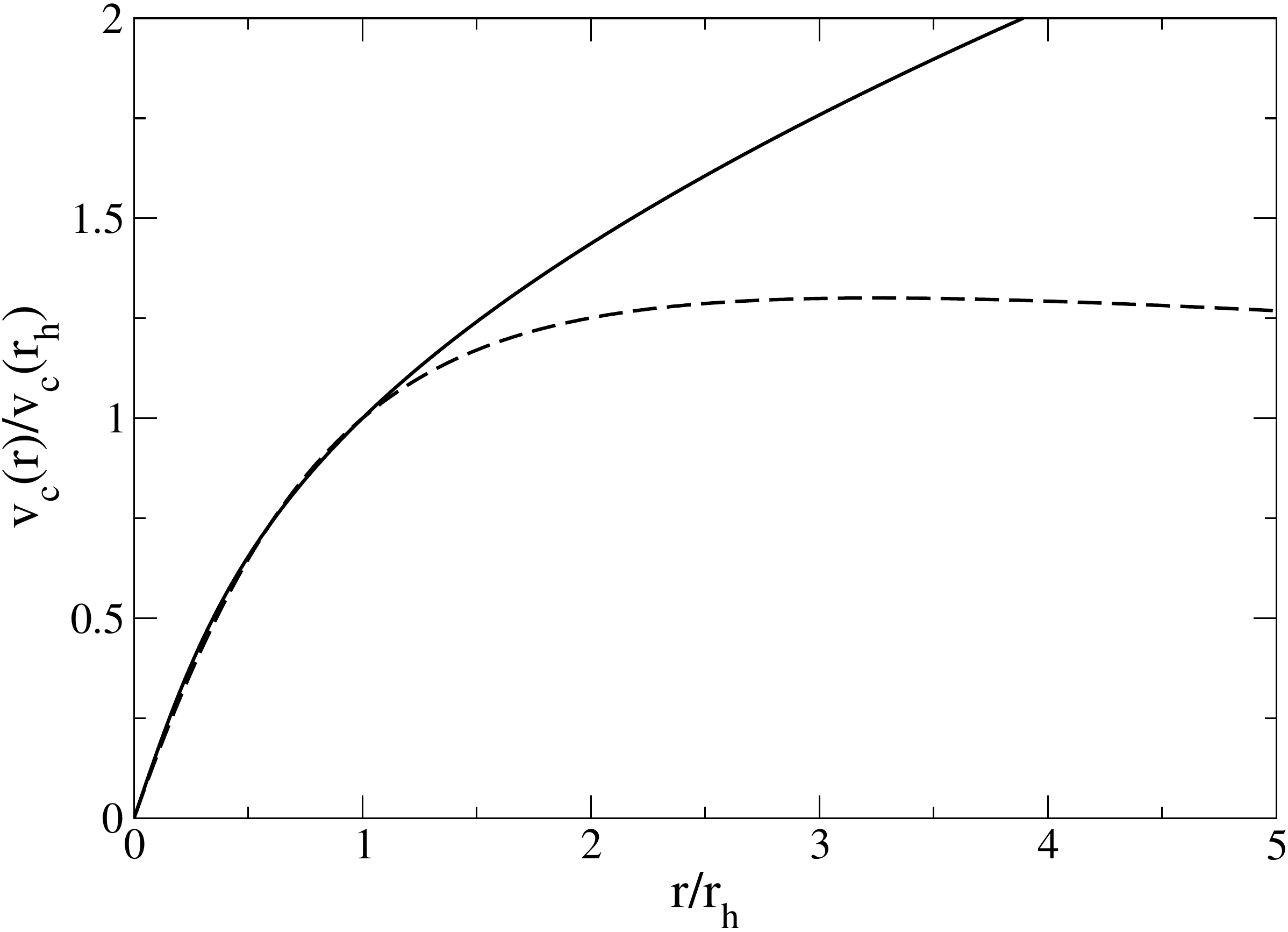}
\caption{Circular velocity profile of a logotropic sphere (solid line) compared with the Burkert profile (dashed line).
\label{vitLOGO}}
\end{figure}

The density and circular velocity profiles of a logotrope are plotted as a 
function of the radial distance normalized by the halo radius in Figs.
\ref{densLOGO} and \ref{vitLOGO}. They are compared to the observational Burkert
profiles of Eqs. (\ref{dm1}) and (\ref{dm3}).  We see that the logotropic model
gives a good agreement with the Burkert profile up to the halo radius, i.e. for
$r\le r_h$. At larger distances, the agreement is less good because, in the
logotropic model, the density decreases too slowly. This implies that the
circular velocity keeps increasing after $r_h$ while the circular velocity
obtained from the Burkert density profile reaches a maximum and decreases at
large distances. However, there exists galaxies such as LSB galaxies presenting
rotation curves in which the circular velocity increases monotonically. On the
other hand, as already indicated, the logotropic model is not expected to be
valid at large distances since it does not take tidal effects or incomplete
violent relaxation into account. Therefore, the fact that it can account for the
observations up to the halo radius is already satisfying. This is sufficient to
determine the physical characteristics of dark matter halos.

The logotropic equation of state has a very interesting property. According to Eq. (\ref{ldm6}), we have
\begin{equation}
\rho_0 r_h=\left (\frac{A}{4\pi G}\right )^{1/2}\xi_h.
\label{ldm15}
\end{equation}
Therefore, if the logotropic temperature $A$ is the same for all the halos (we
claim in Sec. \ref{sec_pred} that $A$ is a universal constant), the surface
density $\Sigma_0=\rho_0 r_h$ is also the same. This is precisely what is
observed. Indeed, it is an empirical fact that the surface density
$\Sigma_0=\rho_0 r_h$ is approximately the same for all the galaxies even if
their sizes and masses vary by several orders of magnitude (up to $14$ orders of
magnitude in luminosity). Its best-fit value is $\Sigma_0=141_{-52}^{+83}\,
M_{\odot}/{\rm pc}^2$ \cite{kormendy,spano,donato}.  To our knowledge, this
observation has not been explained theoretically. It turns out that this result
is a direct consequence of the logotropic equation of state if we assume that
the dark matter halos have the same logotropic constant $A$. This is
particularly appealing if we interpret this constant as a temperature in a
generalized thermodynamical framework (see Appendix \ref{sec_log}). In this
interpretation, the universe is ``isothermal'' except that isothermality does
not refer to a linear equation of state but to a logotropic equation of state. 

The relation (\ref{ldm15}) allows us to determine the logotropic temperature $A$ from the measurement of the surface density $\Sigma_0$. We find
\begin{equation}
A=4\pi G \left (\frac{\Sigma_0}{\xi_h}\right )^2=2.13\times 10^{-9}\, {\rm g}\, {\rm m}^{-1}{\rm s}^{-2}.
\label{ldm16}
\end{equation}
The dimensionless logotropic temperature is then given by
\begin{equation}
B=\frac{4\pi G}{\epsilon_0}\frac{1}{1-\Omega_{m,0}} \left (\frac{\Sigma_0}{\xi_h}\right )^2=3.53\times 10^{-3}.
\label{ldm17}
\end{equation}
We note that this value satisfies the cosmological bound $0\le B\le 0.09425$ obtained in Sec. \ref{sec_vs}, as well as the bounds $0\le B\le 0.0262$ and  $0\le B\le 0.0379$ obtained from the measurement of the shift parameter from CMB in Sec. \ref{sec_shift}. This agreement is particularly rewarding because the value of $B$ obtained in Eq. (\ref{ldm17}) is based on galactic observations that are completely different from the cosmological observations of Sec. \ref{sec_vs} and \ref{sec_shift}.  Therefore, a logotropic equation of state can simultaneously describe dark matter halos and account for cosmological constraints. This may be a hint that dark matter and dark energy are the manifestations of a single dark fluid.

There are interesting consequences of this result.
According to Eq. (\ref{ldm11}), the mass of the halos calculated at the halo
radius $r_h$ is given by  $M_h=1.49\Sigma_0 r_h^2$. Since the surface density of
the dark matter halos is constant, Eq. (\ref{ldm11}) implies that
$M_h/M_{\odot}=210(r_h/{\rm pc})^2\propto
r_h^2$. This scaling is consistent with the observations \cite{vss}. On the
other hand, Strigari {\it et al.} \cite{strigari} have proposed that all  dwarf
spheroidal galaxies (dSphs) of the Milky Way have the same total dark matter
mass contained within a radius of $r_u=300\, {\rm pc}$. From the observations,
they obtain $\log (M_{300}/M_{\odot})=7.0^{+0.3}_{-0.4}$. This result is still
unexplained. We now show how the logotropic equation of state immediately leads
to this result. According to Eq. (\ref{ldm9}), using $\xi_u=r_u/r_0$, we have
\begin{equation}
M_{300}=4\pi \rho_0 r_0 r_u^2 \theta'\left (\frac{r_u}{r_0}\right ).
\label{ldm18}
\end{equation}
Introducing the halo radius $r_h=\xi_h r_0$, we obtain
\begin{equation}
M_{300}=4\pi \rho_0 r_h \frac{r_u^2}{\xi_h} \theta'\left
(\xi_h\frac{r_u}{r_h}\right ).
\label{ldm19}
\end{equation}
As we have already indicated, it is an observational evidence that the surface
density $\rho_0 r_h$ of the halos is a constant.\footnote{In our model, this can
be explained theoretically by the universality of $A$ (see Sec.
\ref{sec_pred}).} In principle, the last term in Eq. (\ref{ldm19}) is
not a constant since it depends on $r_h$ which substantially changes from halo
to halo. However, for the
logotropic distribution, we have the asymptotic result $\theta(\xi)\sim
\xi/\sqrt{2}$ for $\xi\rightarrow +\infty$ \cite{logo}. More precisely,  for
$\xi\ge 8$ (typically), $\theta'(\xi)$ undergoes damped oscillations about
$1/\sqrt{2}$.
Therefore, $\theta'(\xi)$ quickly reaches a {\it constant} value $1/\sqrt{2}$.
For the dSphs considered in \cite{strigari}, $\xi_h r_u/r_h\gg 1$ (see, e.g.,
Table 2 of \cite{destri}), so Eq.
(\ref{ldm19}) may be replaced by
\begin{equation}
M_{300}= \frac{4\pi \Sigma_0 r_u^2}{\xi_h\sqrt{2}},
\label{ldm20}
\end{equation}
which is a constant in agreement with the claim of Strigari {\it et al.}
\cite{strigari}. Furthermore, using $\Sigma_0=141 M_{\odot}/{\rm pc}^2$, the
numerical application gives
\begin{equation}
M_{300}=1.93\times 10^{7}\, M_{\odot},
\label{ldm21}
\end{equation}
leading to $\log (M_{300}/M_{\odot})=7.28$ in very good agreement with the
observational value. It is also relevant to express this result directly in
terms of the logotropic temperature $A$ under the form
\begin{equation}
M_{300}=r_u^2 \left (\frac{2\pi A}{G}\right )^{1/2},
\label{ldm22}
\end{equation}
which shows that the constancy of $M_{300}$ is due to the universality of $A$.

\section{A prediction of the logotropic temperature}
\label{sec_pred}

In the previous section, the values of $A$ and $B$ have been deduced from the
observations.  In this section, we show that they can actually be predicted from
general considerations.

Introducing the cosmological density $\epsilon_{\Lambda}=\rho_{\Lambda}c^2=(1-\Omega_{m,0})\epsilon_0$, which corresponds to the present value of the dark energy term  (internal energy) in Eq. (\ref{ed7}), we can rewrite Eq. (\ref{ed11}) as
\begin{equation}
\frac{\rho_* c^2}{\epsilon_{\Lambda}}=\frac{\Omega_{m,0}}{1-\Omega_{m,0}}e^{1+1/B}.
\label{pred1}
\end{equation}
Since $B$ is expected to be very small, this ratio is huge. This makes us think
of the huge ratio $\rho_P/\rho_{\Lambda}\sim 10^{123}$ between the vacuum
energy, associated to the Planck density $\rho_P=5.16\times 10^{99}\, {\rm g}\,
{\rm m}^{-3}$, and the dark energy (or cosmological constant $\Lambda$),
associated to the cosmological density $\rho_{\Lambda}=6.72\times 10^{-24}\,
{\rm g}\, {\rm m}^{-3}$. In the framework of the $\Lambda$CDM model, this huge
ratio is at the origin of the cosmological constant problem
\cite{weinbergcosmo,paddycosmo}. In the present approach, in which there is no
cosmological constant nor dark energy, it has a completely different
significance. We propose that it determines the dimensionless logotropic
temperature $B$. Therefore, we propose to identify the density $\rho_*$ in the
logotropic equation of state (\ref{ed1}) with the Planck density $\rho_P$. In
that case, Eq. (\ref{pred1}) becomes
\begin{equation}
\frac{\rho_P}{\rho_{\Lambda}}=\frac{\Omega_{m,0}}{1-\Omega_{m,0}}e^{1+1/B}.
\label{pred2}
\end{equation}
This yields approximately $B\simeq 1/\ln(\rho_P/\rho_{\Lambda})\simeq 
1/[123\ln(10)]$, where $123$ is the famous number occurring in
the ratio $\rho_P/\rho_{\Lambda}\sim 10^{123}$. More precisely,
\begin{equation}
B=\frac{1}{\ln\left
(\frac{1-\Omega_{m,0}}{\Omega_{m,0}}\frac{\rho_P}{\rho_{\Lambda}}\right )-1}= 
3.53\times 10^{-3}.
\label{pred3}
\end{equation}
This value turns out to be in perfect agreement with the
value  obtained from the measurement of the surface density of dark matter halos
in Sec. \ref{sec_ldm}. If we accept that this agreement is not fortuitous, or
coincidental, this is a strong argument in favor of the logotropic model.
Therefore, by taking $\rho_*=\rho_P$ in the logotropic equation of state
(\ref{ed1}), we obtain a {\it prediction} of the parameter $B$ which is in
perfect agreement with the observations. With this
value, there is no free parameter in our model, so we can predict all the
measurable quantities that occur in cosmology and in the study of dark matter
halos (see Table I in Sec. \ref{sec_na}). We emphasize that our
estimate of $B$ only depends on the values of $\rho_P$, $\rho_{\Lambda}$ and
$\Omega_{m,0}$ that are known accurately from the observations.

In conclusion, the logotropic equation of state  (\ref{ed1}) can be written as
\begin{equation}
P=B\epsilon_{\Lambda}\ln\left (\frac{\rho}{\rho_P}\right ),
\label{pred4}
\end{equation}
with $\rho_P=5.16\times 10^{99}\, {\rm g}\, {\rm
m}^{-3}$, $\epsilon_{\Lambda}=(1-\Omega_{m,0})\epsilon_0=6.04\times 10^{-7} \,
{\rm g}\, {\rm m}^{-1}\, {\rm s}^{-2}$, and
$B=3.53\times 10^{-3}$. The logotropic temperature is
\begin{equation}
A=B\,\epsilon_{\Lambda}=2.13\times 10^{-9} \, {\rm g}\, {\rm m}^{-1}\, {\rm s}^{-2}.
\label{pred5}
\end{equation}
It is of the order of the cosmological density $\epsilon_{\Lambda}$ divided by
$\ln(\rho_P/\rho_{\Lambda})\simeq [123\ln(10)]$.

The $\Lambda$CDM model is recovered for
$B=0$. This corresponds to $\rho_P\rightarrow +\infty$, hence
$\hbar\rightarrow 0$. Therefore, the fact that $B$ is small but nonzero shows
that quantum mechanics plays a role in the late universe in relation to dark
energy. 

The logotropic equation of state (\ref{pred4}) has several nice properties.

(i) If our approach is correct, there is no dark matter nor dark energy. There
is just one dark fluid described by the equation of state (\ref{pred4}). In that
case, it may be relevant to interpret the logotropic temperature $A=2.13\times
10^{-9} \, {\rm g}\, {\rm m}^{-1}\, {\rm s}^{-2}$ as a fundamental constant
which supersedes the cosmological constant. We note that it depends on all the
fundamental constants of physics $\hbar$, $G$, $c$, and $\Lambda$ [see Eqs.
(\ref{pred3}) and (\ref{pred5})]. Then, Eq. (\ref{ed6}) with
$\rho_*=\rho_P$ may be used to predict $\Omega_{m,0}$ for a given value of the
present energy density $\epsilon_0$ determined from the observations  by the
Hubble constant. In the present interpretation $\Omega_{m,0}$ does not
represent the proportion of dark matter. It simply gives the proportion of the
rest-mass energy of the dark fluid as compared to its total energy (the
remaining energy being internal energy).

(ii) Since dark matter and dark energy are the manifestation of a single dark
fluid, there is no cosmic coincidence problem. On the other hand, the
cosmological constant problem $\rho_P/\rho_{\Lambda}\sim 10^{123}$ is translated
into an equation [Eq. (\ref{pred2})] that determines the logotropic temperature
$B$.

(iii) In the polytropic model (see Sec. \ref{sec_two}), the pressure is negative
because $K<0$, i.e. because the polytropic temperature is negative. In the
logotropic model, the logotropic temperature $A$ is positive, and the pressure
is negative because $\rho<\rho_P$. Positive
temperatures are more satisfying on a physical point of view than negative
temperatures. On the other hand, while the pressure $P(\rho)$ is negative, which
is required at the cosmological scale to produce an acceleration of the
expansion of the universe, its derivative $P'(\rho)$ is positive, which is
required at the scale of dark matter halos to counteract self-gravity
and avoid density cusps.

(iv) It is interesting to see that both the cosmological
density $\rho_{\Lambda}$ and the Planck density $\rho_P$ appear in the
logotropic equation of state (\ref{pred4}). The presence
of the Planck density is rather unexpected, and intriguing, because the
logotropic equation of state only
describes a relatively old universe, dominated by dark matter and dark energy
($\rho_{\Lambda}$) well
after the phase of early inflation dominated by quantum mechanics ($\rho_P$).
The fact that
the Planck density appears as a density scale in the logarithm of
Eq. (\ref{pred4}) is very
interesting because only the logarithm of the ratio of the cosmological density
over the Planck density has a  physical meaning since these quantities differ by
$123$ orders of magnitude. The occurrence of the Planck density suggests
that the logotropic equation of state (\ref{pred4}) may be the limit of a more
general equation of state connecting the very early universe ($\rho_P$) to the
very late universe  ($\rho_{\Lambda}$). In relation to this observation, we note
that the pressure in the logotropic  model vanishes precisely at the Planck
scale, so our treatment breaks down at that scale.

Now that $B$ and $A$ have been derived from theoretical considerations [see
Eqs. (\ref{pred3}) and (\ref{pred5})], we can reverse the arguments of Sec.
\ref{sec_ldm} and {\it predict} the values of $\Sigma_0$ and $M_{300}$ [see
Eqs. (\ref{ldm15}) and (\ref{ldm22})]. We obtain $\Sigma_0=141\, M_{\odot}/{\rm
pc}^2$ and $M_{300}=1.93\times 10^{7}\, M_{\odot}$ in perfect agreement with the
observations, and without any free parameter. We can also obtain an estimate of
the Jeans length $\lambda_J$ at
the beginning of the matter era where perturbations start to grow. We assume
that the matter era starts at $R_i=10^{-4}$, corresponding to the epoch of
matter-radiation equality. In this era, we can make the approximation
$\epsilon=\rho c^2$, so the Jeans wavenumber is given by \cite{weinbergbook}:
\begin{equation}
k_J^2=\frac{4\pi G\rho R^2}{c_s^2}.
\label{pred6}
\end{equation}
From Eqs. (\ref{edadd1}) and (\ref{vs8}), we find $\rho_i=2.54\times 10^{-12}
{\rm g}/{\rm m}^3$ and $(c_s^2/c^2)_i=9.33\times 10^{-15}$. This leads to a
Jeans length $\lambda_J=2\pi/k_J=1.25\times 10^{18}\, {\rm m}=40.4\, {\rm pc}$
which is of the order of magnitude of the smallest known dark matter halos such
as Willman I ($r_h=33\, {\rm pc}$), see, e.g., Table 2 of \cite{destri}. We
predict that there should
not exist halos
of smaller size since the perturbations are stable for $\lambda<\lambda_J$. This
is in agreement with the observations. By contrast, in the CDM model, since
$P=0$, the velocity of sound $c_s=0$. As a result, the Jeans length is zero
($\lambda_J=0$), implying  that the homogeneous
background is unstable at all scales so that halos of any
size should be observed in principle, which is not the case. Therefore, a small
but nonzero value of $B$, yielding a nonzero velocity of sound and a
nonzero  Jeans length, is able to account for
the minimum observed size of dark matter halos. It also puts a cut-off in
the density power spectrum of the perturbations and sharply suppresses
small-scale linear power. Finally, it solves the cusp problem and the missing
satellite problem. On the other hand, if we make the numerical application for
the present-day
universe ($R_0=1$), we find
$\rho_0=2.54\times 10^{-24} {\rm g}/{\rm m}^3$, $(c_s^2/c^2)_0=9.42\times
10^{-3}$ and $\lambda_J=1.25\times 10^{26}\, {\rm m}=4.06\, 10^9\,
{\rm pc}$. The Jeans length is so large (of the order of the horizon
$\lambda_H=c/H_0=1.32\times 10^{26}\, {\rm m}$ since $\lambda_J\sim
c/\sqrt{G\rho_0}\sim c/H_0$)  that there is no Jeans
instability in the present universe. A
detailed study of the perturbations in the logotropic model is beyond the scope
of this paper and will be considered in a future work.

\section{Numerical applications}
\label{sec_na}

In this section, we provide the values of the principal quantities that occur in
cosmology and in the study of DM halos (see Table I). We make the numerical
application for: (i)  $B=0$, corresponding to the $\Lambda$CDM
model, (ii) $B=0.09425$, corresponding to the cosmological constraint of Sec.
\ref{sec_vs}, and (iii) $B=3.53\times 10^{-3}$, corresponding to the prediction
of Sec. \ref{sec_pred}.

\begin{table*}[t]
\centering
\begin{tabular}{|c|c|c|c|c|c|c|}
\hline
 & $B=0$ ($\Lambda$CDM) & $B=0.09425$ & $B=3.53\times 10^{-3}$ \\
\hline
$R_w$ &  & $2.09\times 10^{-2}$ & $7.00\times10^{-42}$ \\
\hline
$(\epsilon/\epsilon_0)_w$ &  & $3.02\times 10^{4}$ & $7.97\times10^{122}$ \\
\hline
$t_w$ (Gyrs) &  & $0.0535$ & $3.29\times 10^{-61}$ \\
\hline
$R_c$ & $0.574$ & $0.576$ & $0.574$ \\
\hline
$(\epsilon/\epsilon_0)_c$ & $2.17$ & $2.05$ & $2.17$ \\
\hline
$t_c$ (Gyrs) & $7.18$ & $7.33$ & $7.19$ \\
\hline
$R_2$ &  $0.723$  & $0.744$ & $0.723$ \\
\hline
$(\epsilon/\epsilon_0)_2$ & $1.45$ & $1.33$ & $1.45$ \\
\hline
$t_2$ (Gyrs) & $9.61$ & $10.1$ & $9.63$ \\
\hline
$t_0$ (Gyrs) & $13.8$ & $14.0$ & $13.8$ \\
\hline
$q_0$ &  $-0.589$ & $-0.692$ & $-0.593$ \\
\hline
$w_0$ & $-0.726$ & $-0.794$ & $-0.729$ \\
\hline
$(c_s^2/c^2)_0$ & $0$ &  $0.331$ &  $9.42\, 10^{-3}$ \\
\hline
$R_s$ & & $1.26$ & $3.77$ \\
\hline
$(\epsilon/\epsilon_0)_s$ &  & $0.910$ & $0.741$ \\
\hline
$t_s$ (Gyrs) & & $17.3$ & $34.5$ \\
\hline
$R_M$ &  & $1.59$ & $4.75$ \\
\hline
$(\epsilon/\epsilon_0)_M$ &  & $0.889$ & $0.7405$ \\
\hline
$t_M$ (Gyrs) &  & $20.7$ & $38.3$ \\
\hline
$R$   & $1.7342$ & $1.7505$ &
$1.7348$ \\
\hline
$A$ (${\rm g}\, {\rm m}^{-1}\, {\rm s}^{-2}$)&  & $5.69\times 10^{-8}$ &
$2.13\times 10^{-9}$ \\
\hline
$\Sigma_0$ ($M_{\odot}/{\rm pc}^2$) &  & $1523$ & $141$ \\
\hline
$M_{300}$ ($M_{\odot}$) &  & $2.08\times 10^8$ & $1.93\times 10^7$ \\
\hline
\end{tabular}
\label{table1}
\caption{Quantities that occur in cosmology and in the study of dark matter
halos for different values of $B$ (see the text for details). Some observational
values are $R_c=0.571\pm 0.013$ \cite{suzuki}, $q_0=-0.53^{+0.17}_{-0.13}$
\cite{giostri}, and $R_2=0.719\pm 0.017$ \cite{suzuki}.}
\end{table*}

We recall that ($R_w$, $\epsilon_w$, $t_w$) refer to the values at
which the pressure of the universe becomes negative,  ($R_c$, $\epsilon_c$,
$t_c$) 
refer to the values at which the universe accelerates,  ($R_2$, $\epsilon_2$,
$t_2$) refer
to the transition between dark matter and dark energy (i.e. when the internal
energy of the dark fluid dominates its rest-mass energy), $t_0$ is the age
of the universe, $q_0$ is the
present value of the deceleration parameter, $w_0$ is the present value of the
equation of state parameter, $c_{s,0}$ is the present velocity of sound, ($R_s$,
$\epsilon_s$, $t_s$) refer to the values
at which the velocity of sound exceeds the speed of light,  ($R_M$, $\epsilon_M$, $t_M$) refer to the values
at which the universe becomes phantom, $R$ is the shift parameter from CMB, $A$
is the logotropic temperature,
$\Sigma_0$ is the surface density of DM halos, and $M_{300}$ is the mass of
dwarf halos enclosed within a sphere of radius $r_u=300\, {\rm pc}$.

We note that for the predicted value of $B=3.53\times 10^{-3}$, the cosmological
parameters are almost the same as in the $\Lambda$CDM model. This  is
satisfactory since the $\Lambda$CDM model works well at the cosmological scale.
In this sense, there is no important difference, from the observational
viewpoint, between the logotropic model ($B=3.53\times 10^{-3}$) and the
$\Lambda$CDM model ($B=0$) at the cosmological scale. Differences will appear in
the remote future, in about $20\, {\rm Gyrs}$, when the velocity of sound
approaches the speed of light and the universe tends to become phantom (the
logotropic model may break down before). However, on a theoretical point of
view, the logotropic model solves the cosmic coincidence problem (since there is
just one fluid) and avoids the cosmological constant problem (it gives an
interpretation to the number $123$ in terms of a logotropic temperature $B$). On
the
other hand, a small but nonzero value of the logotropic temperature
$B=3.53\times 10^{-3}$ is very important at the scale of dark matter halos.  
First, it yields a nonzero velocity of sound and a nonzero Jeans
length that solve the missing satellite problem and the cusp problem.
Furthermore, the value of the Jeans length at the beginning of the
matter-dominated era is consistent with the observed size of the smallest known
halos, and explains therefore the existence of a minimum halo radius $R_{\rm
min}\sim 10\, {\rm pc}$ in the universe. On the other hand, the value
$B=3.53\times 10^{-3}$ of the logotropic
temperature can account remarkably well for the structural properties of dark
matter halos. Not only our model explains why the surface density $\Sigma_0$ and
the mass $M_{300}$ of DM halos are the same for all the halos but it also
predicts their values with high accuracy (see Table I). Furthermore, it shows
that their values are related, through  $A$ and $B$, to the Planck density
$\rho_P$ and to the cosmological density $\rho_{\Lambda}$ [see
Eqs. (\ref{ldm15}), (\ref{ldm22}), (\ref{pred3}), and (\ref{pred5})].
Such a
dependence was unexpected and remains relatively mysterious. Our model
first suggests
that dark matter and dark energy
are the manifestation of a unique entity (dark fluid) since the properties of
dark matter halos such as $\Sigma_0$ and $M_{300}$  are related, through the
logotropic temperature $A$, to cosmological observables such as 
$\rho_{\Lambda}$ (dark energy). On the other hand, the fact that the Planck
density enters in the logotropic equation of state initially designed to model
dark matter and dark energy is intriguing. It suggests that quantum mechanics
manifests itself at the cosmological scale in relation to dark energy. This may
be a hint for a fundamental theory of quantum gravity. This also suggests that
the logotropic equation of state may be the limit of a more general equation of
state providing a possible unification of dark energy ($\rho_{\Lambda}$) in the
late universe and inflation (vacuum energy $\rho_{P}$) in the primordial
universe. This may be a clue for the elaboration of a more general theory
incorporating inflation.

\section{Conclusion}

We have proposed that the universe is made of a single dark fluid described by a
logotropic equation of state. In our model, there is no cosmological constant,
no dark matter, and no dark energy. As a result, the cosmological constant
problem and the cosmic coincidence problem do not arise. What we traditionally
call dark matter, dark energy, and cosmological constant can be related to the
intrinsic properties of the dark fluid. Dark matter corresponds to the rest-mass
energy of the dark fluid and dark energy corresponds to its internal energy. 
The logotropic temperature $A$ of the dark fluid can be interpreted as a
fundamental constant of physics  superseding the  cosmological constant.

We have first determined bounds on the dimensionless  logotropic temperature $B$ by using cosmological constraints:

(i) The condition that the velocity of sound is not always imaginary implies $B\ge 0$.

(ii) The condition that the present velocity of sound is real implies $B\le B_1=0.377$. This condition coincides with the condition that the present universe is non-phantom.

(iii)  The condition that the present velocity of sound is less than the speed of light implies $B\le B_1/2=0.1885$.

(iv)  The condition that the present universe is non-relativistic implies $B\le
B_1/4=0.09425$.

(v) The measurement of the shift parameter from CMB implies $B\le 0.0379$ or $B\le 0.0262$ depending on the estimates.

We have then proposed that dark matter halos are described by the same
logotropic equation of state. For $B>0$, this equation of state leads to flat
cores instead of density cusps, so it solves the cusp-core problem. 
Furthermore, it generates a universal density profile and a universal rotation
curve that agree with the observational Burkert profiles \cite{burkert} up
to the halo radius $r_h$.\footnote{The rotation curve of logotropic dark matter
halos
continues to increase (instead of decreasing) at larger distances. This is
consistent with the rotation curves of certain galaxies, such as low surface
brightness (LSB) galaxies, that are particularly well isolated. For other galaxies,
tidal effects or complex physical processes (e.g. incomplete
relaxation) have to be taken into account at large distances to confine the
system.} More specifically, and more strikingly, a logotropic equation of state
explains naturally (i) the recent observation of Burkert \cite{burkertnew} that
the density of dark matter halos decreases as $r^{-1}$ close to the core radius,
(ii) the observation that the surface density of dark matter halos is the same
for all the halos \cite{kormendy,spano,donato}, and (iii) the observation that
the mass of dark matter halos contained in a fixed radius (e.g. $300$ pc) is the
same for all the dwarf halos \cite{strigari}. Using the measured value of the
surface
density $\Sigma_0=141\, M_{\odot}/{\rm pc}^2$ \cite{kormendy,spano,donato}, we
obtain a value  $B=3.53\times 10^{-3}$ for the logotropic temperature. From this
value, we find that the mass of dark matter halos contained in a radius of 
$300$ pc is $M_{300}=1.93\times 10^{7}\, M_{\odot}$, in agreement with the value
obtained by Strigari {\it et al.} \cite{strigari}.

Finally, we have argued that the reference density $\rho_*$ appearing in the
logotropic equation of state should be identified with the Planck density
$\rho_P$. This {\it predicts} the value of the logotropic temperature to be
$B=3.53\times 10^{-3}$ in perfect agreement with the value deduced from the
observations. The corresponding value of the dimensional logotropic temperature
is $A=B\epsilon_{\Lambda}=2.13\times 10^{-9} \, {\rm g}\, {\rm m}^{-1}\, {\rm
s}^{-2}$. Such a small value of $B$ yields sensibly the same results as the
$\Lambda$CDM model ($B=0$) at the cosmological scale. However, having
$B=3.53\times 10^{-3}$ instead of $B=0$ is crucial at the scale of dark matter
halos to explain their properties, as shown above. Furthermore,
it implies a nonzero velocity of sound and a nonzero Jeans length that is, at
the
beginning of the matter era, of the order of the minimum size ($\sim 10\, {\rm
pc}$) of the observed dark matter
halos. This also solves the missing satellite problem.

The next step is to justify the logotropic equation of state from first
principles. We note that our approach is already very economical. Instead of
having dark matter and dark energy with two different equations of state, we
just have one dark fluid with one equation of state. The logotropic equation of
state involves two parameters, the reference density $\rho_*$ and the logotropic
temperature $A$. They are related to each other by Eq. (\ref{ed6}) which depends
on the known (observed) values of $\epsilon_0$ and $\Omega_{m,0}$. Assuming that
$\rho_*=\rho_P$, this equation determines $A$, so there is no free parameter in
our model. The polytropic equation of state (Sec. \ref{sec_two}) also
involves two parameters, the polytropic index $\gamma$ and the polytropic
temperature $K$. They are related to each other by an equation similar to Eq.
(\ref{ed6}). However, it seems difficult to justify why the polytropic index
should have a particular value such as $\gamma=-0.089$ 
\cite{spyrou}. On the other hand, in the logotropic model, the logotropic
temperature $A$ is positive so it can be related to an energy scale
$A=3.53\times 10^{-3}\epsilon_{\Lambda}$. In the polytropic model,
the polytropic temperature $K$
is negative so its physical interpretation  is not direct.

We have proposed different possibilities to justify the logotropic equation of state from first principles:

(i) The logotropic equation of state is related to Tsallis generalized thermodynamics \cite{tsallisbook}.
It is associated with the Log-entropy \cite{logo} (see Appendix \ref{sec_log}) which is a sort of regularized Tsallis entropy with index $q=0$ (i.e. $\gamma=0$ in the polytropic terminology). This marginal index, where power laws degenerate into a logarithm, may be viewed as a  fixed point, or as a critical point, in the framework of Tsallis generalized thermodynamics. Therefore, if our cosmological model is correct, it would be a nice confirmation of the interest of generalized thermodynamics in physics and astrophysics.

(ii) We have proposed that the dark fluid could be a relativistic
self-interacting scalar field representing, for example,  a self-interacting BEC
at $T=0$. In that context, the logotropic equation of state arises from the
GP equation with an inverted quadratic potential [Eq.
(\ref{log4})], or from the KG equation with a logarithmic potential
[Eq. (\ref{log6})]. These potentials may be simpler to justify from fundamental
physics than power-law potentials with exponents $2(\gamma-1)=-2.18$ [Eq.
(\ref{poly3})] and  $2\gamma=-0.178$  [Eq. (\ref{poly5})] associated with a
polytropic equation of state with  $\gamma=-0.089$ \cite{spyrou}.

(iii) In Sec. \ref{sec_card}, we have mentioned that our approach provides a new
justification of the Cardassian model \cite{freese}. Inversely, the original
justification of the Cardassian model, namely that the term $\nu(\rho)$ 
arising in the modified Friedmann equation (\ref{card}) may result from the
existence of extra-dimensions, could also be a way to justify the logotropic
model corresponding to $\nu(\rho)=-(8\pi GA/3c^2)\left\lbrack
\ln(\rho/\rho_P)+1\right\rbrack$. In Appendix \ref{sec_sa}, we present a simple
argument to justify the logotropic equation of state and we relate this equation
of state to a long-range confining force that could be a fifth
force or an effective description of higher dimensional physics.

In conclusion, the logotropic equation of state may be a good candidate for the
unification of dark matter and dark energy. Indeed,  it provides a good
description of the cosmological evolution of the universe as a whole and, at the
same time, accounts for many properties of dark matter halos, some of them being
until now unexplained. The logotropic equation of state is rather unique in
accounting for all these observational results in a unified manner. It is
related to generalized thermodynamics and corresponds to the Log-entropy which
is a sort of degenerate Tsallis entropy with index $q=0$. It is also
related to the Cardassian model. Finally, it can be given a field theory
interpretation as it arises from a nice simplified form of GP and KG equations
corresponding to an inverted quadratic potential or to a logarithmic potential,
respectively. All these elements suggest that dark matter
and dark energy may be the manifestation of a unique dark fluid with a
logotropic
equation of state. Some developments of this model
will be given in future works.

\appendix

\section{Relativistic thermodynamics}
\label{sec_gr}

The local form of the first law of thermodynamics can be expressed as
\begin{equation}
d\left (\frac{\epsilon}{\rho}\right )=-P d\left (\frac{1}{\rho}\right )+T d\left (\frac{s}{\rho}\right ),
\label{gr1}
\end{equation}
where $\rho=n m$ is the mass density, $n$ is the number density, and $s$ is the
entropy density in the rest frame. For a system at $T=0$, or for an adiabatic
evolution such that $d(s/\rho)=0$,\footnote{The evolution of a perfect fluid is
adiabatic. In particular, it can be explicitly checked that the Friedmann
equations conserve the entropy \cite{weinbergbook}.} the first
law of thermodynamics reduces to
\begin{equation}
d\epsilon=\frac{P+\epsilon}{\rho}d\rho.
\label{gr2}
\end{equation}
For a given equation of state, Eq. (\ref{gr2}) can be integrated to obtain the relation between the energy density $\epsilon$ and the rest-mass density.

If the equation of state is prescribed under the form  $P=P(\epsilon)$, Eq. (\ref{gr2}) can be immediately integrated into
\begin{equation}
\ln\rho=\int \frac{d\epsilon}{P(\epsilon)+\epsilon}.
\label{gr3}
\end{equation}
If, as an example, we consider the ``gamma law'' equation of
state \cite{bondi,htww}:
\begin{equation}
P=(\gamma-1)\epsilon,
\label{gr4}
\end{equation}
we get
\begin{equation}
P=K\rho^{\gamma},\qquad \epsilon=\frac{K}{\gamma-1}\rho^{\gamma},
\label{gr5}
\end{equation}
where $K$ is a constant of integration.

We now assume that the equation of state is prescribed under the form $P=P(\rho)$. In that case, Eq. (\ref{gr2}) reduces to the first order linear differential equation
\begin{equation}
\frac{d\epsilon}{d\rho}-\frac{1}{\rho}\epsilon=\frac{P(\rho)}{\rho}.
\label{gr6}
\end{equation}
Using the method of the variation of the constant, we obtain
\begin{equation}
\epsilon=A\rho c^2+\rho\int^{\rho}\frac{P(\rho')}{{\rho'}^2}\, d\rho',
\label{gr7}
\end{equation}
where $A$ is a constant of integration.

As an example, we consider the polytropic equation of state
\begin{equation}
P=K\rho^{\gamma}, \qquad \gamma=1+\frac{1}{n}.
\label{gr8}
\end{equation}
For $\gamma=1$, we get
\begin{equation}
\epsilon=A\rho c^2+K\rho\ln\rho.
\label{gr9}
\end{equation}
For $\gamma\neq 1$, we obtain
\begin{equation}
\epsilon=A\rho c^2+\frac{K}{\gamma-1}\rho^{\gamma}=A\rho c^2+nP.
\label{gr10}
\end{equation}
Taking $A=0$, we recover Eqs. (\ref{gr4}) and (\ref{gr5}). Taking $A=1$, a choice that we shall make in the following, we obtain Eqs. (\ref{two8}) and (\ref{two9}).

We first assume $n>0$ (i.e. $\gamma>1$). For $\rho\rightarrow 0$,
\begin{equation}
\epsilon\sim \rho c^2,\qquad P\sim K(\epsilon/c^2)^{\gamma},
\label{gr11b}
\end{equation}
and for  $\rho\rightarrow +\infty$,
\begin{equation}
\epsilon\sim n K \rho^{\gamma},\qquad P\sim \epsilon/n\sim (\gamma-1)\epsilon.
\label{gr11c}
\end{equation}

We now assume $n<0$ (i.e. $\gamma<1$). For $\rho\rightarrow 0$,
\begin{equation}
\epsilon\sim n K \rho^{\gamma},\qquad P\sim \epsilon/n\sim (\gamma-1)\epsilon,
\label{gr11cbis}
\end{equation}
and for  $\rho\rightarrow +\infty$,
\begin{equation}
\epsilon\sim \rho c^2,\qquad P\sim K(\epsilon/c^2)^{\gamma}.
\label{gr11bbis}
\end{equation}

For a general equation of state $P(\rho)$, we
obtain
\begin{equation}
\epsilon=\rho c^2+\rho\int^{\rho}\frac{P(\rho')}{{\rho'}^2}\, d\rho'=\rho
c^2+u(\rho),
\label{gr12}
\end{equation}
where the primitive is determined such that $u(\rho)$ does not contain terms in
$\rho c^2$. We note that $u(\rho)$ may be interpreted as an internal energy
density (see
\cite{aaantonov} and Appendix \ref{sec_ef}). Therefore, the energy density
$\epsilon$ is the sum of the rest-mass energy $\rho c^2$ and the internal
energy $u(\rho)$. The rest-mass energy is positive while the internal energy
can be positive or negative. Of course, the total energy $\epsilon=\rho
c^2+u(\rho)$ is always positive. 

{\it Remark:} according to  Eq. (\ref{gr2}), the velocity of sound defined
by $c_s^2/c^2=P'(\epsilon)$ satisfies the identity
\begin{equation}
\frac{c_s^2}{c^2}=\frac{\rho
\frac{d^2\epsilon}{d\rho^2}}{\frac{d\epsilon}{d\rho}}.
\label{gr13}
\end{equation}

\section{A simple argument to justify the logotropic equation of state and its
relation to a fifth force}
\label{sec_sa}

We have seen in Sec. \ref{sec_two} that the energy density $\epsilon$ of a
relativistic 
dark fluid at $T=0$ (or a perfect fluid)  is the sum of two terms: a rest-mass
term $\rho c^2$
mimicking dark matter and an internal energy term 
\begin{equation}
u(\rho)=\rho\int^{\rho}\frac{P(\rho')}{{\rho'}^2}\, d\rho'
\label{sa1}
\end{equation}
mimicking a ``new fluid''. If we want the new
fluid to mimick dark energy, we should have $P=-u$. Substituting the
relation $u(\rho)=-P(\rho)$ in Eq. (\ref{sa1}) and solving the resulting
differential equation, we find $P={\rm cst.}$ which returns the model of Sec.
\ref{sec_cp}, equivalent to the $\Lambda$CDM model. However, we have explained
that a constant pressure does not account for the structure of dark matter
halos. Therefore, a next guess is $P=-u-A$ where $A$ is a constant. Substituting
the relation $u(\rho)=-P(\rho)-A$ in Eq. (\ref{sa1}) and solving the resulting
differential equation, we find the logotropic equation of state
$P=A\ln(\rho/\rho_*)$. We note that $P\sim -u$ as $\rho\rightarrow 0$, similarly
to the equation of state of dark energy. Another possible guess is $P=u/n$ with
$n\rightarrow -1$. Substituting the relation
$u(\rho)=n P(\rho)$ in Eq. (\ref{sa1}) and solving the resulting
differential equation, we find the polytropic equation of state
$P=K\rho^{1+1/n}$. For $n=-1$, we obtain $P=K$. The polytropic  equation of
state with an arbitrary index $n$ is
interesting at a  general level (beyond dark energy). For example, the index
$n=1$, leading to a quadratic equation of state $P=K\rho^2$, corresponds to a
stiff fluid ($P=u$) possibly existing in the very early universe \cite{stiff}.

The pressure $P$ can be associated with a self-interaction potential between
particles. If we consider a power-law interaction of the form 
\begin{equation}
U(r_{ij})={U_0}r_{ij}^{\alpha},
\label{sa2}
\end{equation}
corresponding to a force $F(r_{ij})=-\alpha{U_0}r_{ij}^{\alpha-1}$, the
virial theorem writes (see, e.g., Appendix I of \cite{neq2}):
\begin{equation}
2K-\alpha W=3PV,
\label{sa3}
\end{equation}
where $K$ is the kinetic energy, $W$ is the potential energy, $P$ is the
pressure, and $V$ is the
volume of the system. Taking $K=0$
for a dark fluid at $T=0$ and identifying $W/V$ as the internal energy $u$, we
obtain (see an alternative derivation of this result in the Appendix of
\cite{freese}): 
\begin{equation}
P=-\frac{1}{3}\alpha u.
\label{sa4}
\end{equation}
The equation of state $P=-u$ of dark energy corresponds to $\alpha=3$ yielding
$U=U_0 r_{ij}^3$ and $F=-3U_0 r_{ij}^2$. 
This is a long-range confining force that could be a fifth force or an effective
description of higher dimensional physics \cite{freese}. The
equation of state
$P=u/n$
associated with a polytrope of index $n$ corresponds to $\alpha=-3/n$ yielding
$U=U_0 r_{ij}^{-3/n}$ and $F=(3/n)U_0 r_{ij}^{-3/n-1}$. In particular, the
equation of state of radiation $P=u/3$ associated with a polytrope of index
$n=3$ corresponds to $\alpha=-1$ yielding
$U=U_0 r_{ij}^{-1}$ and $F=U_0 r_{ij}^{-2}$ (Coulombian force).
The stiff equation of state $P=u$ associated with a polytrope of index
$n=1$ corresponds to $\alpha=-3$ yielding
$U=U_0 r_{ij}^{-3}$ and $F=3U_0 r_{ij}^{-4}$ (attractive
force).\footnote{Actually, in Zel'dovich model \cite{zeldovich}, the stiff
equation of state arises from a screened Coulombian potential (see Sec. 4.2.
of \cite{aabh}).} We note that
the dark energy equation of state and the stiff equation of state correspond to
a polytropic index $n=-1$ and $n=+1$ respectively, and to 
a potential with an exponent $+3$ and $-3$. This symmetry may be
related to the
comment following Eq. (\ref{log4}).

For a logarithmic potential of interaction of the form 
\begin{equation}
U(r_{ij})={{\tilde U}_0}\ln r_{ij},
\label{sa5}
\end{equation}
the virial theorem writes (see, e.g., Appendix I of \cite{neq2}):
\begin{equation}
2K-\frac{{\tilde U}_0 N^2}{2}=3PV,
\label{sa6}
\end{equation}
where $N$ is the number of particles in the volume $V$. Therefore, it is
tempting to associate the constant $A$ (logotropic
temperature) in the equation of state  $P=-u-A$ as arising from a logarithmic
potential of interaction of the form of Eq. (\ref{sa5}) such that
\begin{equation}
A=\frac{{\tilde U}_0N^2}{6V}.
\label{sa7}
\end{equation}
The total potential of interaction is therefore $U=U_0 r_{ij}^3+{{\tilde
U}_0}\ln r_{ij}$
and the total force is $F=-3U_0 r_{ij}^2-{{\tilde U}_0}/r_{ij}$. The meaning of
Eq. (\ref{sa7}) is not clear, but it is interesting to note that $A>0$ for an
attractive potential ${\tilde U}_0>0$ (expected for dark energy), which is the
correct sign of the logotropic temperature.

\section{Field theory}
\label{sec_ft}

\subsection{The Klein-Gordon equation}
\label{sec_kg}

The Klein-Gordon (KG) equation for a complex scalar field $\phi$ writes
\begin{equation}
\frac{1}{c^2}\frac{\partial^2\phi}{\partial t^2}-\Delta\phi+\frac{m^2c^2}{\hbar^2}\left (1+\frac{2\Phi}{c^2}\right )\phi+2\frac{dV}{d|\phi|^2}\phi=0,
\label{kg1}
\end{equation}
where  $V=V(|\phi|^2)$ is the self-interaction potential and $\Phi$ is an
external potential that, in a simplified model, can be identified with the
gravitational potential (see \cite{abrilph} for a fully general relativistic
treatment). The energy density and the pressure of the scalar field are given by
\begin{equation}
\epsilon=\frac{1}{2c^2}\left |\frac{\partial\phi}{\partial t}\right
|^2+\frac{1}{2}\left |\nabla\phi\right
|^2+V(|\phi|^2)+\frac{m^2c^2}{2\hbar^2}|\phi|^2,
\label{kg2}
\end{equation}
\begin{equation}
P=\frac{1}{2c^2}\left |\frac{\partial\phi}{\partial t}\right
|^2+\frac{1}{2}\left |\nabla\phi\right
|^2-V(|\phi|^2)-\frac{m^2c^2}{2\hbar^2}|\phi|^2.
\label{kg3}
\end{equation}
The KG equation without self-interaction can be viewed as the
relativistic generalization of the Schr\"odinger equation. Similarly, the
KG equation with a self-interaction can be viewed as the relativistic
generalization of the GP equation. In order to recover the
Schr\"odinger and GP equations in the nonrelativistic limit
$c\rightarrow +\infty$, we make the transformation
\begin{equation}
\phi({\bf r},t)=A e^{-im c^2 t/\hbar}\psi({\bf r},t),
\label{kg4}
\end{equation}
where $A$ is a constant. This constant can be determined by the following argument. Substituting Eq. (\ref{kg4}) in the expression of the energy density (\ref{kg2}), we get
\begin{eqnarray}
\epsilon=\frac{A^2}{2c^2}\left |\frac{\partial\psi}{\partial t}\right |^2+\frac{A^2m^2c^2}{\hbar^2}|\psi|^2+\frac{A^2 m}{\hbar}{\rm Im}\left (\psi\frac{\partial\psi^*}{\partial t}\right )\nonumber\\
+\frac{A^2}{2}\left |\nabla\psi\right
|^2+V(|\psi|^2).\qquad
\label{kg5}
\end{eqnarray}
In the nonrelativistic limit $c\rightarrow +\infty$, we have $\epsilon\sim\rho c^2$ where $\rho$ is the rest-mass density. On the other hand, according to Eq. (\ref{kg5}),
\begin{eqnarray}
\frac{\epsilon}{c^2}\rightarrow \frac{A^2m^2}{\hbar^2}|\psi|^2.
\label{kg6}
\end{eqnarray}
If we interpret $\psi$ as the wavefunction normalized such that $|\psi|^2=\rho$,
we find by identification that
\begin{eqnarray}
A=\frac{\hbar}{m}.
\label{kg7}
\end{eqnarray}
Therefore, Eq. (\ref{kg4}) can be rewritten as
\begin{equation}
\phi({\bf r},t)=\frac{\hbar}{m}e^{-im c^2 t/\hbar}\psi({\bf r},t).
\label{kg8}
\end{equation}
Mathematically, we can always make this change of variables. However, we
emphasize that it is only in the nonrelativistic limit $c\rightarrow +\infty$
that $\psi$ has the interpretation of a wave function, and that $|\psi|^2=\rho$
has the interpretation of a rest-mass density. In the relativistic regime,
$\psi$ and $\rho=|\psi|^2$ do not have a clear physical interpretation. We will
call them ``pseudo wave function'' and ``pseudo rest-mass density''.
Nevertheless, it is perfectly legitimate to work with these
variables \cite{abrilph}.

Substituting Eq. (\ref{kg8})  in the KG equation (\ref{kg1}), we obtain
\begin{equation}
\frac{\hbar^2}{2m c^2}\frac{\partial^2\psi}{\partial t^2}-i\hbar \frac{\partial\psi}{\partial t}-\frac{\hbar^2}{2m}\Delta\psi+m\Phi\psi+m\frac{dV}{d|\psi|^2}\psi=0.
\label{kg9}
\end{equation}
On the other hand, the energy density and the pressure can be written in terms of $\psi$ as
\begin{eqnarray}
\epsilon=\frac{\hbar^2}{2m^2c^2}\left |\frac{\partial\psi}{\partial t}\right |^2+|\psi|^2 c^2+\frac{\hbar}{m}{\rm Im}\left (\psi\frac{\partial\psi^*}{\partial t}\right )\nonumber\\
+\frac{\hbar^2}{2m^2}\left |\nabla\psi\right
|^2+V(|\psi|^2),
\label{kg10}
\end{eqnarray}
\begin{eqnarray}
P=\frac{\hbar^2}{2m^2c^2}\left |\frac{\partial\psi}{\partial t}\right |^2+\frac{\hbar}{m}{\rm Im}\left (\psi\frac{\partial\psi^*}{\partial t}\right )\nonumber\\
+\frac{\hbar^2}{2m^2}\left |\nabla\psi\right
|^2-V(|\psi|^2).
\label{kg11}
\end{eqnarray}

\subsection{The Gross-Pitaevskii equation}
\label{sec_gp}

Taking the nonrelativistic limit $c\rightarrow +\infty$ of the KG equation (\ref{kg9}), we obtain the nonlinear Schr\"odinger equation
\begin{equation}
i\hbar \frac{\partial\psi}{\partial t}=-\frac{\hbar^2}{2m}\Delta\psi+m\Phi\psi+m\frac{dV}{d|\psi|^2}\psi.
\label{gp1}
\end{equation}
It can be written as a GP equation of the form
\begin{equation}
i\hbar \frac{\partial\psi}{\partial t}=-\frac{\hbar^2}{2m}\Delta\psi+m\Phi\psi+m h(|\psi|^2)\psi
\label{gp2}
\end{equation}
with a potential
\begin{equation}
h(|\psi|^2)=\frac{dV}{d|\psi|^2}\qquad {\rm i.e.}\qquad h(\rho)=V'(\rho).
\label{gp3}
\end{equation}
The GP equation describes a  BEC at
$T=0$. The GP equation with a cubic nonlinearity (corresponding to
$h(\rho)\propto \rho$) can be derived from the mean field Schr\"odinger equation
with a pair contact potential (see Sec. II.A. of  \cite{gpp1}). The present
approach shows that the GP equation with an arbitrary potential $h(\rho)$ can be
derived from the KG equation with a self-interaction  potential $V(\rho)$. More
precisely, Eq. (\ref{gp3}) shows that the potential $h(\rho)$ in the GP equation
is equal to the derivative of the potential $V(\rho)$ in the KG equation.
Reciprocally, $V(\rho)$ is a primitive of $h(\rho)$. The primitive of $h(\rho)$
played some role in our previous studies \cite{gpp1}, and  was noted $H(\rho)$.
When the GP equation is derived from the KG equation, we have
\begin{equation}
H(\rho)=V(\rho).
\label{gp4}
\end{equation}
The KG equation can be written in terms of $h$ as
\begin{equation}
\frac{1}{c^2}\frac{\partial^2\phi}{\partial t^2}-\Delta\phi+\frac{m^2c^2}{\hbar^2}\left (1+\frac{2\Phi}{c^2}\right )\phi+2\frac{m^2}{\hbar^2}h\left (\frac{m^2}{\hbar^2}|\phi|^2\right )\phi=0.
\label{gp5}
\end{equation}
We also have
\begin{equation}
V(|\phi|^2)=H\left (\frac{m^2}{\hbar^2}|\phi|^2\right ).
\label{gp6}
\end{equation}

\subsection{The Madelung transformation}
\label{sec_mad}

Using the Madelung \cite{madelung} transformation
\begin{eqnarray}
\label{mad1}
\psi=\sqrt{\rho} e^{iS/\hbar},\qquad {\bf u}=\frac{1}{m}\nabla S,
\end{eqnarray}
where $\rho({\bf r},t)$ is the density, $S({\bf r},t)$ is an action, and ${\bf u}({\bf r},t)$ is interpreted as an irrotational velocity field, we can rewrite the GP equation (\ref{gp2}) in the form of hydrodynamic equations (see, e.g., \cite{gpp1}):
\begin{eqnarray}
\label{mad2}
\frac{\partial\rho}{\partial t}+\nabla\cdot (\rho {\bf u})=0,
\end{eqnarray}
\begin{eqnarray}
\label{mad3}
\frac{\partial {\bf u}}{\partial t}+({\bf u}\cdot \nabla){\bf u}=-\frac{1}{\rho}\nabla P-\nabla\Phi-\frac{1}{m}\nabla Q,
\end{eqnarray}
where
\begin{eqnarray}
\label{mad4}
Q=-\frac{\hbar^2}{2m}\frac{\Delta\sqrt{\rho}}{\sqrt{\rho}}
\end{eqnarray}
is the Bohm quantum potential and $P$ is the pressure. The pressure is  given by a barotropic equation of state $P=P(\rho)$ determined by the potential $h(\rho)$ according to
\begin{eqnarray}
\label{mad5}
h'(\rho)=\frac{P'(\rho)}{\rho}.
\end{eqnarray}
Since $\nabla h=(1/\rho)\nabla P$, the potential $h$ in the GP equation has the
interpretation of an enthalpy in the quantum Euler equation
(\ref{mad3}). We note that the pressure is explicitly given by
\begin{eqnarray}
\label{mad6}
P(\rho)=\rho h(\rho)-H(\rho)=\rho V'(\rho)-V(\rho).
\end{eqnarray}

\subsection{Energy functional}
\label{sec_ef}

The barotropic quantum Euler equations (\ref{mad2}) and (\ref{mad3}) conserve
the mass and the free energy functional\footnote{As a result,  a
minimum of free
energy $F$ at fixed mass $M$ is a stable stationary state of the GP and quantum
Euler equations. It is determined by the Euler-Lagrange equation
$\delta F-\mu\delta M=0$, where $\mu$ is a Lagrange multiplier (chemical
potential), leading to $m\Phi+mh(\rho)+Q=m\mu$, which is equivalent to the
condition of hydrostatic equilibrium $\nabla P+\rho\nabla\Phi+(\rho/m)\nabla
Q={\bf 0}$ \cite{gpp1}.}
\begin{eqnarray}
\label{ef1}
F=E+U,
\end{eqnarray}
where $E=\Theta_c+\Theta_Q+W$ is the sum of the classical kinetic energy $\Theta_c$, the quantum kinetic energy $\Theta_Q$ and the potential energy $W$, and $U$ is the internal energy. They are defined by (see, e.g., \cite{gpp1}):
\begin{eqnarray}
\label{ef2}
\Theta_c=\int\rho\frac{{\bf u}^2}{2}\, d{\bf r},\qquad \Theta_Q=\frac{1}{m}\int\rho Q\, d{\bf r},
\end{eqnarray}
\begin{eqnarray}
\label{ef3}
W=\int\rho\Phi\, d{\bf r},\qquad U=\int\rho \int^{\rho}\frac{P(\rho')}{{\rho'}^2}\, d\rho' d{\bf r}.
\end{eqnarray}
The internal energy can be written as
\begin{eqnarray}
\label{ef4}
U=\int u(\rho) \, d{\bf r},
\end{eqnarray}
where
\begin{eqnarray}
\label{ef4b}
u(\rho)=\rho \int^{\rho}\frac{P(\rho')}{{\rho'}^2}\, d\rho'
\end{eqnarray}
is the internal energy density.

The internal energy density $u(\rho)$ defined by Eq. (\ref{ef4b}) corresponds  precisely to the term that appears in the energy density of Eq. (\ref{gr12}) in addition to the rest-mass density $\rho c^2$. As we have indicated in Sec. \ref{sec_two}, depending on the pressure law $P(\rho)$, this term mimics a ``new fluid'' that adds to ``dark matter''. This new fluid may be an exotic constituent  (e.g. a stiff fluid)  appearing in the early universe \cite{stiff}. It  may also  represent the  ``dark energy'' in the late universe (see Sec. \ref{sec_two}). Using the present formalism, we can make a connection between this new fluid and the potential that appears in the GP and KG equations. Integrating Eq. (\ref{ef4b}) by parts and using Eq. (\ref{mad5}), the internal energy density may be rewritten as
\begin{eqnarray}
\label{ef5}
u(\rho)=\rho h(\rho)-P(\rho).
\end{eqnarray}
According to Eqs. (\ref{gp4}), (\ref{mad6}) and (\ref{ef5}), we obtain
\begin{eqnarray}
\label{ef6}
u(\rho)=H(\rho)=V(\rho),
\end{eqnarray}
or, equivalently,
\begin{eqnarray}
\label{ef7}
u'(\rho)=H'(\rho)=V'(\rho)=h(\rho).
\end{eqnarray}
Therefore, the internal energy density $u(\rho)$, which mimics a ``new fluid''
in Eq. (\ref{gr12}),  is equal to the potential $V(\rho)$ appearing in the KG
equation.\footnote{Actually, this equivalence is valid only in the
nonrelativistic limit $c\rightarrow +\infty$. The relativistic regime is more
complicated to investigate (since $\rho$ is not the rest-mass energy) and will
be considered specifically in a future communication.}

{\it Remark:} We note that $H(\rho)$ and $V(\rho)$ are defined up to a term of
the form $a\rho+b$. Therefore, we can adapt
the coefficients $a$ and $b$ in order to have the simplest expressions of $H$
and $V$. We will use this prescription in the following section.

\subsection{Particular examples}
\label{sec_pe}

The internal energy density  $u(\rho)$  is determined by the equation of state  $P(\rho)$. As we have seen, this equation of state can be obtained from a field theory based on the KG or GP equation. In this section, we consider particular examples of equations of state.

\subsubsection{Isothermal equation of state}
\label{sec_iso}

We consider the isothermal equation of state \cite{chandra}:
\begin{eqnarray}
\label{iso1}
P=\rho \frac{k_B T}{m}.
\end{eqnarray}
In the present context, this equation of state is expected to arise from a self-interaction potential, not from thermal motion. As a result, the temperature  has to be regarded as an effective temperature which can be either positive or negative. The internal energy is
\begin{eqnarray}
\label{iso2b}
u=\frac{k_B T}{m} \rho\ln\rho.
\end{eqnarray}
The potential in the GP equation is
\begin{eqnarray}
\label{iso2}
h=\frac{k_B T}{m}\ln\rho,\qquad H=\frac{k_B T}{m} \rho\ln\rho.
\end{eqnarray}
The GP equation takes the form
\begin{equation}
i\hbar \frac{\partial\psi}{\partial t}=-\frac{\hbar^2}{2m}\Delta\psi+m\Phi\psi+2k_B T \ln|\psi| \psi.
\label{iso3}
\end{equation}
The free energy can be written as $F=E-TS$ where $E$ is the energy, $T$ is the temperature and $S$ is the Boltzmann entropy
\begin{equation}
S=-k_B\int \frac{\rho}{m}\ln\frac{\rho}{m}\, d{\bf r}.
\label{iso4}
\end{equation}
The potential in the KG equation (\ref{kg1}) is
\begin{equation}
V(|\phi|)=2\frac{m k_B T}{\hbar^2}|\phi|^2 \ln|\phi|,
\label{iso5}
\end{equation}
so the KG equation writes
\begin{eqnarray}
\frac{1}{c^2}\frac{\partial^2\phi}{\partial
t^2}-\Delta\phi+\frac{m^2c^2}{\hbar^2}\left (1+\frac{2\Phi}{c^2}\right
)\phi\nonumber\\
+\frac{4 m k_B T}{\hbar^2}\ln\left (\frac{m|\phi|}{\hbar}\right
)\phi=0.
\end{eqnarray}

\subsubsection{Polytropic equation of state}
\label{sec_poly}

We consider the polytropic equation of state \cite{chandra}:
\begin{eqnarray}
\label{poly1}
P=K\rho^{\gamma},\qquad \gamma=1+\frac{1}{n},
\end{eqnarray}
where the polytropic constant $K$ may be positive or negative (for the same
reason as before), and
the
polytropic index $\gamma$ is arbitrary. The internal energy is
\begin{eqnarray}
\label{poly2b}
u=\frac{K}{\gamma-1} \rho^{\gamma}=\frac{P}{\gamma-1}.
\end{eqnarray}
Since $P=(\gamma-1)u$, this component in Eq. (\ref{two2}) can mimic dark
energy when $\gamma\rightarrow 0$ (see Sec. \ref{sec_two}). For $\gamma=0$, the
pressure is constant ($P=K$) and we recover the equation of state  $P=-u$ of
dark energy. The potential in the GP equation is
\begin{eqnarray}
\label{poly2}
h=\frac{K\gamma}{\gamma-1}\rho^{\gamma-1},\qquad H=\frac{K}{\gamma-1} \rho^{\gamma}.
\end{eqnarray}
The GP equation takes the form
\begin{equation}
i\hbar \frac{\partial\psi}{\partial t}=-\frac{\hbar^2}{2m}\Delta\psi+m\Phi\psi+K(n+1)m|\psi|^{2/n}\psi.
\label{poly3}
\end{equation}
The free energy can be written as $F=E-K S_{\gamma}$ where $E$ is the energy, $K$ is the polytropic temperature and  $S_{\gamma}$ is the Tsallis entropy \cite{tsallisbook}:
\begin{equation}
S_{\gamma}=-\frac{1}{\gamma-1}\int (\rho^{\gamma}-\rho)\, d{\bf r}.
\label{poly4}
\end{equation}
The potential in the KG equation (\ref{kg1}) is
\begin{equation}
V(|\phi|)=\frac{K}{\gamma-1}\left (\frac{m}{\hbar}\right
)^{2\gamma}|\phi|^{2\gamma},
\label{poly5}
\end{equation}
so the KG equation writes
\begin{eqnarray}
\frac{1}{c^2}\frac{\partial^2\phi}{\partial
t^2}-\Delta\phi+\frac{m^2c^2}{\hbar^2}\left (1+\frac{2\Phi}{c^2}\right
)\phi\nonumber\\
+\frac{2 m^2 K \gamma}{(\gamma-1)\hbar^2}\left (\frac{m|\phi|}{\hbar}\right
)^{2(\gamma-1)}\phi=0.
\end{eqnarray}

The usual GP equation writes \cite{revuebec}:
\begin{equation}
i\hbar \frac{\partial\psi}{\partial t}=-\frac{\hbar^2}{2m}\Delta\psi+m\Phi\psi+ \frac{4\pi a_s\hbar^2}{m^2}|\psi|^2\psi.
\label{poly8}
\end{equation}
It describes a BEC at $T=0$ with a quadratic self-interaction ($h\propto |\psi|^2$). The GP equation (\ref{poly8}) can be derived from the mean field Schr\"odinger equation with a pair contact potential (see Sec. II.A. of  \cite{gpp1}). In that case, $a_s$ represents the scattering length of the bosons. It is positive when the self-interaction is repulsive and negative when the self-interaction is attractive. The potential in the GP equation (\ref{poly8}) is
\begin{eqnarray}
\label{poly6}
h=\frac{4\pi a_s\hbar^2}{m^3}\rho,\qquad H=\frac{2\pi a_s\hbar^2}{m^3}\rho^2.
\end{eqnarray}
The  pressure is
\begin{eqnarray}
\label{poly7}
P=\frac{2\pi a_s\hbar^2}{m^3}\rho^2.
\end{eqnarray}
It corresponds to a polytrope of index $n=1$ (i.e. $\gamma=2$) and polytropic constant $K={2\pi a_s\hbar^2}/{m^3}$. The internal energy is
\begin{eqnarray}
\label{poly6b}
u=K\rho^2=P.
\end{eqnarray}
Since $P=u$, this component in Eq. (\ref{two2})  can mimic stiff matter
\cite{stiff}. Historically, the stiff equation of state was
introduced by Zel'dovich
\cite{zeldocosmo,zeldovich} in the
context of baryon stars in which the baryons interact through a vector meson
field. The free energy can
be written as $F=E-KS_{2}$ where $E$ is the energy, $K$ is the polytropic
temperature and $S_{2}$ is the Tsallis entropy of index $\gamma=2$:
\begin{equation}
S_2=-\int \rho^{2}\, d{\bf r}.
\label{poly9}
\end{equation}
The potential in the KG equation  (\ref{kg1}) is 
\begin{equation}
V(|\phi|)=\frac{2\pi a_s m}{\hbar^2}|\phi|^4,
\label{poly10}
\end{equation}
so the KG equation writes
\begin{eqnarray}
\frac{1}{c^2}\frac{\partial^2\phi}{\partial
t^2}-\Delta\phi+\frac{m^2c^2}{\hbar^2}\left (1+\frac{2\Phi}{c^2}\right
)\phi\nonumber\\
+\frac{8\pi a_s m}{\hbar^2}|\phi|^2\phi=0.
\end{eqnarray}

\subsubsection{Logotropic equation of state}
\label{sec_log}

We consider the logotropic equation of state \cite{pud,logo}:
\begin{eqnarray}
\label{log1}
P=A\ln(\rho/\rho_*),
\end{eqnarray}
where the logotropic constant $A$ may be positive or negative. The logotropic
equation of state (\ref{log1})  can be viewed as the limiting form of the
polytropic equation of state (\ref{poly1}) when
$\gamma\rightarrow 0$ ($n\rightarrow -1$) and $K\rightarrow \infty$ in such a way  that
$A=K\gamma$ is finite \cite{logo}. The internal energy is
\begin{eqnarray}
\label{log2}
u=-A\ln(\rho/\rho_*)-A=-P-A.
\end{eqnarray}
Since $P=-u-A$, this component in Eq. (\ref{two2}) can mimic  dark energy.
The potential in the GP equation is
\begin{eqnarray}
\label{log3}
h=-\frac{A}{\rho},\qquad H=-A\ln(\rho/\rho_*)-A.
\end{eqnarray}
The GP equation takes the form
\begin{equation}
i\hbar \frac{\partial\psi}{\partial t}=-\frac{\hbar^2}{2m}\Delta\psi+m\Phi\psi-Am\frac{1}{|\psi|^2}\psi.
\label{log4}
\end{equation}
Eq. (\ref{log4}) can be viewed as a GP equation with an inverted quadratic potential, i.e. with the exponent $-2$ instead of $+2$ in the usual GP equation (\ref{poly8}). To our knowledge, this equation has not been introduced before. This equation can be obtained as the limiting form of Eq.
(\ref{poly3}) when $\gamma\rightarrow 0$  ($n\rightarrow -1$) and $K\rightarrow \infty$ with
$A=K\gamma$ finite. However, it cannot be obtained from a polytropic equation of
state with $\gamma=0$ (which corresponds to a constant pressure $P=K$) because a
constant pressure yields $h=0$ (or $h={\rm cst.}$) in the GP equation, which is
different from Eq. (\ref{log4}). This shows the ``regularizing'' property of the
logotropic equation of state when $\gamma\rightarrow 0$, as in the case of the
Lane-Emden equation (see Sec. \ref{sec_ldm}). The free energy can be written as
$F_L=E-AS_{L}$ where $E$ is the energy, $A$ is the logotropic temperature and 
$S_{L}$ is the log-entropy \cite{logo}:
\begin{equation}
S_{L}=\int \ln\rho\, d{\bf r}.
\label{log5}
\end{equation}
The potential in the KG equation (\ref{kg1}) is
\begin{equation}
V(|\phi|)=-2A\ln|\phi|,
\label{log6}
\end{equation}
so the KG equation writes
\begin{eqnarray}
\frac{1}{c^2}\frac{\partial^2\phi}{\partial
t^2}-\Delta\phi+\frac{m^2c^2}{\hbar^2}\left (1+\frac{2\Phi}{c^2}\right
)\phi-\frac{2A}{|\phi|^2}\phi=0.\nonumber\\
\end{eqnarray}

\end{document}